\documentclass[aps,twocolumn,preprintnumbers,amsmath,amssymb,superscriptaddress]{revtex4-1} 

\usepackage[bottom]{footmisc}

\usepackage{adjustbox}			 
\usepackage{dsfont} 			
\usepackage{physics}
\usepackage{tikz}
\usepackage{braket}
\usepackage{graphicx}
\usepackage{dcolumn}
\usepackage{color}
\usepackage{amsmath}
\usepackage{placeins}
\usepackage{ulem} 
\usepackage[breaklinks,colorlinks,bookmarks=false,citecolor=blue,linkcolor=red,urlcolor=blue]{hyperref}

\begin{document}

\title{Quantum robustness of fracton phases}

\author{M.~M\"uhlhauser}
\email{matthias.muehlhauser@fau.de}
\affiliation{Institute for Theoretical Physics, FAU Erlangen-N\"urnberg, Germany}

\author{M.~R.~Walther}
\email{matthias.walther@fau.de}
\affiliation{Institute for Theoretical Physics, FAU Erlangen-N\"urnberg, Germany}

\author{D.~A.~Reiss}
\email{david.reiss@fu-berlin.de}
\affiliation{Dahlem Center for Complex Quantum Systems and Physics Department, Freie Universit\"at Berlin, Germany}

\author{K.~P.~Schmidt}
\email{kai.phillip.schmidt@fau.de}
\affiliation{Institute for Theoretical Physics, FAU Erlangen-N\"urnberg, Germany}

\date{\today}

\begin{abstract}
  The quantum robustness of fracton phases is investigated by studying the influence of quantum fluctuations on the X-Cube model and Haah's code, which realize a type-I and type-II fracton phase, respectively. To this end a finite uniform magnetic field is applied to induce quantum fluctuations in the fracton phase resulting in zero-temperature phase transitions between fracton phases and polarized phases. Using high-order series expansions and a variational approach, all phase transitions are classified as strongly first order, which turns out to be a consequence of the (partial) immobility of fracton excitations. Indeed, single fractons as well as few-fracton composites can hardly lower their excitation energy by delocalization due to the intriguing properties of fracton phases, as demonstrated in this work explicitly in terms of fracton quasi-particles.   
\end{abstract}

\maketitle

\section{Introduction}
\label{sect::intro}
Intrinsic topological order is a prominent and attractive theme in modern physics due to its fascinating physical properties \cite{Wen_1989,Wen_1990,Wen_2004}. Indeed, quantum phases possessing topological order display long-range entangled ground states and a ground-state degeneracy depending on the real-space topology. Elementary excitations in two dimensions are so-called anyons \cite{Leinaas_1977,Wilczek_1982} having a generalized particle statistics distinct from conventional bosons and fermions. Although anyonic point particles are forbidden in three spatial dimensions according to the spin-statistics theorem, the concept of fractional statistics can be generalized to spatially extended objects like membrane excitations in 3D \cite{Hamma_2005,Nussinov_2008,Reiss_2019}. 

The quantum robustness against local decoherence renders intrinsic topological order attractive for future quantum technologies. Anyonic excitations are at the heart of topological quantum computation \cite{Kitaev_2003,Nayak_2008}, while Kitaev has suggested to use the topologically protected ground-state degeneracy for quantum memories \cite{Kitaev_2003}. Unfortunately, two-dimensional topological stabilizer codes like Kitaev's toric code are fragile against thermal fluctuations and it is therefore mandatory to consider topological quantum memories in (at least) three spatial dimensions \cite{Alicki_2009,Castelnovo_2007,Nussinov_2009}. The physical origin of the thermal fragility in two-dimensional stabilizer codes is the finite energy barrier between different topological ground states, e.g., the toric code becomes thermally stable only for dimensions larger three \cite{Bravyi_2009}.
The recently discovered fracton phases reach beyond such topological codes \cite{Chamon_2005,Bravyi_2011,Haah_2011,Bravyi_2013,Yoshida_2013,Vijay_2015,Vijay_2016,Slagle_2017,Slagle17,Ma_2017,Pretko_2017a,Pretko_2017b,Petrova_2017,Halasz_2017,Prem_2019,Williamson_2019,Song_2019,Schmitz_2018,Pai_2019,Yan_2019,Shirley_2019,Fuji_2019,Chamon_2019} and have interesting crosslinks to other domains in physics like elasticity \cite{Pretko_2018,Gromov_2019,Pai_2018,Kumar_2019,Pretko_2019,Radzihovsky_2019}, localization \cite{Chamon_2005,Prem_2017,Pai_2019b}, gravity \cite{Pretko_2017c,Yan_2019b,Yan_2019c}, Majorana fermions \cite{You_2019,Wang_2019}, and deconfined quantum criticality \cite{Ma_2018}. 

The three-dimensional fracton codes display fully localized elementary excitations, so-called fractons, in a translationally invariant system and a sub-extensive ground-state degeneracy as a function of system size. Fracton order can be grouped in two types. In \mbox{type-I} fracton phases, topologically non-trivial composites of fractons are able to move in lower-dimensional subspaces of the system in real space, while for type-II fracton phases any topologically non-trivial assembly of fractons remains localized under any local perturbations. The most prominent representatives of fracton order are the X-Cube model (type-I) \cite{Castelnovo_2010,Vijay_2016} and Haah's code (type-II) \cite{Haah_2011}. However, the type-I fracton order in the X-Cube model is still thermally fragile due to the existence of string-like logical operators resulting in a finite energy barrier between ground states \cite{Bravyi_2009,Haah_2011}. This is different for Haah's code, where any  string-like logical operator is absent. As a consequence, the energy barrier becomes macroscopic scaling logarithmically with the system size leading to a partial protection against thermal fluctuations \cite{Bravyi_2013}.

Apart from understanding the role of thermal fluctuations on fracton codes, it is important and interesting to investigate the breakdown of fracton topological order at zero temperature. For the more conventional topological models like the toric code, color codes, or string-net models very rich phase diagrams with quantum critical behavior have been found when adding perturbations like an external magnetic field \cite{Trebst_2007,Vidal_2009,Tupitsyn_2010,Dusuel_2011,Schulz_2012,Jahromi_2013a,Schulz_2013,Schmidt_2013,Jahromi_2013b,Schulz_2014,Dusuel_2015,Schuler_2016,Reiss_2019,Schotte_2019}. This is widely unexplored for three-dimensional fracton order, which is the main motivation for this work. We want to understand the impact of the intriguing properties of fracton excitations on the quantum critical breakdown of fracton order and we concentrate on potential differences between type-I and type-II fracton order. On general grounds one would expect that type-I fracton phases are more likely to display second-order quantum phase transitions compared to type-II fracton phases due to the enhanced mobility of composite fracton excitations. Here we study the quantum robustness of the X-Cube model and Haah's code by adding an external uniform magnetic field. The application of high-order series expansions about the low- and high-field limit allows to quantitatively locate the phase transitions between the fracton and polarized phases. All the phase transitions are found to be strongly first order, which is confirmed by variational calculations and is in agreement with quantum Monte Carlo (QMC) simulations for the X-Cube in a field \cite{Devakul_2018}. The first-order nature of the phase transitions can be understood in terms of the (partial) immobility of fracton excitations, which is reflected in the properties of fracton quasi-particles.   

The paper is organized as follows: In Sect.~\ref{sect::models} the physical properties of the X-Cube model and Haah's code are discussed. Then both fracton codes in a magnetic field and their exact dualities are investigated in Sect.~\ref{sect::FF}. Technical details about the methods are presented in Sect.~\ref{sect::methods}. The ground-state phase diagrams of the fracton models in a uniform magnetic field are discussed in Sect.~\ref{sect::pd}, while the properties of fracton quasi-particles are explored in Sect.~\ref{sect::fractons}. A conclusion follows in Sect.~\ref{sect::conclusion}.

\section{Fracton models}
\label{sect::models}
In this section we review the known physical properties of Haah's code and the X-Cube model. These are then used to study the effects of additional magnetic fields which are the central focus of this work investigated in the later sections.
\subsection{Haah's code}
\label{ssect::haah}
\begin{figure}
	\centering
	\begin{minipage}{0.31\columnwidth}
		\adjincludegraphics[width = \textwidth, trim={ {.2\width} {.15\height} {.25\width} {.13\height}}, clip]{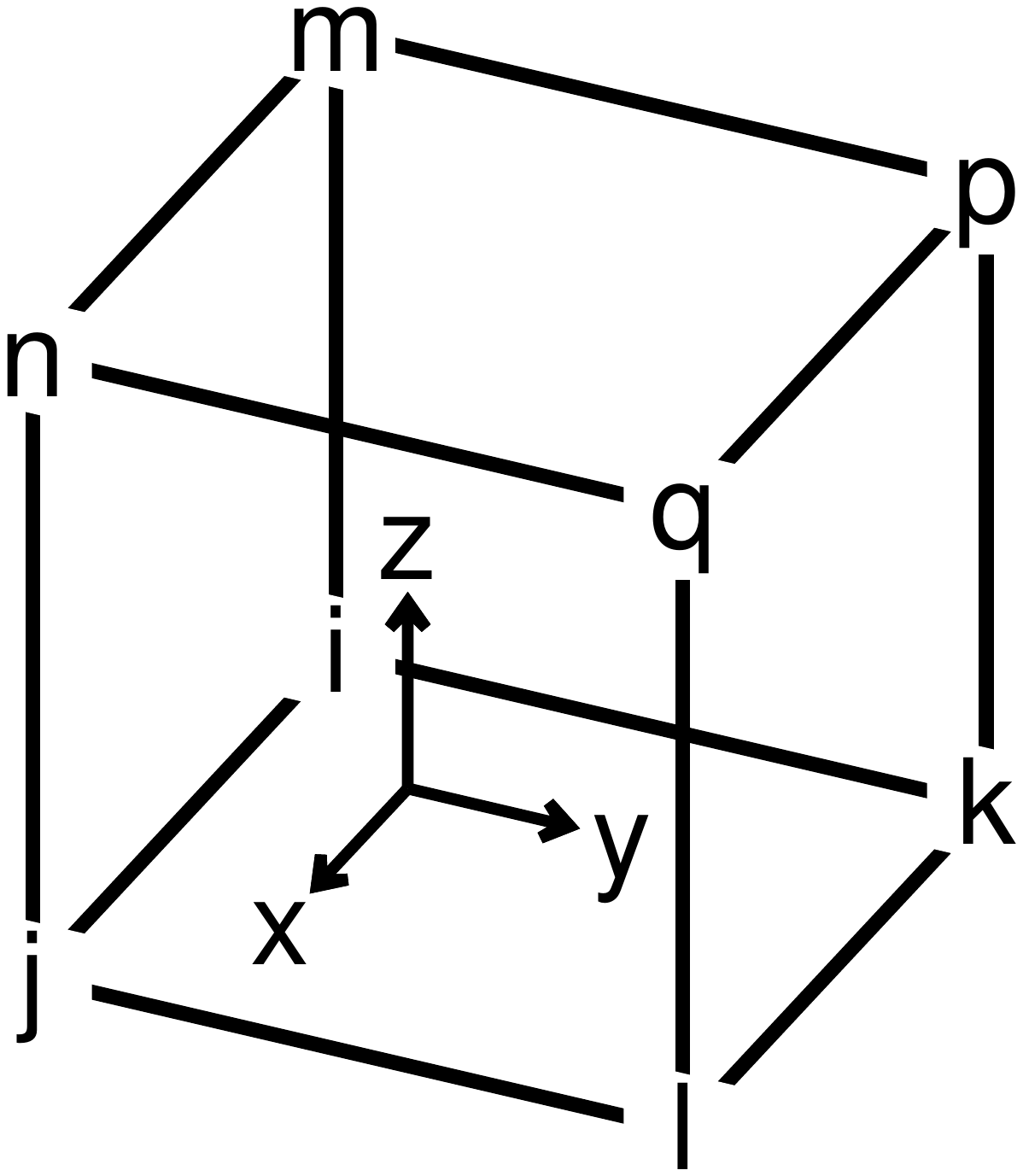}
	\end{minipage}
	\hfill
	\begin{minipage}{0.31\columnwidth}
		\adjincludegraphics[width = \textwidth, trim={ {.2\width} {.15\height} {.25\width} {.13\height}}, clip]{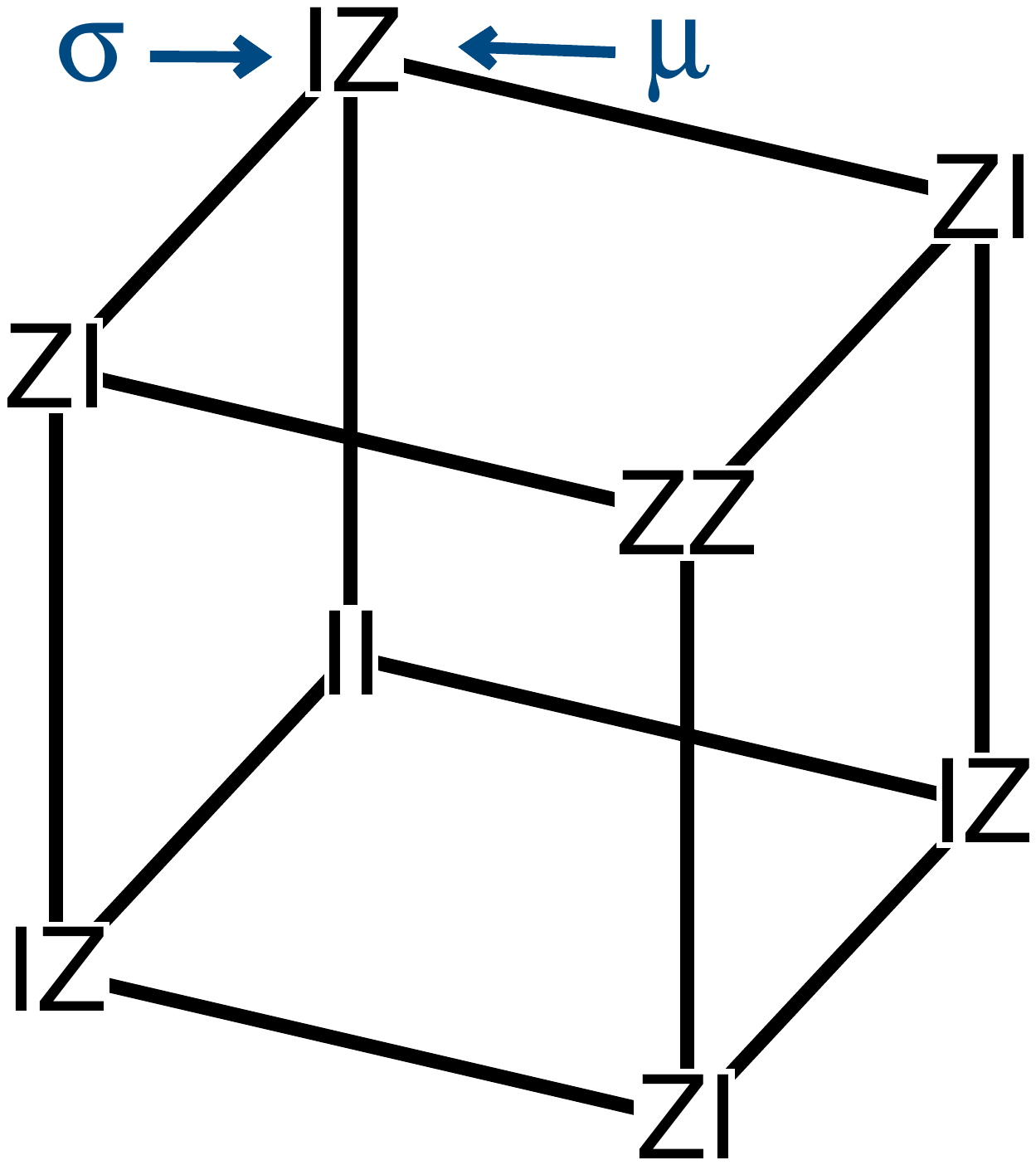}
	\end{minipage}
	\hfill
	\begin{minipage}{0.31\columnwidth}
		\adjincludegraphics[width = \textwidth, trim={ {.2\width} {.15\height} {.25\width} {.13\height}}, clip]{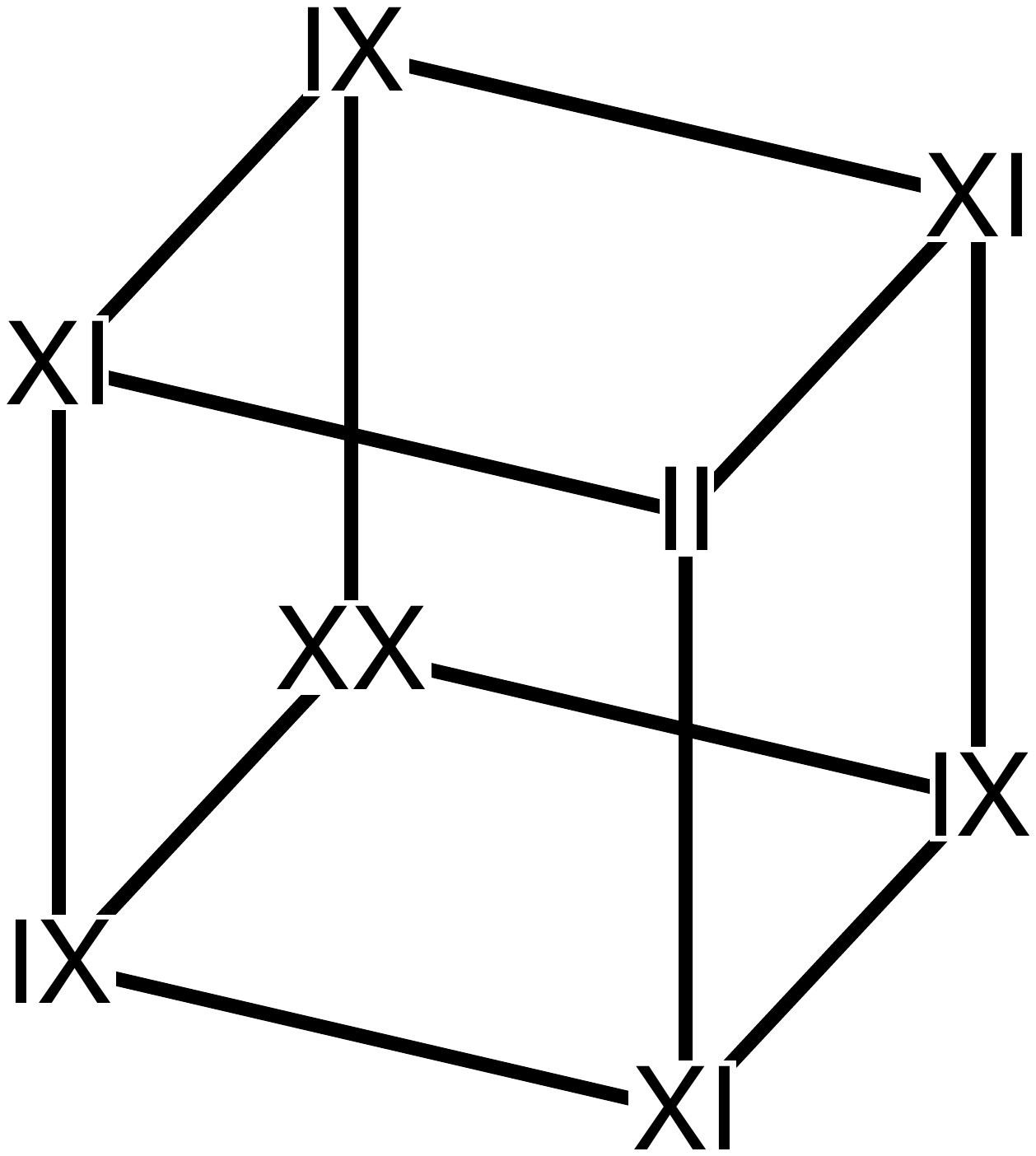}
	\end{minipage}
	\caption{Illustration of the $\hat{A}_c$ and  $\hat{B}_c$ operators that define Haah's code. The cube on the left shows the chosen labeling of the sites of cube $c$ and the chosen coordinate system used in this work. Two spins-1/2 called $\sigma$ and $\mu$ are located on each vertex of the lattice. The cube in the middle illustrates the action of $\hat{A}_c$ on the spins of cube $c$. Here IZ represents the tensor product $\mathds{1} \otimes  \mu^z$, where the operator on the left acts on the $\sigma$-spin and the operator on the right on the $\mu$-spin. Analogously, the right cube illustrates the action of $\hat{B}_c$ on the spins of cube $c$.}
	\label{fig::HaahsCode::Definition}
\end{figure}

Haah's code is defined on the cubic lattice where two distinct spin-1/2 degrees of freedom
$\sigma$ and $\mu$ are placed on each vertex. The Hamiltonian of Haah's code \cite{Haah_2011} is given as
\begin{equation}
\hat{\mathcal{H}}_{\rm Haah} = - J\sum_c \left( \hat{A}_c + \hat{B}_c \right)\, ,
\label{equ::HaahsCode::Hamiltonian}
\end{equation}
with 
\begin{eqnarray*}
	\hat{A}_c :=  \mu^z_j  \mu^z_k  \sigma^z_l  \mu^z_m   \sigma^z_n   \sigma^z_p  \sigma^z_q   \mu^z_q\,, \\
	\hat{B}_c :=  \sigma^x_i   \mu^x_i  \mu^x_j  \mu^x_k  \sigma^x_l  \mu^x_m   \sigma^x_n   \sigma^x_p\,.
\end{eqnarray*}
Here we labeled the sites of each cube $c$ as indicated in Fig.~\ref{fig::HaahsCode::Definition}. The operators $\sigma^\alpha_i$ ($\mu^\alpha_i$) with $\alpha \in \{x,y,z\}$ are the conventional Pauli matrices acting on the $\sigma$-spin ($\mu$-spin) on site $i$. We consider $J > 0$ without loss of generality.

Haah's code is a stabilizer code, i.e., it is a sum of commuting operators: 
\begin{equation}
\Big[ \hat A_c, \hat B_{c'} \Big] = \Big[ \hat A_c, \hat A_{c'} \Big]= \Big[ \hat B_c, \hat B_{c'} \Big] = 0 \quad \forall c, c' 
\end{equation}
and
\begin{equation}
\hat A_c^2 = \hat B_c^2 = \mathds{1} \quad \forall c.
\end{equation}
The eigenvalues $a_c=\pm 1$ of the operators $\hat{A}_c$ and $b_c=\pm 1$ of the operators $\hat{B}_c$ are therefore conserved quantities, which is essential for the exact solvability of Haah's code. 

\subsubsection*{Ground states}
\label{sssect::haah_gs}
Ground states of Haah's code are characterized by \mbox{$a_c = b_c = +1$} for all $c$ and have energy $E_0=-2JN_{\rm c}$ with $N_{\rm c}$ the number of cubes, which equals the number of sites $N$. In each unit cell of the cubic lattice we have two spin-1/2 degrees of freedom each with a two-dimensional Hilbert space and two operators each with two-dimensional eigenspace. If all eigenvalues $a_c$ and $b_c$ can be chosen independently, which is the case for open boundary conditions, one can construct the ground state $\ket{0}$ uniquely using projectors as
\begin{equation}
\ket{0} = \prod_c \left( \frac{\mathds{1} + \hat A_c}{2} \right)\left( \frac{\mathds{1} + \hat B_c}{2} \right) \ket{\Uparrow}\, ,
\label{equ::HaahsCode::Groundstate}
\end{equation}
where $\ket{\Uparrow}\equiv\ket{\Uparrow}_\sigma \ket{\Uparrow}_\mu$ denotes the fully polarized state of $\sigma$- and $\mu$-spins pointing w.l.o.g.~in $z$-direction. Since the operators $\hat A_c$ and $\hat B_c$ are rather complicated, it is not easy to visualize this ground state in the same fashion as the analogously constructed ground state of the toric code in terms of a loop soup. Nevertheless, it can be useful to think of a \textit{fractal soup} for Haah's code, i.e., the ground state can be seen as an infinite equal-weight superposition of spin product states in the thermodynamic limit resulting from the action of any combination of operators $\hat{A}_c$ and $\hat{B}_c$ on the fully polarized state $\ket{\Uparrow}$.   

For periodic boundary conditions of an \mbox{$L \times L \times L$} cubic lattice, one finds at least one constraint for each type of operators, i.e., the product of all operators $\hat{A}_c$ ($\hat B_c$) equals the identity. The reason for this is that each $ \sigma^z_i$ and $ \mu^z_i$ ($ \sigma^x_i$ and $ \mu^x_i$) appears four times in the product of all operators $\hat{A}_c$ ($\hat{B}_c$). One can therefore determine the eigenvalue of a single operator $\hat{A}_c (\hat{B}_c)$ by the eigenvalues of all the other operators of the same type
\begin{eqnarray}
\prod_c \hat{A}_c = \mathds{1} &\implies& \hat{A}_c = \prod_{c', c' \ne c} \hat{A}_{c'}\, , \\
\prod_c \hat{B}_c = \mathds{1} &\implies& \hat{B}_c = \prod_{c', c' \ne c} \hat{B}_{c'}\, .
\end{eqnarray}
As a consequence, one cannot fix all $2N_c$ spin degrees of freedom, because the eigenvectors of the operators span only an eigenspace of dimension $2N_c-2$ in total. Hence, the lower bound for the ground-state degeneracy of Haah's code is $2^2 = 4$ in this case. An upper bound for the ground-state degeneracy of Haah's code is given by $2^{4L}$ corresponding to $4L$ encoded qubits, because there might exist further constraints depending on the specific $L$ under consideration \cite{Haah_2011}.

\subsubsection*{Excitations}
\label{sssect::haah_excitations}
The elementary excitations of Haah's code are called fractons. Each fracton corresponds either to an eigenvalue $a_c=-1$ ($a$-fracton) or $b_c=-1$ ($b$-fracton) located at the center of cube $c$ with excitation energy $2J$ above $E_0$. Haah's code has therefore a ladder spectrum, i.e., the total energy of any state depends only on the total number of fractons, determined by the set of all eigenvalues $a_c$ and $b_c$. One can then write the Hamiltonian of Haah's code as
\begin{equation}
 \frac{\hat{\mathcal{H}}_{\rm Haah}}{J} = - 2N_{\rm c}+2\hat{\mathcal{Q}}_{\rm Haah}\, ,
\label{equ::HaahsCode::Hamiltonian_counting}
\end{equation}
where we have introduced the fracton counting operator $\hat{\mathcal{Q}}_{\rm Haah}$. We stress that the creation of $n$ fractons with $1\leq n\leq 3$ is impossible by any operator with local support. Nevertheless, in such cases physical $n$-fracton wave functions, which belong to different superselection sectors of the Hilbert space compared to the groundstate, can be defined via projectors. As an example, we state explicitly the one-fracton states using open boundary conditions in the thermodynamic limit:
\begin{eqnarray}
\ket{1_c^{(a)}} &= \frac{\mathds{1} - \hat A_c}{2} \prod_{c',c\neq c'} \left( \frac{\mathds{1} + \hat A_{c'}}{2} \right)\left( \frac{\mathds{1} + \hat B_{c'}}{2} \right) \ket{\Rightarrow}\, ,\\
\ket{1_c^{(b)}} &= \frac{\mathds{1} - \hat B_c}{2} \prod_{c',c\neq c'} \left( \frac{\mathds{1} + \hat A_{c'}}{2} \right)\left( \frac{\mathds{1} + \hat B_{c'}}{2} \right) \ket{\Uparrow}\, ,
\label{equ::HaahsCode::1qp}
\end{eqnarray}
where the product of projectors amounts to a global operator. The generalization to $n$-fracton states with $n>1$ is straightforward. The lowest number of fractons, which can be created locally, is four. This is achieved by acting with $\hat{\sigma}^{\alpha}_i$ or $\hat{\mu}^{\alpha}_i$ with $\alpha\in\{x,z\}$ on the ground state $\ket{0}$ so that either four $a$- or four $b$-fractons are created, respectively.  

\subsubsection*{Symmetries and fractal character}
\label{sssect::haah_sym}
The most useful symmetry of Haah's code is that any $\hat{B}_c$ operator can be mapped to the corresponding $\hat{A}_c$ operator by three steps: i) lattice inversion with the center of a cube as center of inversion, ii) renaming $\sigma$ to $\mu$ and vice versa, and iii) rotating the Hilbert space such that Pauli matrices $x \rightarrow z$ and $z \rightarrow -x$. Obviously, with a similar sequence of operations, one can perform the inverse transformation from $\hat{A}_c$ to $\hat{B}_c$ \cite{Nandkishore18}. This symmetry is very useful as it allows in many cases to investigate just one of the operator types while the properties of the other one can be deduced directly. Additionally, Haah's code has a three-fold rotational symmetry around the $(1,1,1)^T$-axis. Note that none of these symmetries were required in Haah's construction of the model \cite{Haah_2011}. 

\begin{figure}
	\centering
	\begin{minipage}{0.31\columnwidth}
		\adjincludegraphics[width = \textwidth, trim={ {.2\width} {.15\height} {.25\width} {.13\height}}, clip]{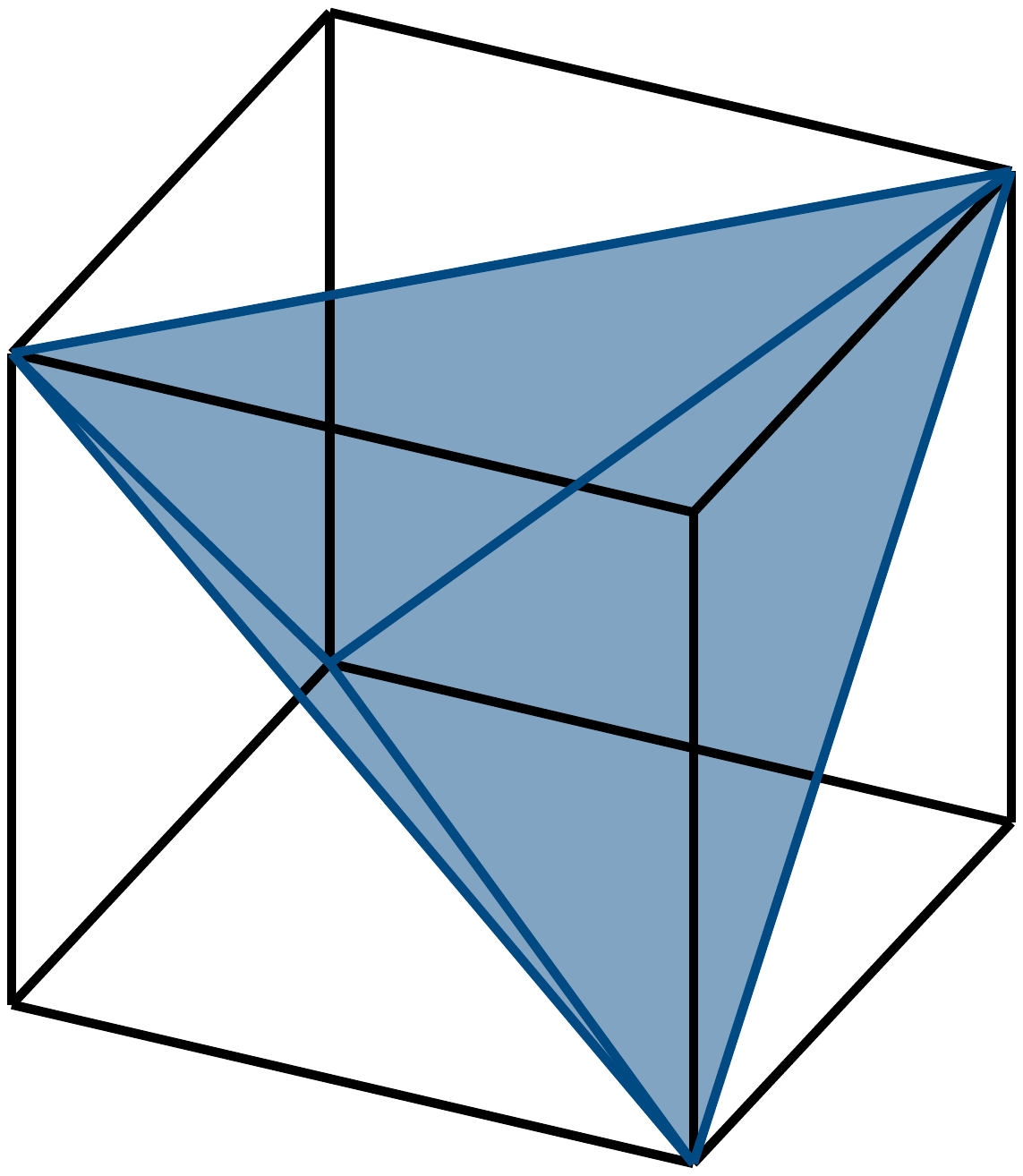}
	\end{minipage}
	\hfill
	\begin{minipage}{0.31\columnwidth}
		\adjincludegraphics[width = \textwidth, trim={ {.2\width} {.15\height} {.25\width} {.13\height}}, clip]{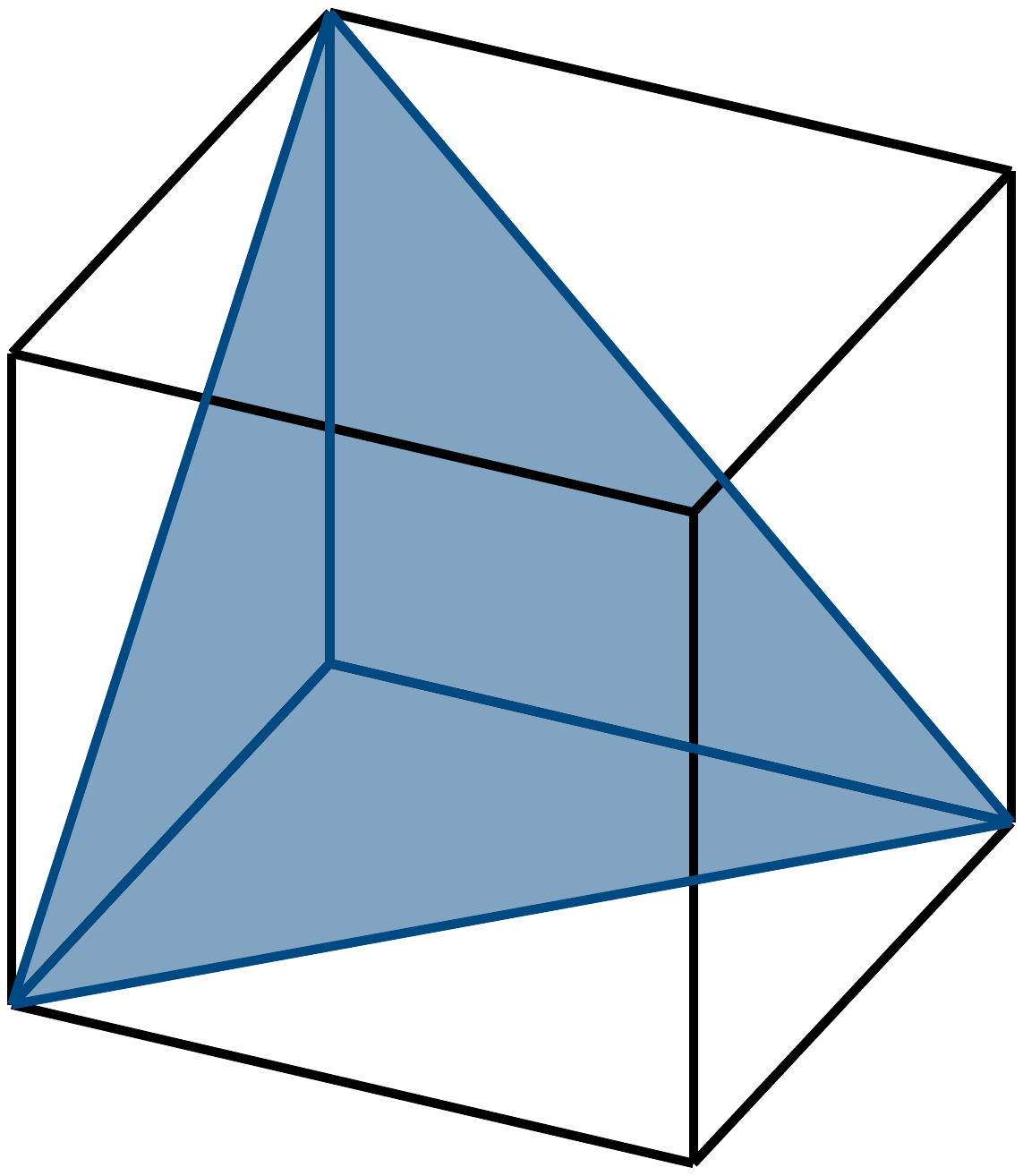}
	\end{minipage}
	\hfill
	\begin{minipage}{0.31\columnwidth}
		\adjincludegraphics[width = \textwidth, trim={ {.2\width} {.15\height} {.25\width} {.13\height}}, clip]{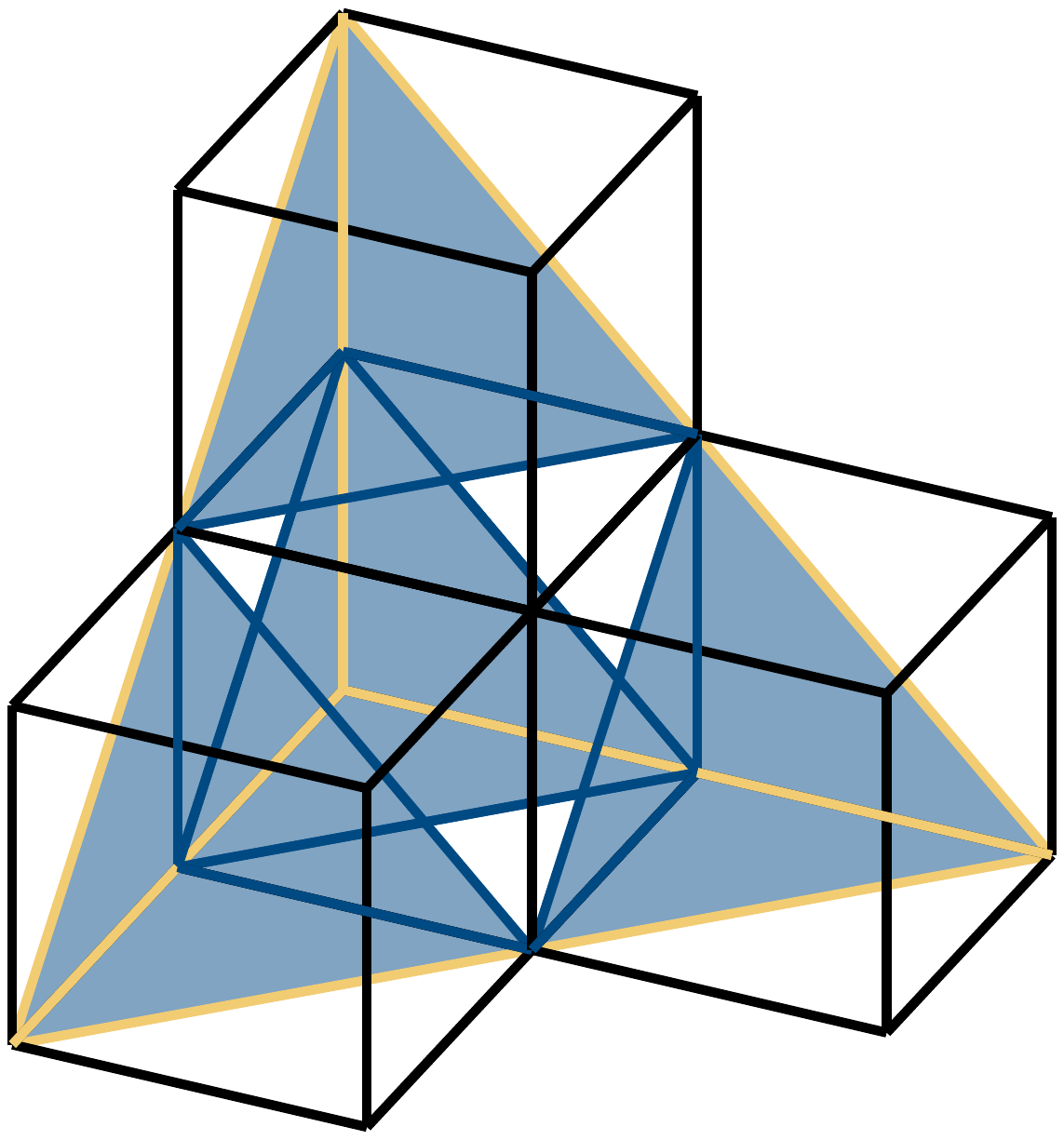}
	\end{minipage}
	\caption{Illustration of the fractal character of the operators $\hat{B}_c$ in Haah's code. The action of a single $\hat{B}_c$ operator on the $\sigma$-spins can geometrically be illustrated as a tetrahedron. This is shown on the left side. The figure in the center illustrates the action of a single $\hat{B}_c$ operator on the $\mu$-spins, again, forming a tetrahedron. The fractal character can be seen when putting four tetrahedrons together. These form a larger version of the same tetrahedron in a self-similar way, which is shown on the right. The same is true for the $\hat{A}_c$ operators by symmetry.}
	\label{fig::HaahsCode::FractalCharacterOperator}
\end{figure}

Haah's code is realizing type-II fracton order, i.e., the elementary fracton excitations are immobile and are located at the corners of fractal operators. This can be illustrated by the action of a $\hat{B}_c$ operator on the $\sigma$- and $\mu$-spins. In Fig.~\ref{fig::HaahsCode::FractalCharacterOperator}, one can see that the operators $\hat{B}_c$ act non-trivially on sites forming a Sierpinski tetrahedron with a self-similar character for both types of spin. Therefore, $\hat{B}_c$ is a fractal operator. The same is true for the $\hat{A}_c$ operators by symmetry. 
\par

\subsection{X-Cube}
\label{ssect::xcube}

\begin{figure}[!ht]
\includegraphics[width = 0.7\columnwidth]{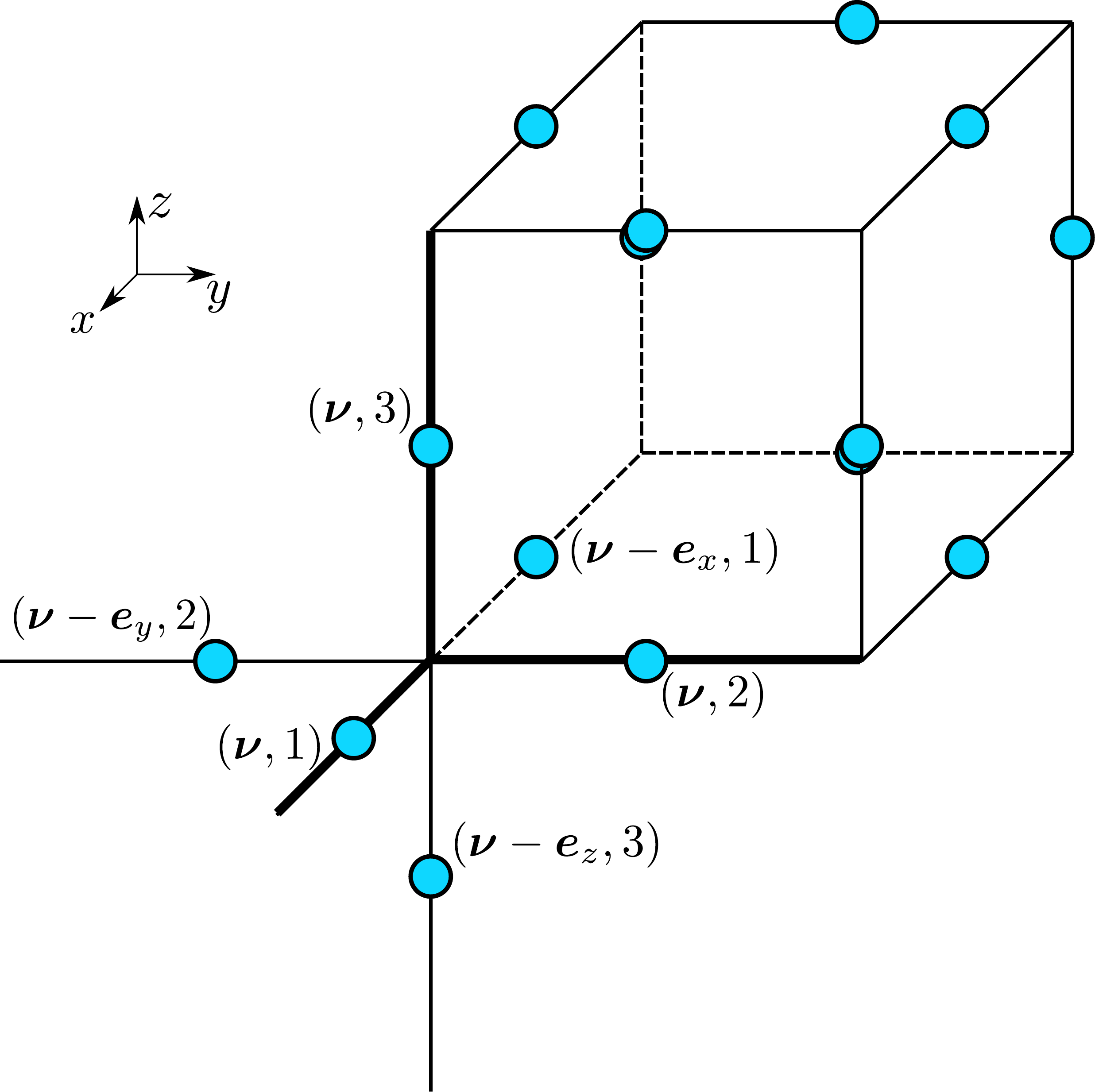}
\caption{The spin-1/2 degrees of freedom of the \mbox{X-Cube} model are located on the edges of a cubic lattice and are illustrated as filled circles. The $\hat{A}_c$ operator is the product of the twelve $\sigma^x$ operators acting on the twelve edges of the cube $c$. The three types of $\hat{B}^{(\kappa)}_{s}$ operators 
are given by $B^{(xy)}_{s}=\sigma^z_{(\boldsymbol{\nu},2)} \sigma^z_{(\boldsymbol{\nu},1)} \sigma^z_{(\boldsymbol{\nu}-\boldsymbol{e}_x,1)} \sigma^z_{(\boldsymbol{\nu}-\boldsymbol{e}_y,2)}$, 
$B^{(xz)}_{s}=\sigma^z_{(\boldsymbol{\nu},3)} \sigma^z_{(\boldsymbol{\nu},1)} \sigma^z_{(\boldsymbol{\nu}-\boldsymbol{e}_x,1)} \sigma^z_{(\boldsymbol{\nu}-\boldsymbol{e}_z,3)}$, and $B^{(yz)}_{s}=\sigma^z_{(\boldsymbol{\nu},2)} \sigma^z_{(\boldsymbol{\nu},3)} \sigma^z_{(\boldsymbol{\nu}-\boldsymbol{e}_z,3)} \sigma^z_{(\boldsymbol{\nu}-\boldsymbol{e}_y,2)}$.\label{Sketch}} 
\end{figure}

The X-Cube model, as introduced in \cite{Vijay16}, is defined on a cubic lattice (for generalization see Refs.~\cite{Slagle_2018,Shirley18}), where a single spin-$1/2$ degree of freedom is placed on every edge. The Hamiltonian of the model reads
 \begin{equation} 
 \hat{\mathcal{H}}_{\text{X-Cube}} = - J \sum\limits_{c} \hat{A}_c- J \sum\limits_{s,\kappa} \hat{B}^{(\kappa)}_{s} \, ,
\label{equ::XCube::Hamiltonian}
 \end{equation}
 where $\hat{B}^{(\kappa)}_{s}$ is the product of four $\sigma^z$-Pauli matrices acting on the four spins closest to the vertex $s$ in the $\kappa$-plane with \mbox{$\kappa \in \{ xy,xz,yz \} $}. 
 In contrast, $\hat{A}_c$ is the product of twelve $\sigma^x$ operators acting on the twelve spins on the edges of cube $c$. Both operator types constituting the X-Cube model are illustrated in Fig.~\ref{Sketch}. In the following we denote again the number of cubes by $N_{\rm c}$, which equals the number of vertices. One therefore has $N_{\rm c}$ operators $\hat{A}_c$, $3N_{\rm c}$ operators $\hat{B}_s^{(\kappa)}$, and $3N_{\rm c}$ spin-1/2 degrees of freedom. All $\hat{A}_c$ and $\hat{B}^{(\kappa)}_{s}$ mutually commute and square to the identity; hence their eigenvalues $a_c$ and $b_s^{(\kappa)}$ equal $\pm 1$. As for Haah's code, these operators are stabilizer operators \cite{Nandkishore18}.

%
%
  
\subsubsection*{Ground states}
\label{sssect::xcube_gs}
Ground states correspond to all states with eigenvalues $a_c=b_s^{(\kappa)}=+1$ for all $c$, $s$, and $\kappa$. The ground-state energy is given by $E_0=-4N_{\rm c}J$. The number of ground states depends on the geometry and the topology of the system in real space \cite{Slagle17, Slagle_2018}. For open boundary conditions, the ground state is unique and can be written as   
 \begin{eqnarray}
   \ket{0}_{\text{X-Cube}} &=&  \prod\limits_{s,\kappa} \left( \frac{\mathds{1}+\hat{B}^{(\kappa)}_{s}}{2}\right) \prod\limits_c \left( \frac{\mathds{1}+\hat{A}_c}{2}\right) \ket{\Uparrow} \nonumber\\
                      &=& \prod\limits_c \left( \frac{\mathds{1}+\hat{A}_c}{2}\right)\ket{\Uparrow} \label{GS1} \, ,
\end{eqnarray}
where $\ket{\Uparrow}$ is the fully polarized state in $z$-direction being trivially an eigenstate of all $\hat{B}_s^{(\kappa)}$ operators with eigenvalues $b_s^{(\kappa)}=+1$. The state $\ket{0}_{\text{X-Cube}}$ corresponds to an infinite equal-amplitude superposition of spin product states in the thermodynamic limit resulting from the action of an arbitrary combination of cube operators on the fully polarized state $\ket{\Uparrow}$. This state can therefore be visualized as a generalized loop soup of flipped spins on rectangular prisms \cite{Shirley_2019b}. 

The ground state can alternatively be written 
 \begin{eqnarray}
   \ket{0}_{\text{X-Cube}} &=&  \prod\limits_{s,\kappa} \left( \frac{1+\hat{B}^{(\kappa)}_{s}}{2}\right) \prod\limits_c \left( \frac{1+\hat{A}_c}{2}\right) \ket{\Rightarrow} \nonumber\\
                      &=&  \prod\limits_{s,\kappa}\left( \frac{1+\hat{B}^{(\kappa)}_{s}}{2}\right) \ket{\Rightarrow} \label{GS2}\quad ,
\end{eqnarray}
where the product runs over all vertices $s$ and orientations $\kappa$ and $\ket{\Rightarrow}$ is the fully polarized state in $x$-direction. Again, the state $\ket{0}_{\text{X-Cube}}$ corresponds to an infinite equal-amplitude superposition of spin product states in the thermodynamic limit resulting from the action of an arbitrary combination of $B_s^{(\kappa)}$ operators on the fully polarized state $\ket{\Rightarrow}$.

For periodic boundary conditions of an $L \times L \times L$-cluster (3-torus), the ground-state degeneracy is $2^{6L-3}$ \cite{Nandkishore18, Slagle17}. The ground states are indistinguishable by local measurements and thus the model is topologically ordered \cite{Nandkishore18}. The operators which distinguish these ground states are non-local loop operators, which we discuss after introducing the elementary excitations of the X-Cube model in the next paragraph.

\subsubsection*{Excitations}
\label{sssect::xcube_excitations}
Excitations in the X-Cube model correspond either to an eigenvalue $a_c=-1$ or $b_s^{(\kappa )}=-1$ with excitation energy $2J$ above $E_0$. Therefore, the X-Cube model has also a ladder spectrum, since the total energy of any state is proportional to the total number of negative eigenvalues. One can thus express the Hamiltonian of the X-Cube model as 
\begin{equation}
\frac{\hat{\mathcal{H}}_{\text{X-Cube}}}{J} = -4N_{\rm c}+2\hat{\mathcal{Q}}_{\text{X-Cube}}\, ,
\label{equ::XCUBE::Hamiltonian_counting}
\end{equation}
where we have introduced the counting operator $\hat{\mathcal{Q}}_{\text{X-Cube}}$ of negative eigenvalues. In the following we discuss the physical properties of cube excitations with $a_c= -1$ and vertex excitations with $b_s^{(\kappa )}= -1$ separately.

\begin{figure}[ht]
  \begin{center}
 \includegraphics[width=0.2 \textwidth]{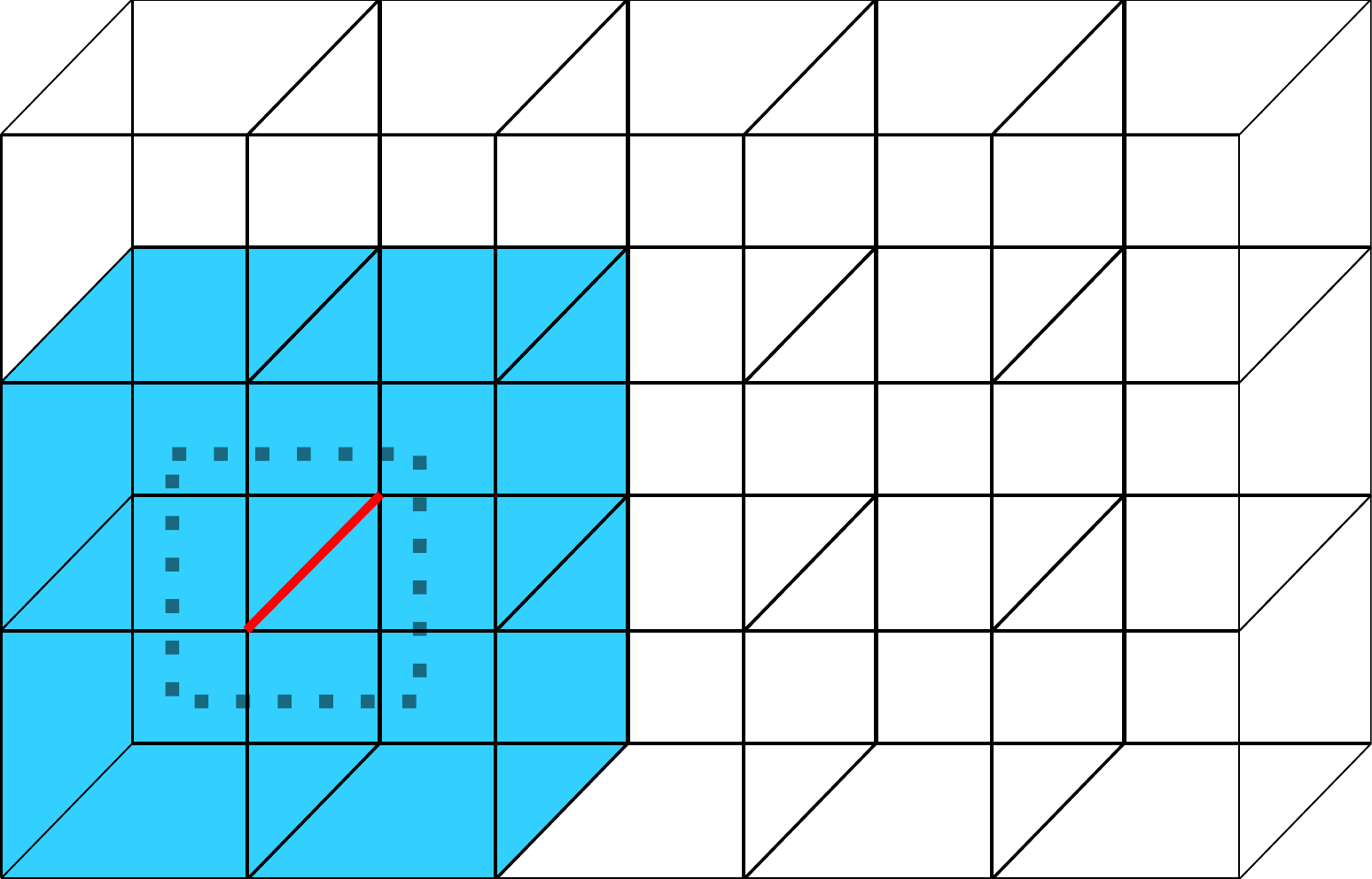} \hspace{1cm}
 \includegraphics[width=0.2 \textwidth]{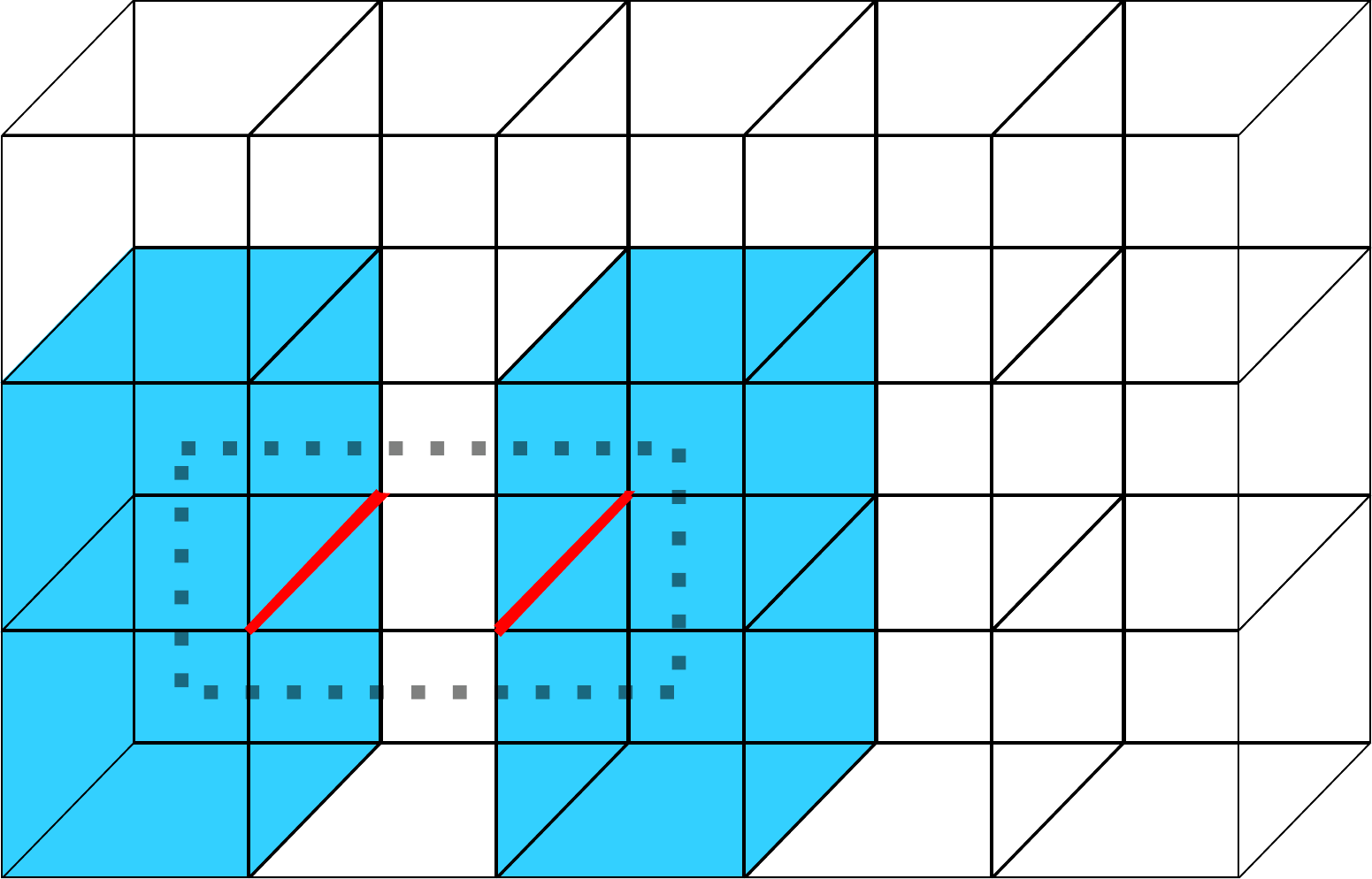} \\ \vspace{0.5cm}
 \includegraphics[width=0.2 \textwidth]{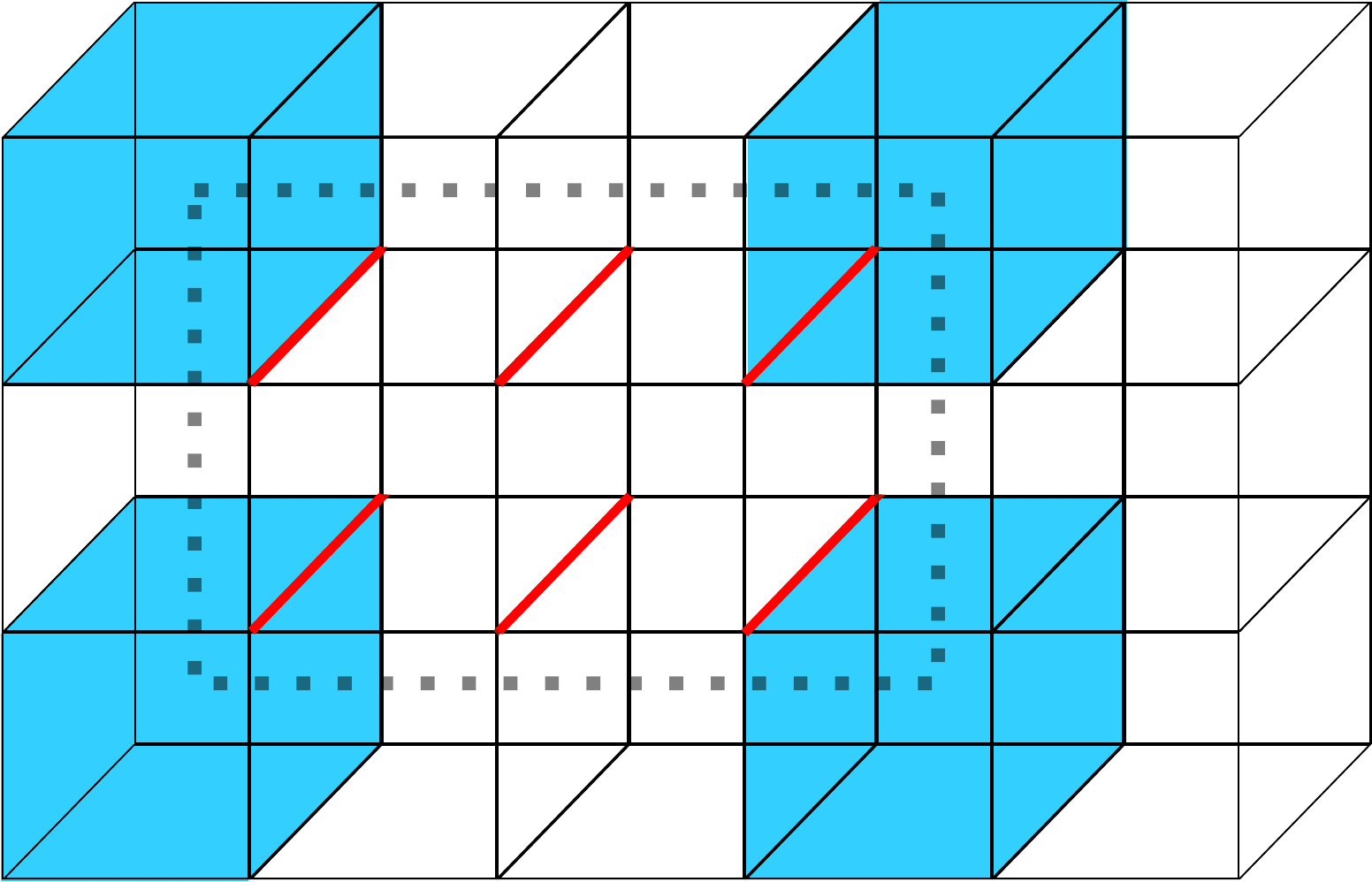} \hspace{1cm}
 \includegraphics[width=0.2 \textwidth]{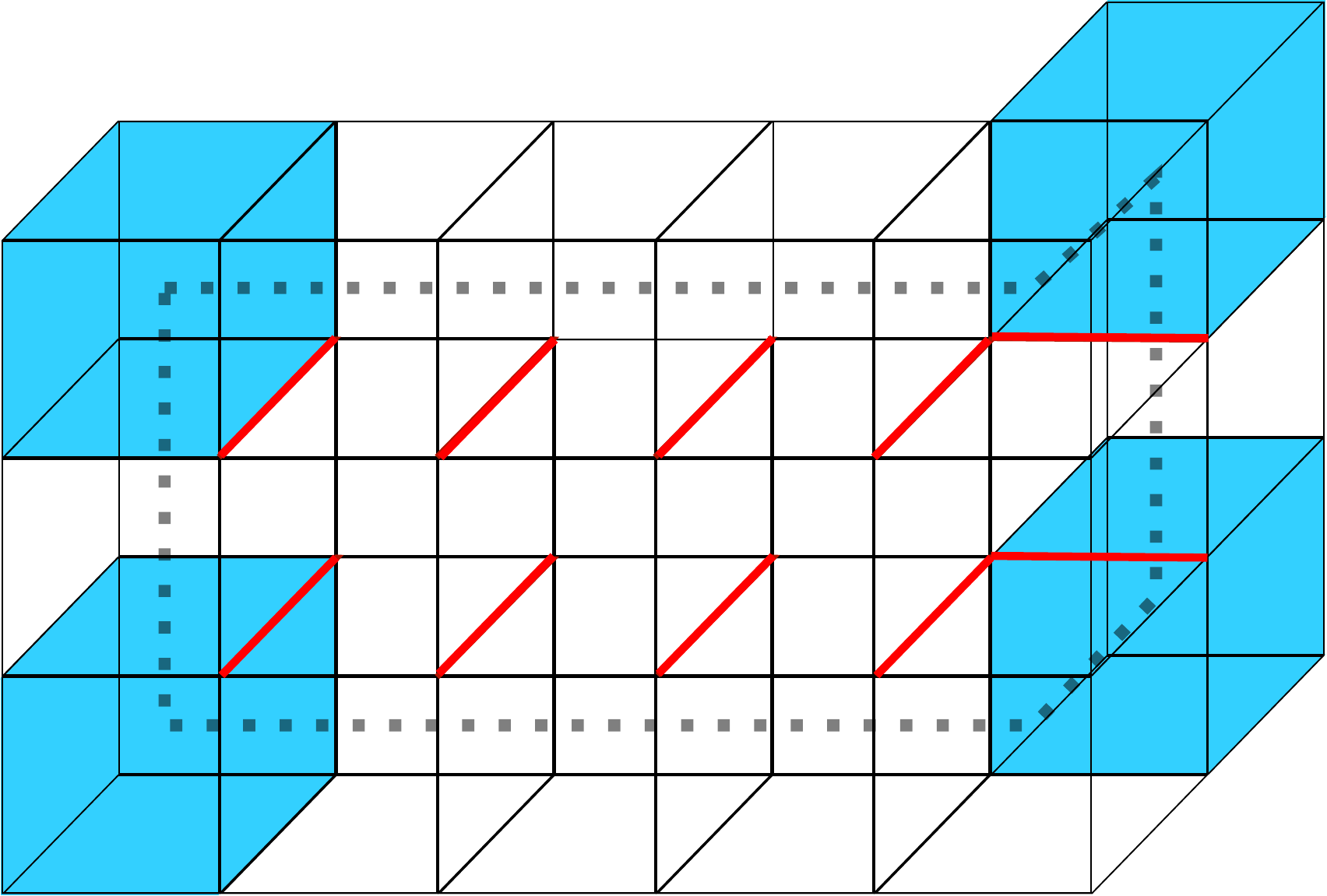} 
  \end{center}
  \caption{From top left to bottom right: First a fractonic four-cube excitation is created by acting with one $\sigma^z$ operator. The action of $\sigma^z$ operators is indicated by 
  red lines. In the following a planon hops to the right by the action of a $\sigma^z$ operator on the adjacent link. Then by acting with several such Pauli operators on adjacent links the membrane is extended. 
  In the end one of the edges of the membrane is moved around the corner.\label{CubesMove}}
 \end{figure}

{\it Cube excitations}: The simplest way to create, to move, and to annihilate cube excitations is by acting with $\sigma^z$ operators \cite{Vijay16}, e.g.,
\begin{equation}
  \sigma^z_{(\boldsymbol{\nu},n)} \ket{0}_{\text{X-Cube}}  = \sigma^z_{(\boldsymbol{\nu},n)} \prod\limits_c \left( \frac{1+\hat{A}_c}{2}\right) \ket{\Uparrow}
\end{equation}
creates the excited state with four-cube excitations 
\begin{equation}
    \prod\limits_{c \not\ni (\boldsymbol{\nu},n) }\left( \frac{1+\hat{A}_c}{2} \right)  \prod\limits_{c^\prime \ni (\boldsymbol{\nu},n) }\left( \frac{1-\hat{A}_{c^\prime}}{2} \right) \ket{\Uparrow},
\end{equation}
where $c^\prime$ are the four cubes with $a_{c'}=-1$ containing the site $(\boldsymbol{\nu},n)$. In a similar way kinetic processes of cube excitations can result from the action of $\sigma^z$ operators on excited states by an appropriate flipping of $a_c$ eigenvalues. Note that the kinetics of cube excitations takes place on the (dual) cubic lattice constituted by the centers of the cubes. As expected for a fracton phase, a single cube excitation is a fracton, i.e., it is immobile in the sense that it cannot be moved without creating additional cube excitations \cite{Vijay16}. If two cube excitations share one cartesian coordinate in terms of the dual lattice, then these aligned two-cube excitations are mobile in a two-dimensional plane, and are thus called planons. The configurations of four-cube excitations, which originate from a local action of a single $\sigma^z$ operator, are constrained to combinations of rectangular membranes \cite{Vijay16, Prem_2017, Shirley_2019b} as illustrated in Fig.~\ref{CubesMove}. 

 \FloatBarrier

A single rectangular membrane of four cube excitations in the $yz$-plane can be created by acting with the membrane operator 
\begin{equation} \hat{M}_{\boldsymbol{\nu}}^{x,\boldsymbol{d}} = \prod\limits_{\substack{\mu_x=0 \\ \mu_y \in [0,d_y] \\ \mu_z \in [0,d_z]}} \sigma^z_{\boldsymbol{\nu}+\boldsymbol{\mu},1}
\end{equation}
on the ground state. The notation $M_{\boldsymbol{\nu}}^{x,\boldsymbol{d}}$ is explained as follows: The lower index $\boldsymbol{\nu}$ specifies the reference point, the first upper index $x$ the direction orthogonal to the membrane, and $\boldsymbol{d}=(0,d_y,d_z)$ gives the extension of the membrane in the $yz$-plane. This membrane forms a rectangular structure with four fractonic cube excitations at the corners as illustrated in Fig.~\ref{CubesMove}. 
It is notable that it is not possible to create a single excitation or a planon from the vacuum. In fact, the minimum amount of created cube excitations from the ground state is four. 
%
%
\begin{figure}[ht]
  \begin{center}
 \includegraphics[width=0.2 \textwidth]{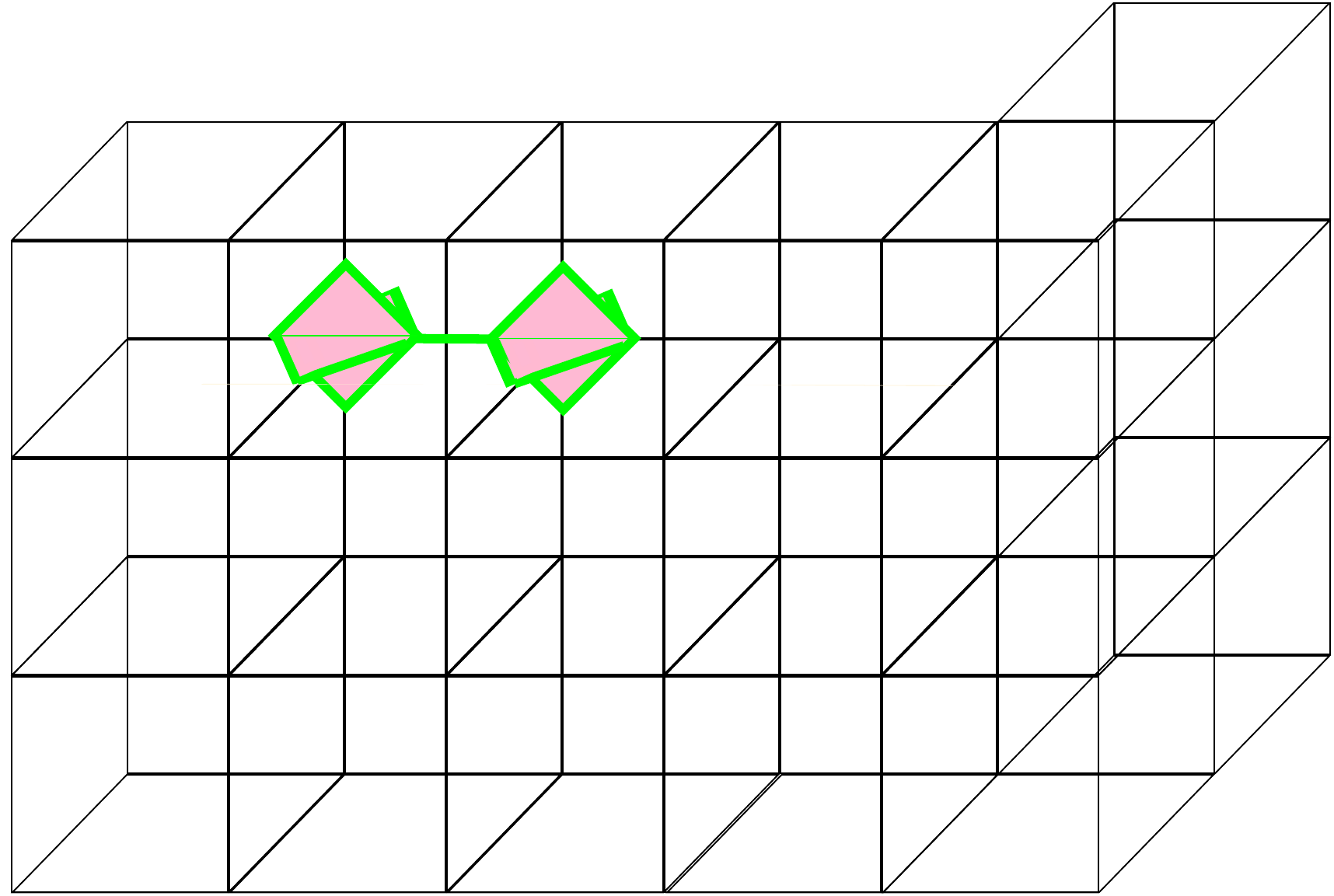} \hspace{0cm}
 \includegraphics[width=0.2 \textwidth]{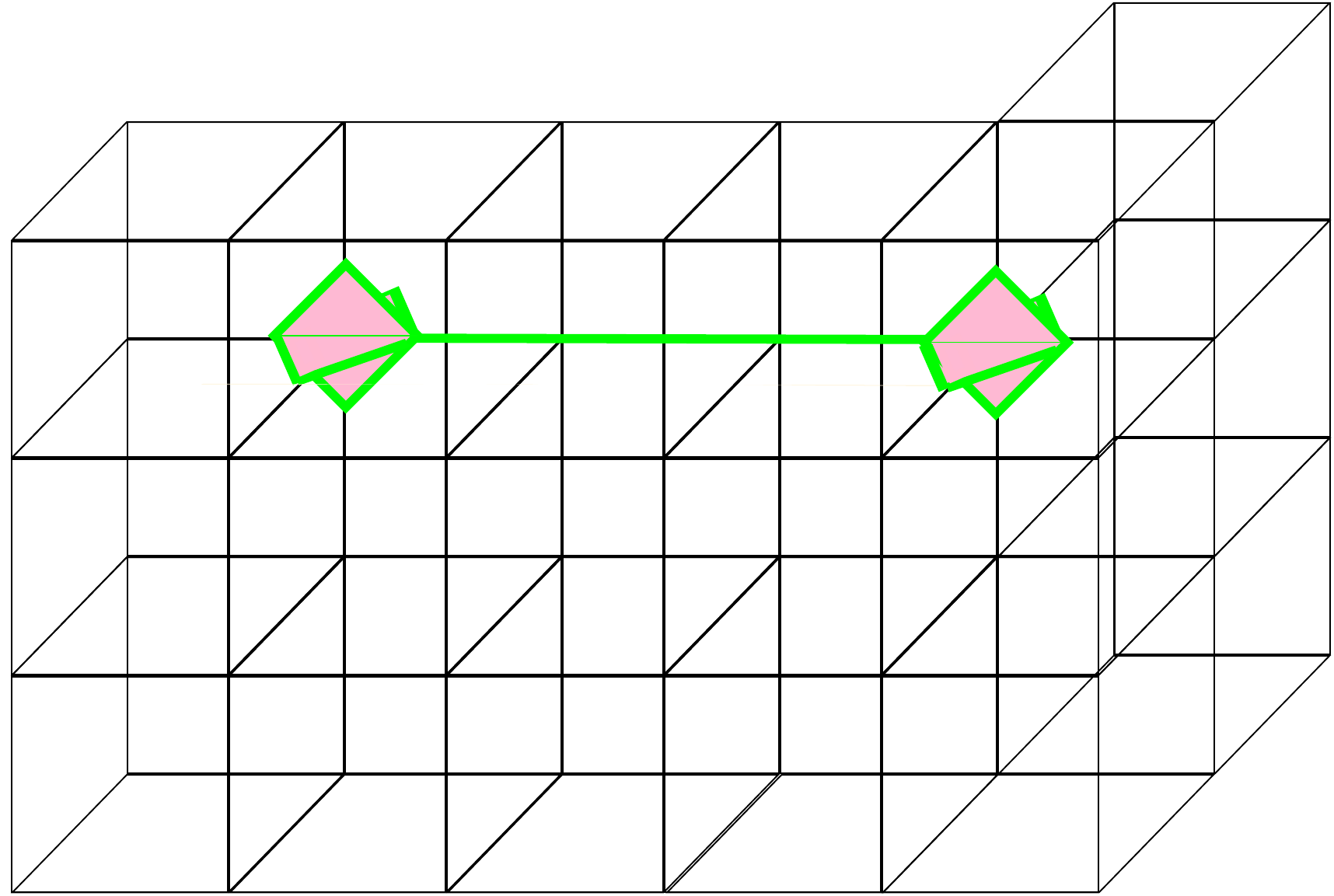} \\ \vspace{0.5cm}
 \includegraphics[width=0.2 \textwidth]{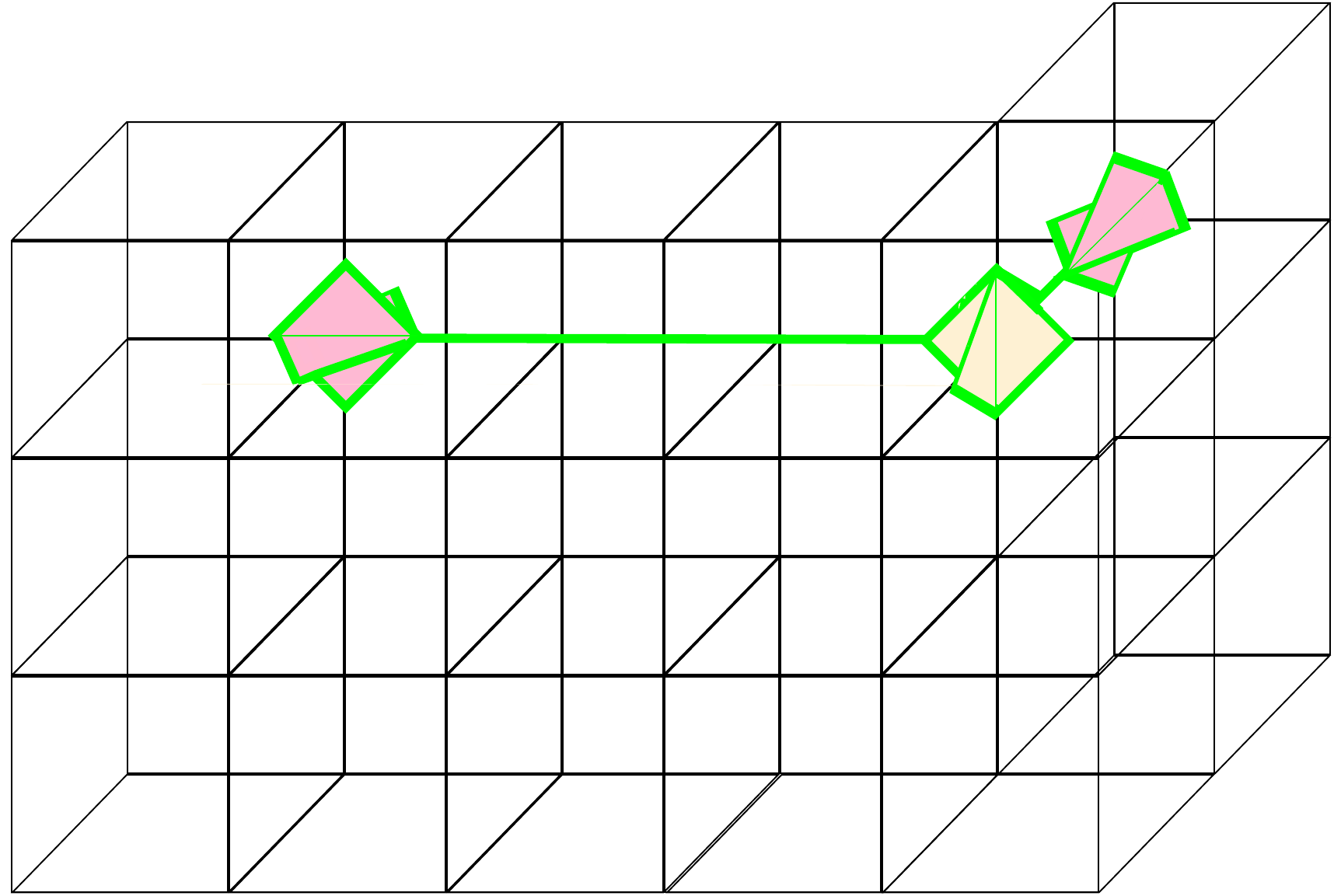} \hspace{0cm}
  \includegraphics[width=0.2 \textwidth]{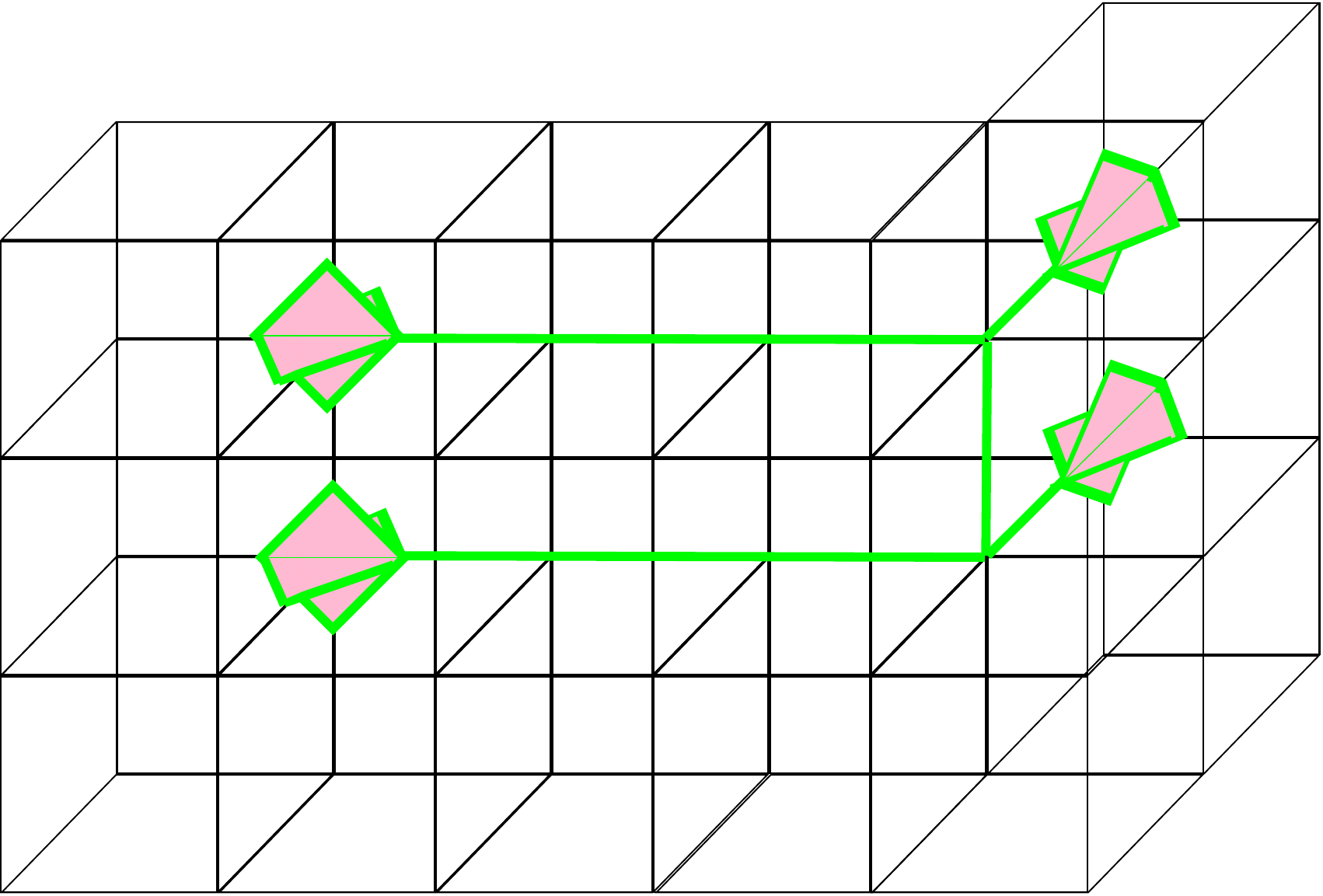}  
  \end{center}
  \caption{\label{LineonsMove}The rhombs correspond to $b_s^{(\kappa)}=-1$ where the orientation $\kappa$ corresponds to the orientation of the rhombs. The green lines symbolize the action of $\sigma^x$ operators in order to move these excitations. From top left to bottom right: First two lineons (each corresponding to a two-vertex excitation) are created as nearest-neighbors in $y$-direction by acting with one $\sigma^x$ operator. By acting with a straight string of $\sigma^x$ operators on the adjacent links in the same direction, one of the two lineons is moved to the right without creating any further excitations (upper right figure). In contrast, an additional lineon excitation is always created at a finite energy cost when moving a lineon around a corner (bottom left figure) demonstrating the restricted one-dimensional mobility of lineon excitations. However, a pair of parallel lineons has a two-dimensional mobility, since the pair is allowed to move around corners without exciting additional excitations (botton right). This is enabled by to annihilating such parallel pairs of lineon excitations. The corresponding operator of this process is called wireframe operator (see Fig.~\ref{Wireframe}).}
\end{figure}

{\it Vertex excitations} The vertex excitations correspond to $b_s^{(\kappa)}=-1$  \cite{Nandkishore18}. The easiest way to manipulate them is by acting with $\sigma^x$ on suitable links \cite{Vijay16}. Note that the different orientations $\kappa$ are important as only $b_s^{(\kappa)}$ eigenvalues, where the link carrying the site $(\boldsymbol{\nu},n)$ is parallel to the $\kappa$-plane, are changed by $\sigma_{(\boldsymbol{\nu},n)}^x$.
Due to the fact that the operators $B_s^{(\kappa)}$ obey the local constraint $\prod_\kappa \hat{B}_{s}^{(\kappa)} = \mathds{1}$ \cite{Ma17Entropy,Lake} at each $s$, 
the energetically lowest excitation corresponds to a pair of negative $b_s^{(\kappa)}$ at the same vertex $s$. This two-particle excitation is called lineon, as it can only move on a one-dimensional line \cite{Vijay16, Shirley_2019b}.
The action of $\sigma_{(\boldsymbol{\nu},n)}^x$ on the ground state creates two lineons as there are exactly four $\hat{B}^{(\kappa)}_{s}$ that do not commute with $\sigma^x_{\boldsymbol{\nu},n}$. Kinetic processes such as the hopping of lineons, as illustrated in Fig.~\ref{LineonsMove}, can be understood as $\sigma^x$ flipping $b_s^{(\kappa)}$ in an appropriate way. 

 \begin{figure}[htbp]
  \includegraphics[width=0.4\textwidth]{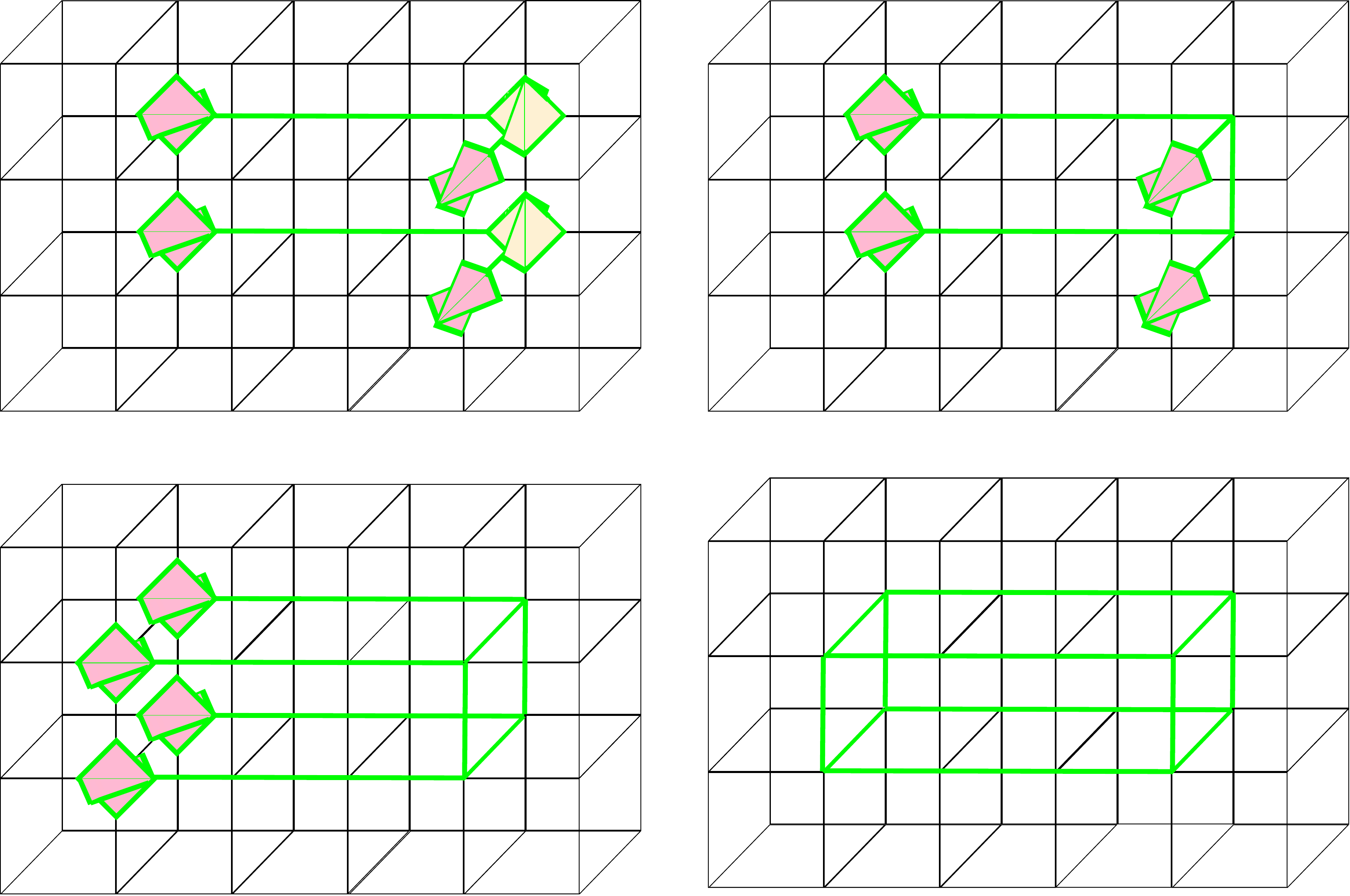}
  \caption{Construction of a wireframe operator. In the figure on the top left a pair of lineons moves around a corner. This creates two additional excitations. These can be annihilated by a further $\sigma^x$ operator as depicted on the top right. On the bottom left these excitations move back to the left. Then they annihilate with the initial excitations. On the bottom right no excitations are left and the wireframe operator as a product of 
  $\sigma^x$ operators is visible. \label{Wireframe}}
 \end{figure}
 
In order to describe these excitations mathematically, we introduce line operators $\hat{L}^{\nu_{\alpha} \in [k_1,k_2]}_{(i,j)}$ with $\alpha\in\{x,y,z\}$ and the unit cell as depicted in Fig.~\ref{Sketch}. For a line operator in $z$-direction one has specifically
\begin{equation} 
  \hat{L}^{\nu_z \in [k_1,k_2]}_{(i,j)} = \prod\limits_{\nu_z \in [k_1,k_2]} \sigma^x_{(\boldsymbol{\nu},3)}. 
\end{equation}
 The direction of the line is indicated by the first letter in the upper index (here $\nu_z$). The lower indices $(i,j)$ refer to the two coordinate directions orthogonal to the one specified in the upper index and state the position in the lattice in alphabetical order; here the first index specifies the discrete $\nu_x$ coordinate $i$ and the second index specifies the discrete $\nu_y$ coordinate $j$.
 
 The line operator $\hat{L}^{\nu_{\alpha} \in [k_1,k_2]}_{(i,j)}$ with $\alpha\in\{x,y,z\}$ acting on the ground state therefore creates one lineon excitation at vertex $(i,j,k_1)$ and another one at $(i,j,k_2)$ that both can move in $\alpha$-direction.
 When a product of line operators with different orientations $\alpha$ is formed such that they share an endpoint, one has to distinguish two situations: 
 In case of two line operators with two distinct orientations $\alpha_1$ and $\alpha_2$, the two lineons at the corner fuse to a single
 lineon $\ell_{\alpha_3}$ which is mobile in the $\alpha_3$-direction with $\alpha_3\neq \alpha_{1,2}$ (see Fig.~\ref{Wireframe}).
 In contrast, the local fusion of three different lineons annihilates all three of them \cite{Shirley_2019b,Ma17}. Technically, this can be expressed in the following local fusion rules \cite{Shirley_2019b,Ma17}
 \begin{align}
  \ell_{\alpha_1} \times \ell_{\alpha_2} &= \ell_{\alpha_3} 	 &\text{for }\, \alpha_{1,2} \neq\alpha_{3}\,,\quad \alpha_1 \neq \alpha_2 \,;\\
  \ell_{\alpha_1} \times \ell_{\alpha_2} \times \ell_{\alpha_3} &= 1  &\text{for } \, \alpha_{1,2} \neq\alpha_{3}\,, \quad \alpha_1 \neq \alpha_2 \,; \\
  \ell_{\alpha_1} \times \ell_{\alpha_2} &= 1 &\text{for  } \,  \alpha_1 =\alpha_2 
  \phantom{\,\,\,, \quad \alpha_1 \neq \alpha_2 \,\,\,;}
 \end{align}
 at one vertex $s$.
 \begin{figure}[htbp]
  \begin{center} 
  \includegraphics[width=0.4\textwidth]{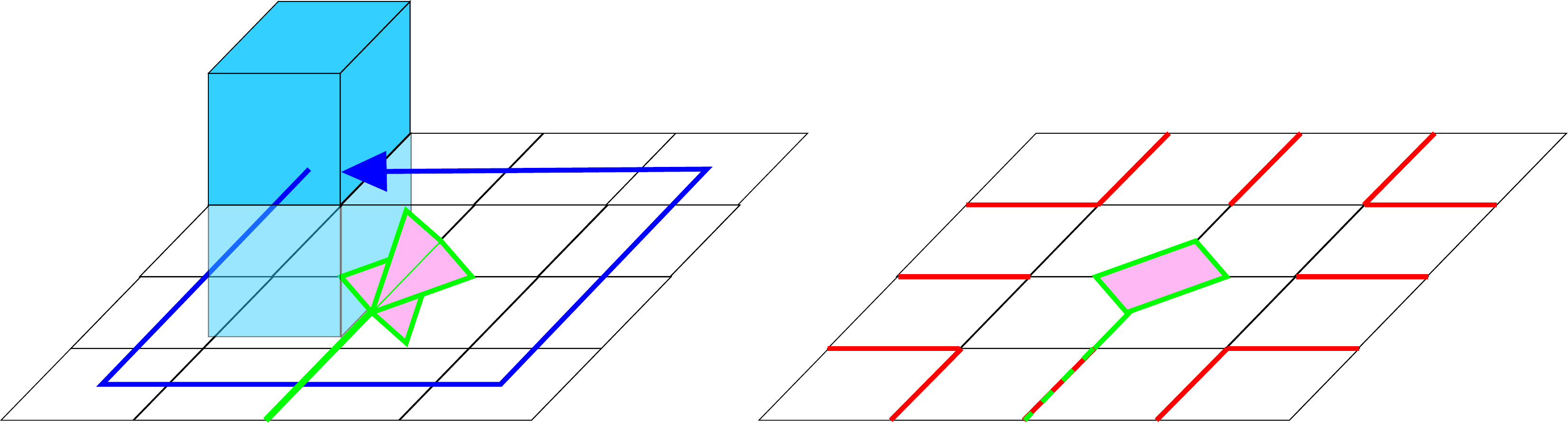} 
  \end{center}
  \caption{Illustration of the winding of a planon around a lineon with flavor $x$ in the $xy$-plane (left). On the right this process is depicted by projecting on the $xy$-plane, thus the only visible part of the planon loop are the red links where $\sigma^z$ operators have acted on. The lineon is located at the end of a string of $\sigma^x$ operators (depicted in green). The string and the loop share a single spin, here depicted as red-green dashed string at the bottom. The commutation relation of Pauli operators results in an additional minus sign as phase for the wave function showing the semionic statistics of planon and lineon excitations.}
 \label{Gauss}
 \end{figure}
 
 Note that the third line represents the fact that two lineons of equal orientation annihilate each other.

 Interestingly, for pairs of lineons with the same orientation $\alpha$ sharing one cartesian coordinate $\nu_\beta$ with $\beta\neq\alpha$, it is possible to move in the two-dimensional plane orthogonal to $\beta$ as a consequence of the fusion rules. In this way it is possible to construct self-annihilating loops called wireframe operators as illustrated in Fig.~\ref{Wireframe}.

{\it Generalized mutual exchange statistics}\label{Statistics}
After this discussion of excitations, we want to discuss the anyonic exchange statistics is present in the fractonic X-Cube model. An intuitive way to understand the quasi-particle statistics is to write down the necessary loop- and string-like operators representing the winding of quasi-particles around each other in terms of Pauli operators. Then one commutes these Pauli operators until the respective loop operator acts trivially on the ground state. As a consequence, an overall sign change of the wave function might result due to the commutation relations. 
Alternatively, one makes use of the fact that contractible wireframe operators are products of $\hat{A}_c$ operators, and contractible membrane loop operators are products of appropriately oriented $\hat{B}_s^{(\kappa)}$ operators in a single plane or in a stack of planes: The generalized mutual statistics becomes directly visible, as one excitation inside the loop results in changing a single sign in the product. The loop operator therefore gives a minus sign for any odd number of enclosed excitations.

To be concrete, a contractible planon loop without an enclosed excitation acts trivially on the ground state as this loop can be written as a product of $B_s^{(\kappa)}$ stabilizers. In contrast, a planon loop around a lineon excitation effectively measures modulo 2 the appropriately oriented $\hat{B}_s^{(\kappa)}$ excitations inside the planon loop as illustrated in Fig.~\ref{Gauss} \cite{Slagle17}. Appropriately oriented means that the links of the membrane loop are in the same plane as the $\hat{B}_s^{(\kappa)}$ excitations. Analogously, a wireframe operator can be written as a product of $\hat{A}_c$ operators, and thus measures the parity of the $\hat{A}_c$ excitations inside the wireframe. 

Overall, in contrast to Haah's code, the X-Cube model features non-trivial mutual exchange statistics, e.g., this manifests itself in the mutual semionic exchange statistics of planons and lineons.  

\subsubsection*{Logical operators}
\label{sssect::xcube_lo}
Similar to the toric code, the ground-state degeneracy can be understood in terms of noncontractible loop operators. In principle two types of such logical loop operators exist:
One corresponds to a lineon pair, of which one lineon is wound around the torus, and annihilates with the other lineon. The other logical operator corresponds to the same process with two planons \cite{Slagle17}. However, pairs of noncontractible loops are not all independent, since for every noncontractible planon loop one can find a noncontractible lineon loop, such that their commutator is non-zero. Hence, it is sufficient to count the independent noncontractible lineon loops \cite{Slagle17}.
In our notation a noncontractible lineon loop operator in the $xy$-plane is written as
\begin{equation}
\hat{L}^{\nu_z \in [0,L]}_{(i,j)} = \prod\limits_{\nu_z \in [0,L]} \sigma^x_{((i,j,\nu_z),3)},
\end{equation}
where $L$ is the circumference of the three-torus. A noncontractible tube of $\hat{A}_c$ operators can be written as a product of four such noncontractible lineon loop operators \cite{Shirley18}
\begin{equation}
 \prod_{c \in \rm tube} \hat{A}_c = \hat{L}^{\nu_z \in [0,L]}_{(i,j)} \hat{L}^{\nu_z \in [0,L]}_{(i+1,j)} \hat{L}^{\nu_z \in [0,L]}_{(i,j+1)} \hat{L}^{\nu_z \in [0,L]}_{(i+1,j+1)}.
\end{equation}
Using a product of such tubes with a rectangular base leads to the following prism with corners at $(i,j)$, \mbox{$(i+n,j)$}, $(i,j+m)$, and $(i+n,j+m)$ 
\begin{equation}
  \label{eq:pri}
\prod_{c \in \rm prism} \hat{A}_c = \hat{L}^{\nu_z \in [0,L]}_{(i,j)} \hat{L}^{\nu_z \in [0,L]}_{(i+n,j)} \hat{L}^{\nu_z \in [0,L]}_{(i,j+m)} \hat{L}^{\nu_z \in [0,L]}_{(i+n,j+m)}
\end{equation}
where $n,m \in \mathds{Z}, \, n \in [1,L] , \, m \in [1,L] $.
This means that the operators involved in the product on the right side of Eq.~\eqref{eq:pri} can not all be contained in our complete set of commuting operators of stabilizers and logical operators:
From three noncontractible lineon loop operators and the product of all operators $\hat{A}_c$ in the prism we can construct the loop operator
\begin{equation}
\hat{L}^{\nu_z \in [0,L]}_{(i+n,j+m)} = \prod_{c \in \rm prism} \hat{A}_c \hat{L}^{\nu_z \in [0,L]}_{(i,j)} \hat{L}^{\nu_z \in [0,L]}_{(i+n,j)} \hat{L}^{\nu_z \in [0,L]}_{(i,j+m)}\,.
\end{equation}
Hence from the product of the operators passing through $(0,0)$, $(0,m)$, and $(0,n)$ we get the operator passing through $(n,m)$ and
we find $L+L-1$ independent logical operators passing through the $xy$-plane. For all three different kind of planes this reasoning arrives at $6L-3$ independent operators in total. We have not shown that we have indeed found all the logical operators.
However, counting independent planon loops \cite{Ma17} or applying a rigorous procedure using algebraic geometry (for odd $L$) \cite{Vijay16} yields the same result.

\section{Fracton models in a field}
\label{sect::FF}
Haah's code and the X-cube model, as introduced in the last section, are exactly solvable codes having a ground state with non-trivial topological fracton order, which is certainly robust under small local perturbations. However, it is an interesting question how the fracton order breaks down at zero temperature, when strong enough competing terms are added to these models. Furthermore, we want to inestigate whether the two types of fracton order behave differently in this case. In order to answer these questions, we add a uniform magnetic field representing one of the simplest perturbations which ultimately causes a phase transition between the fracton phase at small fields and a topologically trivial polarized phase stabilized at large magnetic fields. Here we study exactly this kind of phase transition as well as the physical properties of fracton excitations in the fracton phase for the case of Haah's code and the X-cube model in a uniform magnetic field. Furthermore, we present several exact dualities of fracton codes in specific field directions to isospectral models. These dual models are on the one hand useful for practical calculations and, on the other hand, interesting on their own. We stress that all presented results for ground-state energies as well as excitation energies are directly valid for these dual models.

\subsubsection*{Haah's code in a homogeneous magnetic field}
The Hamiltonian of Haah's code in the presence of a homogeneous magnetic field pointing in an arbitrary direction reads
\begin{equation} \label{equ::HaahsCode::HamiltonianMag}
\hat{\mathcal{H}}_{\rm Haah}^{h} = - J \sum_c  \hat{A}_c - J \sum_c \hat B_c - \sum_i \left( \boldsymbol{h}^\sigma \cdot  \boldsymbol{\sigma}_i +  \boldsymbol{h}^\mu \cdot \boldsymbol{\mu}_i \right),
\end{equation}
where $\boldsymbol{h}^\sigma:=(h_x^\sigma,h_y^\sigma,h_z^\sigma)$,  $\boldsymbol{h}^\mu:=(h_x^\mu,h_y^\mu,h_z^\mu)$, \mbox{$\boldsymbol{\sigma}_i:=(\sigma^x_i,\sigma^y_i,\sigma^z_i)$}, and $\boldsymbol{\mu}_i:=(\mu^x_i,\mu^y_i,\mu^z_i)$ are vectors representing all possible field directions.
From now on, we will call $\boldsymbol{h}^\sigma$ the $\sigma$-type field and $\boldsymbol{h}^\mu$ the $\mu$-type field.

The two limiting cases for Haah's code in a homogeneous magnetic field are the low- and the high-field limit. In the low-field limit $ J \gg \abs{\boldsymbol{h}^\sigma},\abs{\boldsymbol{h}^\mu}$ the system is in the fracton phase as described in Sect.~\ref{ssect::haah}. In the opposite case $ J \ll \abs{\boldsymbol{h}^\sigma},\abs{\boldsymbol{h}^\mu}$ the system realizes a topologically trivial polarized phase. In between there must be a quantum phase transition which we investigate for certain field directions.

\begin{figure}
	\centering
	\begin{minipage}{0.31\columnwidth}
		\includegraphics[width = \textwidth]{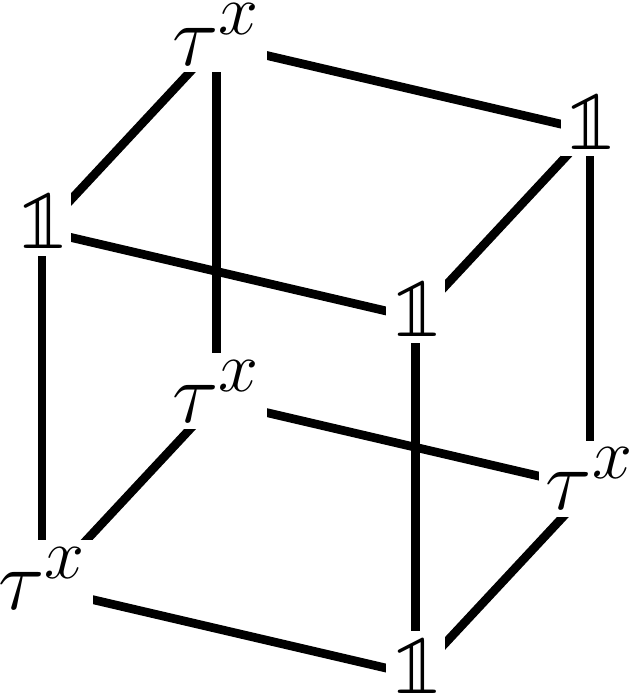}
		$\tau^x_{a_1}\tau^x_{a_2} \tau^x_{a_3} \tau^x_{a_4}$
	\end{minipage}
	\hfill
	\begin{minipage}{0.31\columnwidth}
		\includegraphics[width = \textwidth]{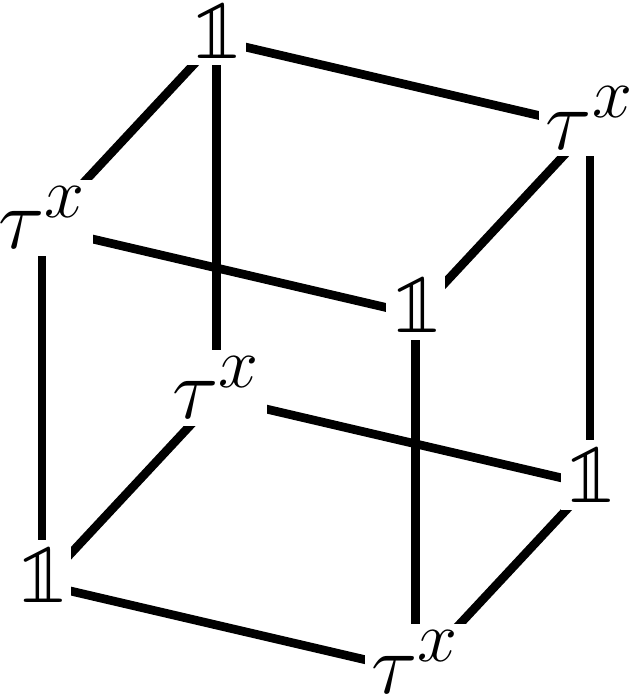}
		$ \tau^x_{b_1}\tau^x_{b_2} \tau^x_{b_3} \tau^x_{b_4}$
	\end{minipage}
	\hfill
	\begin{minipage}{0.31\columnwidth}
		\includegraphics[width = \textwidth]{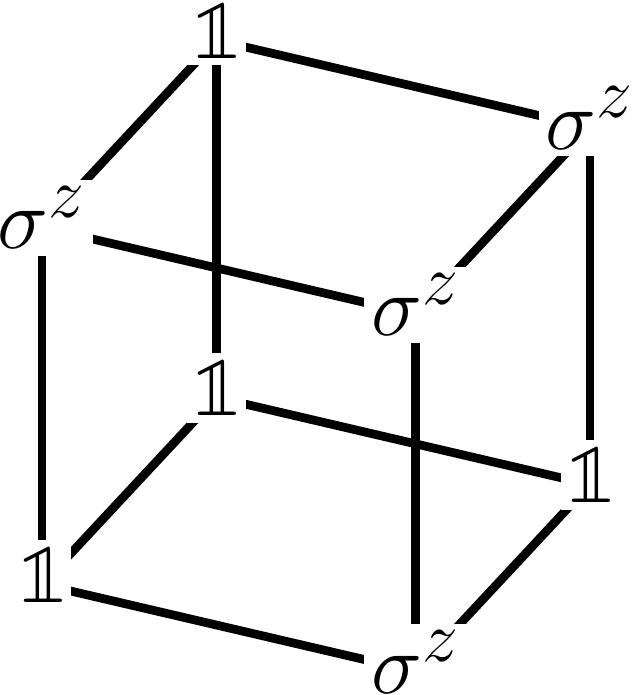}
		$\hat{B}_c^{(\sigma)}$
	\end{minipage}
	\caption{The figure on the left and in the center illustrate the action of $\sum_{{{\rm tetrahedra},a=(a_1,a_2,a_3,a_4)}} \tau^x_{a_1}\tau^x_{a_2} \tau^x_{a_3} \tau^x_{a_4}$ and $\sum_{{{\rm tetrahedra},b=(b_1,b_2,b_3,b_4)}} \tau^x_{b_1}\tau^x_{b_2} \tau^x_{b_3} \tau^x_{b_4}$ on the pseudo-spins on the dual cubic lattice. The figure on the right shows the action of a $\hat{B}_c$ operator on the $\sigma$-spins in the original formulation of Haah's code (the operator is depicted as $\hat{B}_c^{(\sigma)}$). Note that the action of a $\hat{B}_c$ operator on the $\sigma$-spins and the action of a $\sigma^x$ operator on the $\tau$ pseudo-spins are identical up to an inversion about the center of the cube. As these two actions correspond to the perturbation of the high- and low-field limit in the single-type case, respectively, this case is self-dual.}
	\label{HaahsCode_InMag_SelfDuality}
\end{figure}

In the \textit{two-type parallel case}, $\boldsymbol{h}^\sigma = \boldsymbol{h}^\mu \ne 0$ and they point in $x$- or $z$-direction, where both directions yield the same physics due to the equivalence of operators $\hat{A}_c$ and $\hat{B}_c$. Here we set \mbox{$\boldsymbol{h}^\sigma=\boldsymbol{h}^\mu=(h_x,0,0)$}: 
%
\begin{equation}
  \label{eq:haah_two_type}
  \begin{aligned}  \hat{\mathcal{H}}^{\rm 2type,\parallel}_{\text{Haah}} &= - J \sum_c \left( \hat{A}_c +\hat B_c \right)- h_x \sum_i\left( \sigma_i^x + \mu_i^x\right) \, . 
		  \end{aligned}
\end{equation}
It is then possible to introduce pseudo-spins 1/2 with Pauli matrices $\boldsymbol{\tau}_{\boldsymbol{c}}$ located at the center of cubes $c$ so that the diagonal entries of $\tau^z_{\boldsymbol{c}}$  correspond to $a_c=\pm 1$. Setting $b_c=+1$ for all $c$ in order to focus on the relevant low-energy sector, one obtains the following dual formulation as single-type transverse-field fractal Ising model \cite{Vijay_2016}
%
\begin{equation}
  \label{eq:haah_two_type_duality}
  \begin{aligned}  \hat{\mathcal{H}}^{\rm 2type,\parallel,dual}_{\text{Haah}} =&  -J N_{\rm c}-J\sum_{\boldsymbol{c}} \tau_{\boldsymbol{c}}^z\\
                  &   -h_x\sum_{\substack{{\rm tetrahedra}\\a=(a_1,a_2,a_3,a_4)}}
                 \tau^x_{a_1}\tau^x_{a_2} \tau^x_{a_3} \tau^x_{a_4} \\
                  &   -h_x\sum_{\substack{{\rm tetrahedra}\\b=(b_1,b_2,b_3,b_4)}}
                 \tau^x_{b_1}\tau^x_{b_2} \tau^x_{b_3} \tau^x_{b_4}
		    \, ,   \end{aligned}
\end{equation}
where the sums run over the two sets of tetrahedra as illustrated in Fig.~\ref{HaahsCode_InMag_SelfDuality}, left and center. Note that all our results discussed below are therefore also valid for the dual transverse-field fractal Ising model. 

In the \textit{two-type orthogonal case}, $\boldsymbol{h}^\sigma$ and $\boldsymbol{h}^\mu$ are both nonzero, but one of them points in $x$-direction and the other in $z$-direction. Here we choose \mbox{$\boldsymbol{h}^\sigma=(0,0,h_z^\sigma)$} and \mbox{$\boldsymbol{h}^\mu=(h_x^\mu,0,0)$} so that Eq.~\eqref{equ::HaahsCode::HamiltonianMag} can be split into two independent parts $\hat{\mathcal{H}}_{A,\mu}$ and $\hat{\mathcal{H}}_{B,\sigma}$ with
%
\begin{eqnarray}
  \hat{\mathcal{H}}_{A,\mu} &=& -J\sum_c \hat{A}_c-h_x^\mu\sum_i\hat{\mu}_i^x  \\ 
  \hat{\mathcal{H}}_{B,\sigma} &=& -J\sum_c \hat{B}_c-h_z^\sigma\sum_i\hat{\sigma}_i^z   
\label{equ::HaahsCode::Two-type_orthogonal}
\end{eqnarray}
and $[\hat{\mathcal{H}}_{A,\mu},\hat{\mathcal{H}}_{B,\sigma}]=0$. The two-type orthogonal case therefore reduces to the two independent parts $\hat{\mathcal{H}}_{A,\mu}$ and $\hat{\mathcal{H}}_{B,\sigma}$, which we call single-type cases as the field terms only depend on one type of spin operators. Below we show that both single-type cases possess each an exact self-duality.

In the \textit{single-type} case where $\boldsymbol{h}^\sigma$ or $\boldsymbol{h}^\mu$ is zero and the other field points in $x$- or $z$-direction, note that both field directions feature exactly the same physics due to the duality of $\hat{A}_c$- and $\hat{B}_c$ operators. We consider specifically \mbox{$\boldsymbol{h}^\sigma=(h_x^\sigma,0,0)$} and $\boldsymbol{h}^\mu=0$ so that the $\hat{B}_c$ operators remain conserved quantities and the Hamiltonian reads
%
\begin{equation}
  \label{eq:haah_single_type}
  \begin{aligned}  \hat{\mathcal{H}}^{\rm single}_{\text{Haah}} &= - J \sum_c  \hat{A}_c - J \sum_c \hat B_c - h_x^\sigma \sum_i \sigma_i^x\, . 
		     \end{aligned}
\end{equation}

As in the two-type parallel case, the eigenvalues $b_c$ remain conserved quantities and the introduction of the same kind of pseudo-spins 1/2 yields the single-type transverse-field fractal Ising model \cite{Vijay_2016}
%
\begin{equation}
  \label{eq:haah_single_type_duality}
  \begin{aligned}  \hat{\mathcal{H}}^{\rm single,dual}_{\text{Haah}} =&  -J N_{\rm c}-J\sum_{\boldsymbol{c}} \tau_{\boldsymbol{c}}^z\\
                  &   -h_x^\sigma \sum_{\substack{{\rm tetrahedra}\\a=(a_1,a_2,a_3,a_4)}}
                 \tau^x_{a_1}\tau^x_{a_2} \tau^x_{a_3} \tau^x_{a_4}  
		    \, ,   \end{aligned}
\end{equation}
where the second sum runs over the four sites $a_j$ of a tetrahedron as illustrated in Fig.~\ref{HaahsCode_InMag_SelfDuality}, left. As a consequence, the lattice topology corresponds to a Sierpinski lattice having a fractal dimension. Keeping in mind that the $\mu$-spins do not play any role for this specific single-type field, one observes that the action of the $\hat{A}_c$ operators in Eq.~\eqref{eq:haah_single_type} on the $\sigma$-spins is identical to the four-spin interaction in Eq.~\eqref{eq:haah_single_type_duality}. As a consequence, Haah's code in a single-type field in $x$-direction as well as its dual single-type transverse-field fractal Ising model are self-dual, e.g., one has for the ground-state energy per cube as a function of interaction strengths $J$ and $h_x^\sigma$ the relation $\epsilon (J,h_x^\sigma)=\epsilon (h_x^\sigma, J)$.  

\subsubsection*{The X-Cube model in a homogeneous magnetic field}
The Hamiltonian of the X-Cube model in the presence of a homogeneous magnetic field pointing in an arbitrary direction reads
\begin{equation*}
\mathcal{\hat H} = - J \sum_c \hat{A}_c - J \sum_{s, \kappa} \hat{B}_{s}^{(\kappa)} - \sum_i \left( \boldsymbol{h}\cdot  \boldsymbol{\hat \sigma}_i \right),
\label{equ::XCube::HamiltonianMag}
\end{equation*}
where $\boldsymbol{h}:=(h_x,h_y,h_z)$ and $\boldsymbol{\sigma}_i:=(\sigma^x_i,\sigma^y_i,\sigma^z_i)$. Again, a quantum phase transition must occur between the fracton low-field limit with $ J \gg \abs{\boldsymbol{h}}$ and the high-field limit with $ J \ll \abs{\boldsymbol{h}}$, where the system is in a trivial polarized phase.

In this work we concentrate on the two specific single-field directions $(h_x,0,0)$ and $(0,0,h_z)$, which have also been investigated by recent QMC simulations \cite{Devakul_2018} finding first-order phase transitions between the fracton and the polarized phase. Here either the operators $\hat{A}_c$ or the operators $\hat{B}^{(\kappa)}_s$ trivially commute with the field term, i.e., the corresponding eigenvalues remain conserved quantities at any field strength. As a consequence, a field in $x$- or $z$-direction modifies only the eigenvalues $b_s^{(\kappa)}$ or $a_c$, respectively. Furthermore, for these two cases it is possible to find the following duality transformations to isospectral models.

First we consider the dual model of the X-Cube model in a magnetic field $(0,0,h_z)$. In this case the eigenvalues $b_s^{(\kappa)}$ are conserved quantities so that the Hilbert space decouples into independent sectors. The relevant low-energy physics takes place in the sector with $b_s^{(\kappa)}=+1$ for all $s$ and $\kappa$ (see Ref.~[\onlinecite{Vijay_2016}] for the general mapping to a generalized gauge theory in all sectors). It is then possible to introduce pseudo-spins 1/2 located at the center of cubes $c$ and onto which Pauli matrices $\boldsymbol{\tau}_{\boldsymbol{c}}$ act so that $\tau^z_{\boldsymbol{c}}$ has diagonal entries corresponding to $a_c=\pm 1$. The bare X-Cube model is then mapped to an effective field term. In contrast, the local action of the $z$-field on the pseudo-spins is to flip four pseudo-spins forming a local plaquette. Altogether, one obtains \cite{Vijay_2016,Williamson_2016,Nandkishore18} as dual a transverse-field plaquette Ising model 
%
\begin{equation}
  \label{eq:xcube_z_duality}
  \begin{aligned} 
  \hat{\mathcal{H}}^{z,{\rm dual}}_{\text{X-Cube}} =&  -3J N_{\rm c}-J\sum_{\boldsymbol{c}} \tau_{\boldsymbol{c}}^z\\
                  &   -h_z \sum_{\boldsymbol{c}}\sum_{(\alpha,\beta)} \tau^x_{\boldsymbol{c}}\tau^x_{\boldsymbol{c}+ \boldsymbol{\mathrm{e}}_\alpha}  \tau^x_{\boldsymbol{c}+ \boldsymbol{\mathrm{e}}_\beta} \tau^x_{\boldsymbol{c}+ \boldsymbol{\mathrm{e}}_\alpha+\boldsymbol{\mathrm{e}}_\beta}  
		    \, ,   \end{aligned}
\end{equation}
where the second sum runs over the tuples \mbox{$(\alpha,\beta) \in \{(x,y), (x,z), (y,z) \}$}. Note that the conditions $b_s^{(\kappa)}=+1$ for all $s$ and $\kappa$ represent constraints on the allowed states of the Hilbert space in the dual pseudo-spin formulation. Furthermore, the model has an important subsystem symmetry:
flipping $\tau^x\rightarrow -\tau^x$ in any plane of the cubic lattice leaves the model invariant \cite{footnote1}. 

For the case of the X-Cube model in a field in $x$-direction, it is also possible to formulate a dual description, which is, however, more involved. Here the eigenvalues $a_c$ are conserved quantities so that again decoupled Hilbert space sectors exist. The relevant low-energy physics takes place in the sector with $a_c=+1$ for all $c$, which we focus on in the following. Keeping in mind the local constraint $\prod_\kappa \hat{B}_{s}^{(\kappa)} = \mathds{1}$, the local Hilbert space at each vertex $s$ is four-dimensional, consisting of the vacuum configuration $b_s^{(\kappa)}=+1$ for all $\kappa$ and the three types $\alpha\in\{x,y,z\}$ of lineon excitations $l_\alpha$. We therefore introduce hardcore boson creation (annihilation) operators $t^\dagger_{\alpha,s}$ ($t^{\phantom{\dagger}}_{\alpha,s}$) creating (annihilating) a lineon with flavor $\alpha$ on site $s$. In contrast to the case of a field in $z$-direction, the relevant pseudo-spin degrees of freedom are rather dimer states like singlet and triplet operators in valence bond solids. The dual Hamiltonian can then be written as 
\begin{equation}
  \label{eq:xcube_x_duality}
 \begin{aligned}
\mathcal{H}^{x,\text{dual}}_{\text{X-Cube}} = & -(4N_s J/3) - 2 J \sum_{s,\alpha=x,y,z} t^\dagger_{\alpha,s} t_{\alpha,s}^{\phantom{\dagger}}\\ & - 
h_x \sum\limits_{\substack{ \alpha = x,y,z \\ <s,s^\prime>_\alpha}} \Bigg[ \left( t^\dagger_{\alpha,s} + t_{\beta,s}^{\dagger} t_{\gamma,s}^{\phantom{\dagger}} + \mathrm{h.c.}  \right)  \\ & \otimes \left( t^\dagger_{\alpha,s^\prime} + t_{\beta,s^\prime}^{\dagger} t_{\gamma,s^\prime}^{\phantom{\dagger}}+ \mathrm{h.c.}  \right)  \Bigg]
\, , 
  \end{aligned}
 \end{equation}
where $\alpha, \beta, \gamma$ are always cyclic and $\langle s,s'\rangle_\alpha$ denotes nearest neighbors in $\alpha$-direction. 
To the best of our knowledge, this model has not been presented and studied before. 
It also possesses a subsystem symmetry in the form of the parity $\prod_{s \in \text{plane}, \kappa \parallel \text{plane}} B_s^{(\kappa)} = \mathds{1}$ of the $b_s^{(\kappa)}$ for each plane. This symmetry is, however, not obvious in the dual formulation, since it is related to the orientation of the operators $\hat{B}^{(\kappa)}_s$ in real space. The restricted mobility of a lineon can then be seen as a consequence of this parity conservation.

\section{Methods}
\label{sect::methods}

In order to study the quantum robustness of fracton models in a magnetic field as introduced in Sect.~\ref{sect::FF}, we apply the method of perturbative continuous unitary transformations (pCUT) and variational calculations. The latter technique gives qualitative insights in the ground-state phase diagrams. The phase transitions are then located quantitatively by high-order series expansions of the ground-state energy using the pCUT method. Furthermore, this method is used to calculate the single- and multi-fracton excitation energies within the fracton phase at finite fields. In the following we give the relevant technical information for both approaches. 

\subsection{pCUT method}
\label{ssect::pCUT}

We perform high-order series expansions for both fracton models in a field. Technically, these high-order linked-cluster expansions can be gained with the help of pCUT \cite{Knetter_2000,Knetter_2003}, whose generic aspects are described in the following.

One can always rewrite any lattice Hamiltonian ${\cal H}$ exactly as
\begin{equation}
\label{Eq:Hami}
\hat{{\cal H}}=\hat{{\cal H}}_0+\sum_j \lambda_j \hat{{\cal V}}^{(j)}\quad ,
\end{equation}
where the $\lambda_j$ are the perturbative parameters and the unperturbed part $\hat{\mathcal{H}}_0$ is diagonal in appropriate supersites. For the conventional high-field expansion we use single spins as supersites, while for the low-field expansion inside the fracton phase, pseudo-spins 1/2 referring to the eigenvalues of the stabilizer operators are the relevant degrees of freedom. In both cases one can express $\hat{\mathcal{H}}_0$ in appropriate units as 
\begin{eqnarray}\label{h_0_q}
\hat{\mathcal{H}}_0 &=& E_0+\hat{\mathcal{Q}} \quad ,
\end{eqnarray}
where $E_0$ denotes a constant and $\hat{\mathcal{Q}}$ is an operator counting the local excitations. This decomposition of $\hat{\mathcal{H}}_0$ is always possible, since the local spectra of the supersites is equidistant in all considered cases.

Supersites interact via the perturbation \mbox{$\mathcal{V}\equiv\sum_j\lambda_j\hat{\mathcal{V}}^{(j)}$}. For the high- and low-field expansion several supersites are linked simultaneously by the perturbation. As a consequence of Eq.~\eqref{h_0_q}, one can rewrite Eq.~\eqref{Eq:Hami} as
\begin{equation}
\label{Eq:Hami_final}
\hat{{\cal H}}=\hat{{\cal H}}_0+ \sum_{n=-N}^N \hat{T}_n \quad ,
\end{equation}
so that $[\hat{\mathcal{Q}},\hat{T}_n]=n\hat{T}_n$. Physically, the operator \mbox{$\hat{T}_n \equiv\sum_j \lambda_j \hat{T}^{(j)}_n$} corresponds to all processes where the change of energy quanta with respect to $\hat{\mathcal{H}}_0$ is exactly $n$. The maximal (finite) change in energy quanta is called $\pm N$. For Haah's code in a field, $N=4$ holds for all low- and high-field expansions performed. For the X-cube model in a field, both low-field expansions in $\lambda_x:=h_x/J$ and $\lambda_z:=h_z/J$ as well as the high-field expansion in $\tilde{\lambda}_x:=1/\lambda_x$ also feature $N=4$, while $N=12$ for the high-field expansion in $\tilde{\lambda}_z:=1/\lambda_z$. Note, however, that in all considered cases all $n$ are even, which reduces the complexity of the pCUT calculation.

In the pCUT method, Hamiltonian \eqref{Eq:Hami_final} is mapped model-independently up to high orders in the perturbations to an effective Hamiltonian $\hat{\mathcal{H}}_\text{eff}$ with $[\hat{\mathcal{H}}_{\rm eff},\hat{\mathcal{Q}}]=0$. The general structure of $\hat{\mathcal{H}}_{\rm eff}$ is then a weighted sum of operator products $\hat{T}_{n_1}\cdots \hat{T}_{n_k}$ in perturbation theory of order $k$. The block-diagonal $\hat{\mathcal{H}}_\text{eff}$ conserves the number of quasi-particles (qp). This represents a major simplification of the quantum many-body problem, since one can treat each quasi-particle block separately, corresponding to a few-body problem. For example the 0qp-block is given by a single matrix element representing the ground-state energy in all considered cases. Physically, the quasi-particles in the fracton phase correspond to dressed fractons (or composites of fractons), while the elementary quasi-particles of the polarized phase are conventional dressed spin-flip excitations.

The more demanding part of the pCUT method is model-dependent and corresponds to a normal-ordering of $\hat{\mathcal{H}}_\text{eff}$ for which the explicit processes have to be specified. This can be either done via a full graph decomposition in linked graphs using the linked-cluster theorem and an appropriate embedding scheme afterwards \cite{Coester_2015} or by calculations on large enough finite clusters, which include all relevant virtual processes. 

Next we discuss the specific pCUT implementation for both fracton codes in a field and detail how we extrapolate the obtained series.

\subsubsection{pCUT for Haah's code in a field}
\label{sssect::pCUT_haah}
All pCUT calculations for Haah's code in a field are performed on sufficiently large clusters. This allows to determine the ground-state energy per cube $\epsilon_{\rm LF}$ ($\epsilon_{\rm HF}$) up to order six in $\lambda:=|\boldsymbol{h}|/J$ ($\tilde{\lambda}:=1/\lambda$) for the low-field fracton (high-field polarized) phase. 

Furthermore, for the low-field fracton phase, pCUT calculations in the one-, two-, and four-fracton sector have been conducted up to order six in $\lambda$ except for the four-fracton sector in the two-type parallel case where order four has been achieved. Here the restricted mobility of fractons is advantageous for the pCUT calculation, since the number of kinetic processes as well as interactions between fractons is highly reduced. Indeed, a single fracton is not allowed to move at all so that the dispersion is completely flat in momentum space and is solely characterized by a local hopping amplitude, which can be calculated as a high-order series expansion in $\lambda$ yielding directly the one-fracton gap $\Delta_{\rm lf}^{\rm 1qp}$.

For the two-fracton and four-fracton sector we find several states connected via non-zero transition probability amplitudes in a given perturbative order. To find all contributing states we start from an initial state $\ket{\boldsymbol{K}, \alpha = 0 }$, where $\boldsymbol{K}$ is the center of mass momentum and $\alpha$ is an arbitrary but fixed enumeration of contributing states.
	As we evaluate $\hat{\mathcal{H}}_\text{eff}$ in position space, we apply a Fourier transformation to $\ket{\boldsymbol{K}, \alpha}$ before applying the effective Hamiltonian. This application results in new states $\ket{\boldsymbol{R}_\beta, \beta}$ of a potentially different multi-particle configuration with a different center of mass $\boldsymbol{R}_\beta$ and a pCUT coefficient $C_\beta$. Finally, we recover a representation in momentum space by a second Fourier transform, where we get an additional factor of $\exp(i \boldsymbol{K} \Delta\boldsymbol{R}_\beta )$ to account for the difference of the centers of mass $\Delta \boldsymbol{R}_\beta = \boldsymbol{R}-\boldsymbol{R}_\beta$. More formally
	\begin{eqnarray*}
		\hat{\mathcal{H}}_\text{eff} \ket{\boldsymbol{K}, \alpha } &=& \frac{1}{\sqrt{N}} \sum_{\boldsymbol{R}} \text{e}^{-i\boldsymbol{K} \boldsymbol{R}} \hat{\mathcal{H}}_\text{eff} \ket{\boldsymbol{R}, \alpha } \\
		& = &  \frac{1}{\sqrt{N}} \sum_{\boldsymbol{R}} \text{e}^{-i\boldsymbol{K} \boldsymbol{R}} \sum_\beta C_\beta\ket{\boldsymbol{R}_\beta, \beta } \\
		& = &  \sum_\beta C_\beta \cdot \text{e}^{-i\boldsymbol{K} \Delta \boldsymbol{R}_\beta } \ket{\boldsymbol{K}, \beta}.
	\end{eqnarray*}
	We iterate this for all new states $\ket{\boldsymbol{K},\beta}$ from the previous step until no new states are found. Due to the fractal character of operators in Haah's code, we end up with a finite number of states and hence a finite-dimensional matrix in a given perturbative order. Additionally, for the two-fracton sector in both single- and two-type cases, we do not find a hopping of any qp. Hence, the dispersion in these cases is flat.
        
	Finally, we want to extrapolate the gap of the lowest mode of each case up to the critical point to discuss the nature of the phase transition. In order to do so, we have to find expressions of the gaps in terms of series in $\lambda$. In the single-type case, we find two contributing states for both the two- and four-fracton sector.
	Hence, we can analytically diagonalize the resulting $2\times 2$ matrices and expand all expressions in a Taylor series.
	For the two-type parallel case, we have larger matrices such that an analytical diagonalization is not possible.
	In these cases we diagonalized the matrix numerically for $\lambda \ll 1$.
	In this regime the lowest contributing order dominates and we can make a line-fit to the double logarithmic plot of the data. 
	The slope of such a line-fit is the exponent of the lowest contributing order and the $y$-intersect is the logarithm of its coefficient.
	As a next step, we subtract the best line-fit from the numerical data and repeat the steps described before.
	In that way, we can fit a series order by order with very high accuracy.
	In order to reduce the effect of numerical noise, we used a multi-precision library with up to 1000 digits in machine precision. 
	Note that this is possible, because all numbers in the original matrices are rational or integer numbers.

\subsubsection{pCUT for the X-Cube in a field}
\label{sssect::pCUT_xcube}
The pCUT calculations for the ground-state energy per link have been calculated via a full-graph decomposition in both phases. This allows to determine the ground-state energy per site $\epsilon_{\rm LF}$ ($\epsilon_{\rm HF}$) up to order eight in $\lambda_x$ ($\tilde{\lambda}_x$) and order six (eight) in $\lambda_z$ ($\tilde{\lambda}_z$). In the low-field (high-field) expansion this corresponds to a total number of 133 (29) of contributing distinct graphs for $\lambda_x$ ($\tilde{\lambda}_x$). For the low-field expansion in $\lambda_z$ no graph decomposition has been used, while for the high-field expansion in $\tilde{\lambda}_z$ 31 graphs contribute due to a double-touch property.

Furthermore, we have calculated the fracton excitation energy of a single cube excitation with  $a_c=-1$ (1qp) up to order six in $\lambda_z$. This excitation is strictly local and the local hopping element does directly correspond to the one-fracton gap $\Delta_{\rm fracton}$. Let us stress again that a single fracton with $b_s^{(\kappa)}=-1$ is not allowed due to the local constraint $\prod_\kappa \hat{B}_{s}^{(\kappa)} = \mathds{1}$.

In the 2qp sector, the dispersion $\omega^{\alpha}_{\rm lineon}(\boldsymbol{k})$ of $\alpha$-lineons with $\alpha\in\{x,y,z\}$ and the dispersion $\omega^{\beta}_{\rm planon}(\boldsymbol{k})$ of $\beta$-planons with $\beta\in\{xy,xz,yz\}$ and $\boldsymbol{k}=(k_x,k_y,k_z)$ have been determined up to order $10$ in $\lambda_x$ and order $7$ in $\lambda_z$, respectively. This was achieved by determining the associated hopping amplitudes on sufficiently large clusters and diagonalizing the effective one-particle hopping Hamiltonian by an appropriate Fourier transformation. 

As discussed above, lineon excitations are only allowed to hop in one dimension. The dispersion of an $\alpha$-lineon therefore only depends on $k_\alpha$ and is flat in the other two momentum directions. Moreover, by symmetry, one has $\omega^{x}_{\rm lineon}(k_x,k_y,k_z)=\omega^{y}_{\rm lineon}(k_y,k_x,k_z)=\omega^{z}_{\rm lineon}(k_z,k_y,k_x)$. The gap $\Delta_{\rm lineon}$ of the $\alpha$-lineon is always located at $k_\alpha=0$. 

The planon excitation is allowed to hop in two-dimensional planes. The dispersion of the $\beta$-planon therefore only depends symmetrically on two momenta and is flat in the third momentum component. The following relations hold by symmetry: $\omega^{xy}_{\rm planon}(k_x,k_y,k_z)=\omega^{xz}_{\rm planon}(k_x,k_z,k_y)=\omega^{yz}_{\rm planon}(k_z,k_y,k_x)$. The gap $\Delta_{\rm planon}$ of the $\beta$-planon is always located at zero $\beta$-components of the momentum.

\subsubsection{Extrapolation}
\label{sssect::extrapol}
Pad\'{e} and Dlog-Pad\'{e} extrapolation schemes are standard methods in the field of series expansions as they allow for an evaluation of the series beyond their original radius of convergence \cite{Guttmann1989}. Pad\'{e} extrapolation is generally used for extrapolating ground-state energies \cite{Roechner2016}, especially when critical fluctuations are absent as in the case of first-order phase transitions found for the perturbed fracton codes studied in this work. We therefore locate the phase-transition point and investigate the convergence of (multi-)fracton excitations energies via Pad\'{e} extrapolations. To this end the perturbation series 
\begin{equation}
	F(\lambda)=\sum_{m = 0}^{r} c_m \lambda^m=c_0+c_1\lambda+c_2\lambda^2+\dots c_{r}\lambda^r,
\end{equation}
with $\lambda\in \mathbb{R}$ and $c_m \in \mathbb{R}$ is interpreted as a Taylor expansion of the Pad\'{e} extrapolant
\begin{equation}
	G^{L/M}(\lambda) = \frac{p_0 + p_1 \lambda + p_2 \lambda^2 +\cdots+p_L \lambda^L}{1 + q_1 \lambda + q_2 \lambda^2 +\cdots+q_M \lambda^M}~.
\end{equation}
Comparing the Taylor expansion of $G^{L/M}(\lambda)$ about \mbox{$\lambda=0$} with the original series $F(\lambda)$, one obtains a linear system of equations that can be solved for a given parameter set $(L,M)$ with $L,M\in\mathbb{N}$ fulfilling the condition $L+M=r$ for an extrapolation in order $r$. Typically the extrapolations with $|L-M|$ being small give the best results.  Extrapolations with unphysical singularities need to be sorted out, as well as defective Pad\'{e} extrapolants that have a singularity at the same point in the numerator and denominator which effectively cancels out. 

A first-order phase transition is located at the critical value $\lambda_{\rm c}$ for which $\epsilon_{\rm lf}=\epsilon_{\rm hf}$. This can be determined by using several non-defective Pad\'e extrapolants $(L,M)$ with $L,M\geq 2$ and maximal order $r$ from the low- and high-field expansion. In the later sections we show the mean and standard deviation of $\lambda_{\rm c}$ obtained by all combinations of (non-defective) low- and high-field Pad\'e extrapolants. The sample standard deviation serves as an uncertainty measure of the extrapolation and does not represent a numerical error bar. The same criteria are applied to the extrapolations of the fracton excitation energies.

\subsection{Variational approach}
\label{ssect::variational}
Another approach to approximately locate the quantum phase transition between the fracton and the polarized phases is to use a variational ansatz
which contains both limits exactly. In the past, this has been exploited successfully for two- and three-dimensional topological codes in a magnetic field in Refs.~[\onlinecite{Dusuel_2015}] and [\onlinecite{Reiss_2019}], respectively.

For both fracton codes in a field, the ground state in the high-field limit is the state where all spins are polarized in the direction of the magnetic field. We denote this state with $\ket{\boldsymbol{h}}$. The ground state in the low-field limit is the topological state with fracton order that can be written using projectors as in Eqs.~\eqref{equ::HaahsCode::Groundstate} and \eqref{GS1}.
	For now we will write the ground state as
	\begin{equation*}
	\ket{0} = \prod_s \left( \frac{\mathds{1}+ \hat A_s}{2}\right) \prod_p  \left(\frac{\mathds{1}+ \hat B_p}{2} \right) \ket{\boldsymbol{h}}
	\end{equation*}
	where the operators $\hat{A}_s$ and $\hat{B}_p$ are either the operators $\hat{A}_c$ and $\hat{B}_c$ for Haah's code or $\hat{A}_c$ and $\hat{B}^{(\kappa)}_s$ (with \mbox{$\prod_p=\prod_s\prod_\kappa$}) for the X-Cube model. This definition is valid provided that $\ket{\boldsymbol{h}}$ is not orthogonal to the ground state. \par
	
The idea of the variational ansatz introduced in \cite{Dusuel_2015} is to define a state $\ket{\alpha, \beta}$ that can be tuned between the exact low- and high-field ground state using two variational parameters $\alpha$ and $\beta$. This variational state reads
	\begin{equation}
	\ket{\alpha, \beta} = \mathcal{N}(\alpha,\beta) \prod_s (\mathds{1}+ \alpha \hat{A}_s) \prod_p (\mathds{1} + \beta \hat B_p) \ket{\boldsymbol{h}},
	\label{equ::VariatioalAnsatz::VariationalState}
	\end{equation}
where $\mathcal{N}(\alpha,\beta)$ is a normalization factor depending on $\alpha$ and $\beta$. Setting $\alpha = \beta = 0$ yields the polarized state $\ket{\boldsymbol{h}}$ and $\alpha = \beta = 1$ results in the fracton state $\ket{0}$. 
	
The variational energy $E(\alpha ,\beta)$ can then be calculated as a function of $\alpha$ and $\beta$ by evaluating
	\begin{equation*}
	E(\alpha ,\beta) = \bra{\alpha ,\beta} \hat{\mathcal{H}} \ket{\alpha ,\beta}.
	\end{equation*}
Finally, minimizing $E(\alpha ,\beta)$ with respect to $\alpha$ and $\beta$ for different magnetic fields $\boldsymbol{h}$ allows to determine an approximate variational phase diagram. 
	
\section{Phase diagrams}
\label{sect::pd}
In this section we locate quantitatively the phase transition between the fracton order and the polarized phase by comparing the low- and high-field expansion of the ground-state energy per site for both fracton codes in a field. The phase transition point corresponds to the crossing point of both expansions, which are found to be very well converged up to the crossing on each side of the transition signaling a (strong) first-order phase transition in all studied cases in agreement with our variational calculations as well as with QMC simulations for the X-Cube model in a field \cite{Devakul_2018}. This finding is further strongly supported by the energetic properties of (multi-)fracton excitations in the fracton phase presented in Sect.~\ref{sect::fractons}.

\subsection{Haah's code in a field}
\label{ssect::haah_pd}
In the following we determine the ground-state phase diagram for Haah's code in a field for the single- and two-type cases.

\subsubsection*{Single-type case}
As a reference point, we start with the single-type case by choosing \mbox{$\boldsymbol{h}^\sigma=(h_x^\sigma,0,0)$} and $\boldsymbol{h}^\mu=0$ as discussed in Sect~\ref{sect::FF}. Consequently, the $\hat{B}_c$ operators remain conserved quantities and we assume $b_c=+1$ for all $c$, which is the relevant sector to study the ground-state phase diagram. Here we know the phase transition point $\lambda_{x,{\rm c}}\equiv (h_x^\sigma/J)_{\rm c}=1$ exactly due to the self-duality $\epsilon(J,h_x^\sigma)=\epsilon(h_x^\sigma,J)$ as demonstrated in Sec.~\ref{sect::FF}. This case is therefore perfectly suited to gauge our results.

\begin{figure}
	\centering
	\adjincludegraphics[width = \columnwidth]{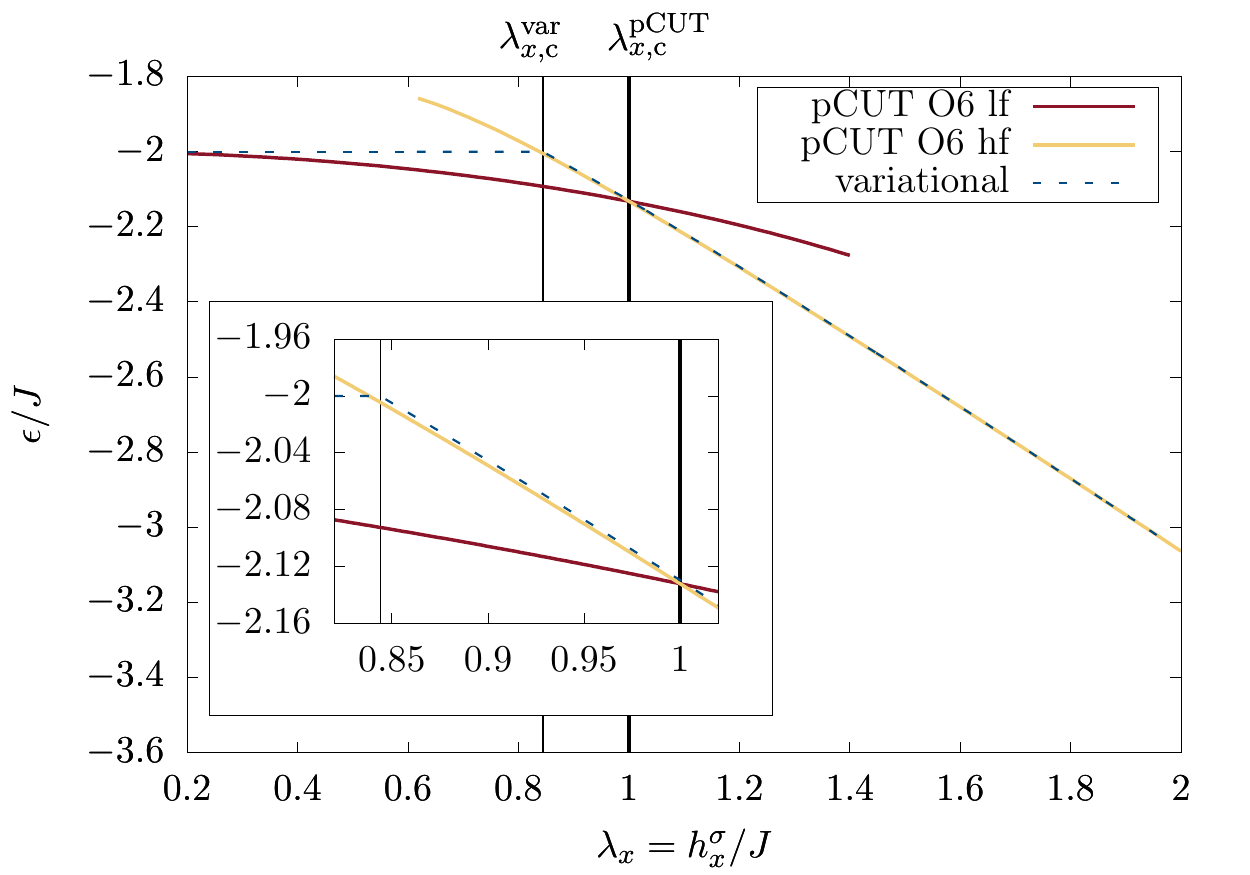}
	\caption{Ground-state energy per site $\epsilon/J$ of Haah's code in the single-type case as a function of $\lambda_x=h_x^\sigma/J$ obtained by pCUT (solid lines) and from the variational ansatz (dashed lines). The high- and low-field expansion for pCUT intersect exactly at $\lambda_{x,\text{c}}^{\rm pCUT} = 1$ due to the self-duality. The variational phase transition is located at $\lambda_{x,\text{c}}^{\rm var} \approx 0.844$. The variational phase transition point is at smaller values of $\lambda_x$, since the variational energy of the fracton phase does not depend on the magnetic field.}
	\label{fig::haah_pd::singleField}
\end{figure}

Concerning the series expansion, the self-duality implies that the high-field expansion $\epsilon_\text{hf}(\tilde \lambda )$ can be obtained by taking the low-field expansion $\epsilon_\text{lf}(\lambda )$ and interchanging $\lambda_x=h_x^\sigma/J$ by $\tilde{\lambda}_x\equiv \lambda_x^{-1}$. We checked that this holds for our expansion up to order six using the pCUT method. The ground-state energy per cube in the low-field expansion for the single-type case reads
		\begin{eqnarray}
			\frac{\epsilon_\text{lf}+J}{J} &=& -1 -\frac{1}{8}\lambda_x^2  -  \frac{3}{512}\lambda_x^4 -\frac{559}{491520}\lambda_x^6\, ,
			\label{equ::BreakdownOfTopPhase::HomogenousInZ::pCUT::LowFieldResult}
		\end{eqnarray}
while the corresponding series for the high-field expansion is 
		\begin{eqnarray}
			\frac{\epsilon_\text{hf}+J}{h_x^\sigma}  &=& -1  - \frac{1}{8}{\tilde \lambda}_x^2 -  \frac{3}{512}{\tilde \lambda}_x^4 -\frac{559}{491520} {\tilde \lambda}_x^6\, .
			\label{equ::BreakdownOfTopPhase::HomogenousInZ::pCUT::HighFieldResult}
		\end{eqnarray}
Here we have put the contribution $-J\sum_c \hat{A}_c=-JN_c$ to the energy on the left side so that the self-duality becomes apparent. The pCUT results are shown in Fig.~\ref{fig::haah_pd::singleField} together with our results using the variational ansatz introduced in Subsect.~\ref{ssect::variational}. Obviously, the level crossing of the low- and high-field expansion is exactly at $\lambda_{x,{\rm c}}=1$ due to the self-duality. Further, we see that the series expansion is well converged up to the phase transition point and the kink implies a first-order phase transition. The pCUT approach appears therefore to be well suited to locate first-order phase transitions quantitatively for three-dimensional perturbed fracton phases. The first-order nature of the phase transition is also found by the variational calculation, which, however, does not respect the self-duality and a first-order phase transition at $\lambda_{x,{\rm c}}\approx 0.844$ is detected.

\subsubsection*{Two-type parallel case}

In the two-type parallel case we choose \mbox{$\boldsymbol{h}^\sigma=\boldsymbol{h}^\mu=(h_x,0,0)$} as introduced int Sec.~\ref{sect::FF} so that the single parameter $\lambda_x\equiv h_x/J$ can be tuned. Again, the $\hat{B}_c$ operators remain conserved quantities. The corresponding conserved eigenvalues $b_c$ are set to one for all $c$, since these fracton excitations are static and play no role for the low-energy physics. In contrast to the single-type case, the location of the phase transition between the fracton and the polarized phase is not known exactly. 

\begin{figure}
	\centering
	\adjincludegraphics[width = \columnwidth]{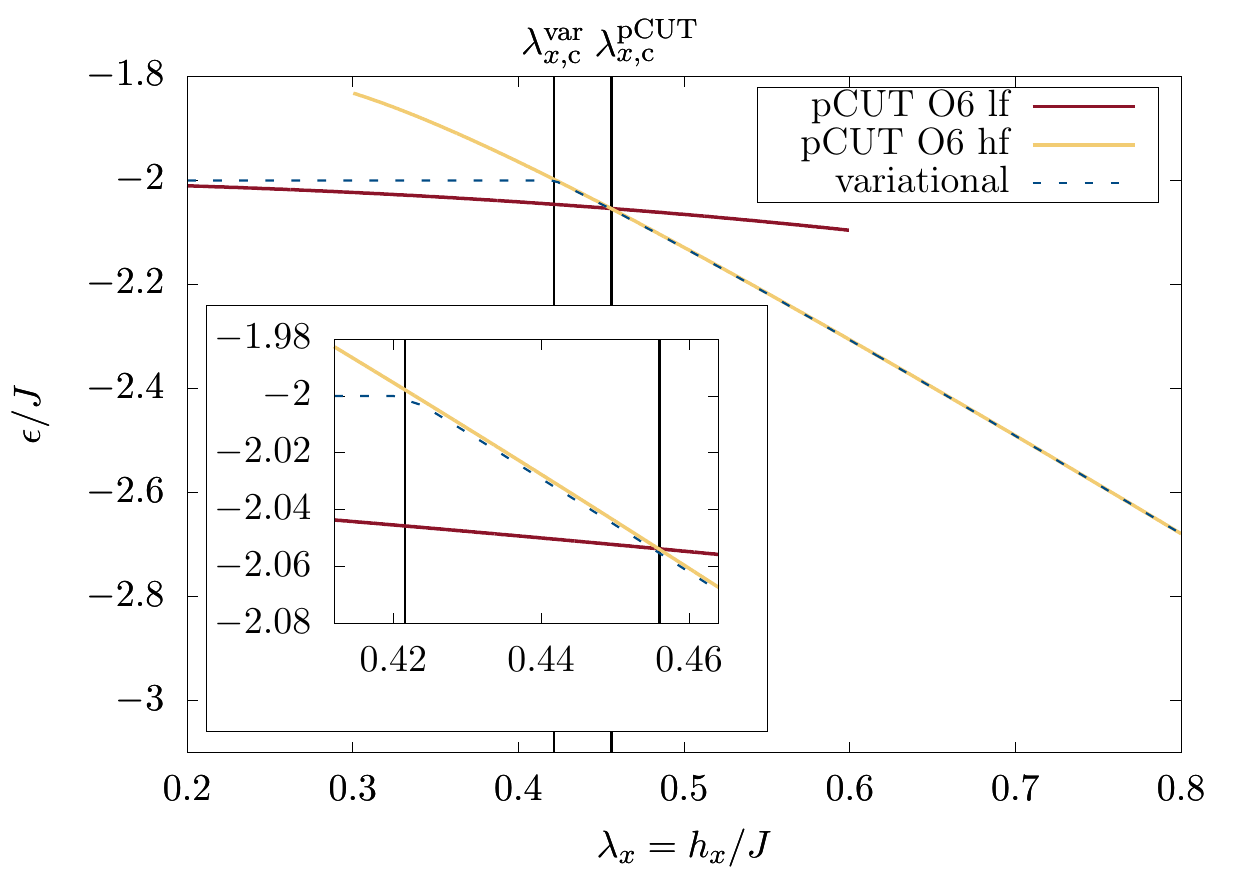}
	\caption{Ground-state energy per site $\epsilon/J$ of Haah's code in the two-type parallel case as a function of $\lambda_x=h_x/J$ obtained by pCUT (solid lines) and from the variational ansatz (dashed lines). The high- and low-field expansion for pCUT intersect at $\lambda_{x,{\rm c}}^{\rm pCUT} = 0.456 \pm 0.006$. The uncertainty represents the standard deviation of the energies at which the relevant Pad\'e extrapolants of the low- and high-field expansion $\epsilon_{\rm lf}$ and $\epsilon_{\rm hf}$ intersect. The variational phase transition point is located at $\lambda_\text{c,var} \approx 0.422$.}
	\label{fig::haah_pd::twoField}
\end{figure}

As in the single-type case, we have determined the low- and high-field expansion of the ground-state energy per cube up to order six using the pCUT method. The obtained series are given by
		\begin{eqnarray}
			\frac{\epsilon_\text{lf}+J}{J} &=& -1 -\frac{1}{4}\lambda_x^2  -  \frac{7}{192} \lambda_x^4-  \frac{18907}{983040}\lambda_x^6
			\label{equ::BreakdownOfTopPhase::HomogenousInZ::pCUT::HighFieldResultSolver}
		\end{eqnarray}
		and
		\begin{eqnarray}
		\frac{\epsilon_\text{hf} + J}{h} &=&-2 -\frac{1}{16} {\tilde \lambda}_x^2 -\frac{19}{57344} {\tilde \lambda}_x^4\nonumber\\
		                                 && - \frac{373249}{13872660480} {\tilde \lambda}_x^6 .
		\label{equ::BreakdownOfTopPhase::SigmaAndMu::pCUT::HighField::Energies}
		\end{eqnarray}
The pCUT and variational results for the two-type parallel case are shown in Fig.~\ref{fig::haah_pd::twoField}. Again, the series expansions are well converged up to the phase transition point and a Pad\'{e} analysis locates the phase transition at 
$\lambda_{x,{\rm c}}^{\rm pCUT} = 0.456 \pm 0.006$ 
with rather small uncertainty. The variational calculation detects the phase transition at slightly smaller values $\lambda_{x,\text{c}}^{\rm var} \approx 0.422$. The nature of the phase transition is found to be strongly first order with both approaches. This can be interpreted as the fracton phase being less robust in the two-type parallel case compared to the single-type case, since there are more fluctuations in the fracton phase and less fluctuations per spin in the polarized phase due to the two types of fields. 
		
\subsubsection*{Two-type orthogonal case}
\begin{figure}
	\centering
	\adjincludegraphics[width = \columnwidth, trim={ {.0\width} {.0\height} {.0\width} {.0\height}}, clip]{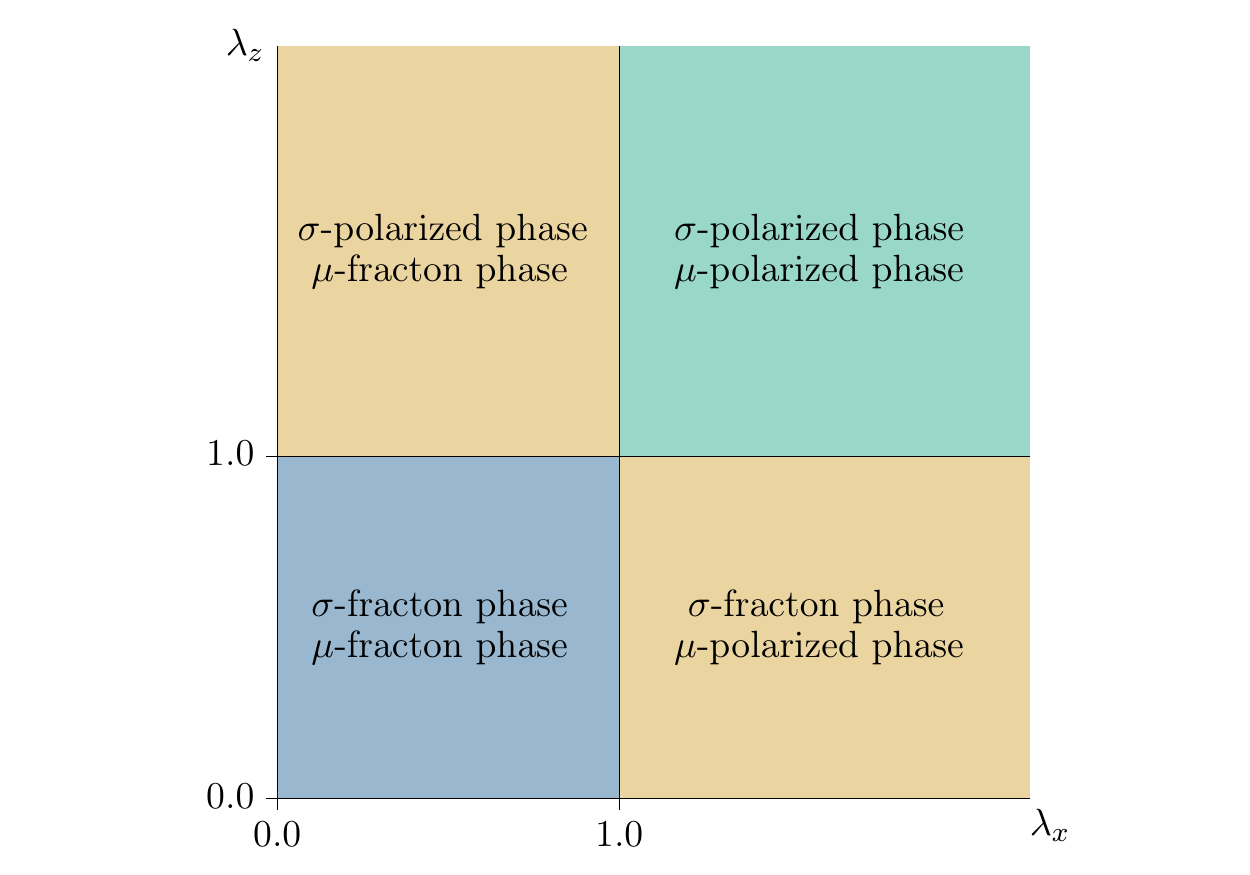}
	\caption{The exact ground-state phase diagram of Haah's code in the two-type orthogonal case as a function of $\lambda_x$ and $\lambda_z$.}
	\label{fig::haah_pd::mixedField}
\end{figure}

In the two-type orthogonal case we choose \mbox{$\boldsymbol{h}^\sigma=(0,0,h_z^\sigma)$} and \mbox{$\boldsymbol{h}^\mu=(h_x^\mu,0,0)$} and we introduce the two parameters $\lambda_z\equiv h_z^\sigma/J$ and $\lambda_x\equiv h_x^\mu/J$. As described in Subsect.~\ref{sect::FF}, Haah's code in this field configuration can be split into two independent Hamiltonians $\hat{\mathcal{H}}_{A,\mu}$ and $\hat{\mathcal{H}}_{B,\sigma}$ with $[\mathcal{H}_{A,\mu},\mathcal{H}_{B,\sigma}]=0$. This case therefore reduces to two single-type cases which can be treated independently. Furthermore, $\hat{\mathcal{H}}_{A,\mu}$ and $\hat{\mathcal{H}}_{B,\sigma}$ are self-dual which implies a first-order phase transition at $\lambda_{\alpha,{\rm c}}=1$ with $\alpha\in\{x,z\}$. The exact two-dimensional ground-state phase diagram of the two-type orthogonal case is displayed in Fig.~\ref{fig::haah_pd::mixedField} as a function of $\lambda_x$ and $\lambda_z$. It contains four distinct phases, since each subsystem can be either in a polarized or in a fracton phase.

\subsection{X-Cube model in a field}
\label{ssect::xcube_pd}
We now turn to the X-Cube model in a field and investigate its ground-state phase diagram for the two single-field cases using series expansions and the variational approach. In both cases no exact information on the phase transition is known, but we compare to QMC simulations \cite{Devakul_2018}. 

\subsubsection*{Single $x$-field}
We set \mbox{$\boldsymbol{h}=(h_x,0,0)$} and use the parameter \mbox{$\lambda_x=h_x/J$}. The eigenvalues $a_c$ of the $\hat{A}_c$ operators remain conserved quantities and the low-energy physics takes place in the sector with $a_c=+1$ for all $c$. In this sector the model is dual to the hardcore-boson model of Eq.~\eqref{eq:xcube_x_duality} as described in Sect.~\ref{sect::FF}.  

\begin{figure}
 \includegraphics[width= 0.9\columnwidth]{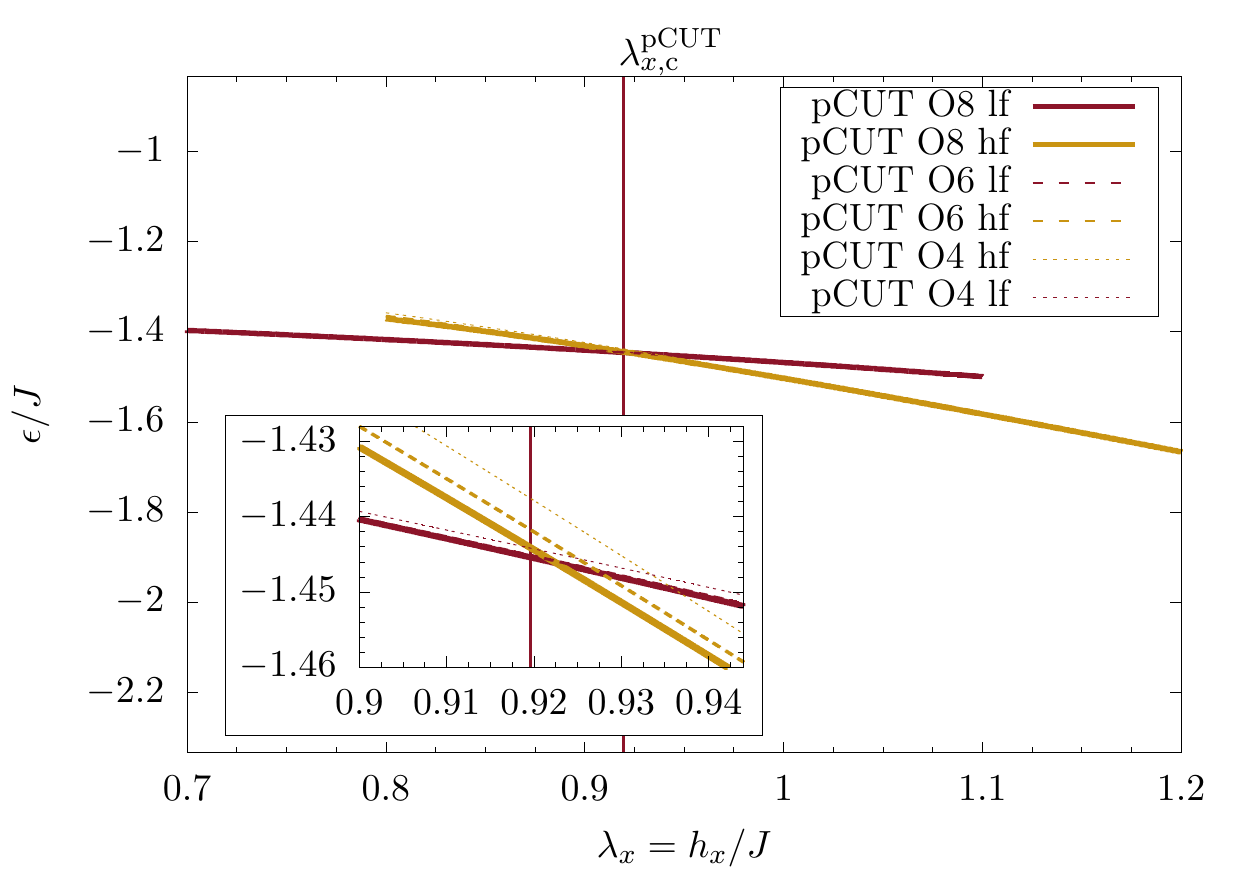}
 \includegraphics[width= 0.9\columnwidth]{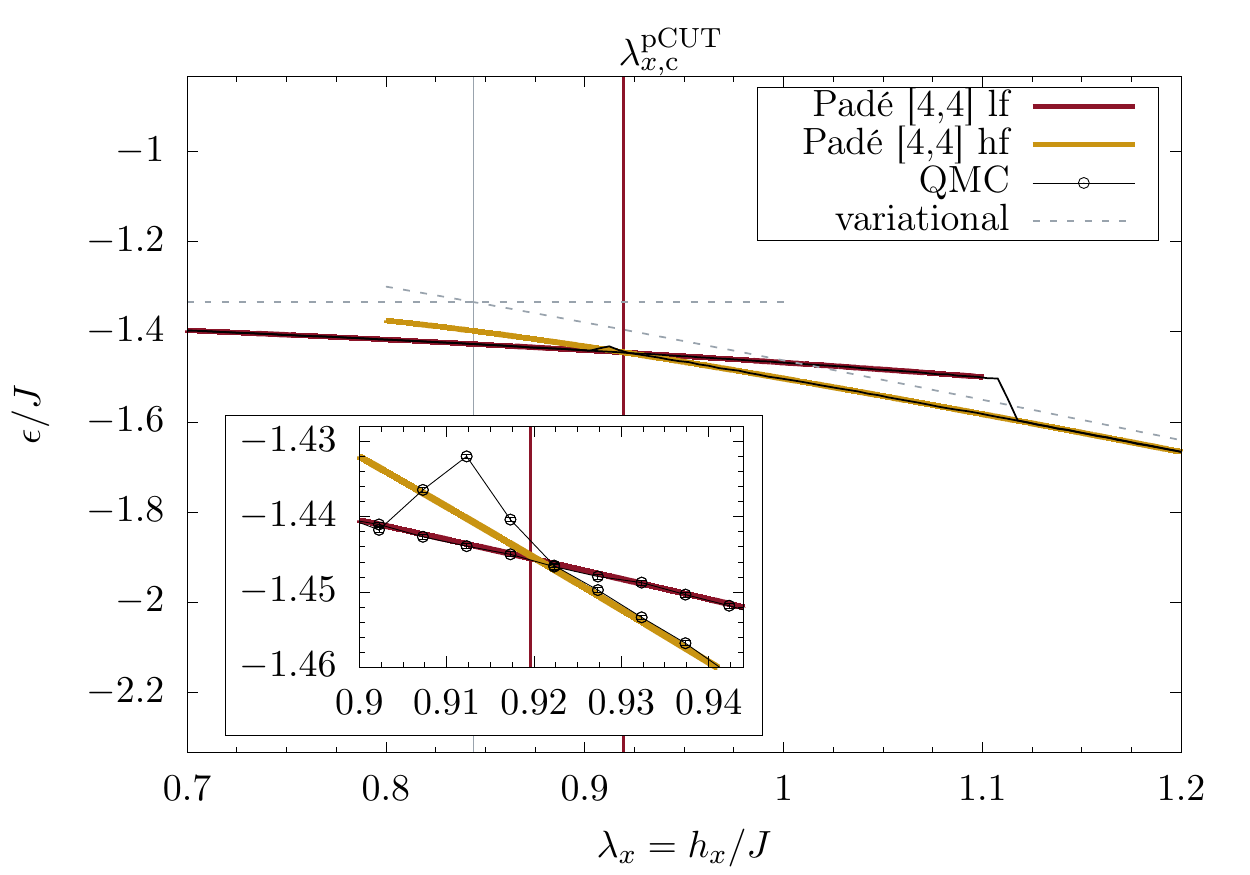}
 \caption{The ground-state energy $\epsilon/J$ of the X-Cube model per link as a function of the parameter $\lambda_x$. {\it Upper panel}: Different bare series of the low-field (darker lines) and high-field (lighter lines) expansion in order 4, 6, and 8. {\it Lower panel}: Comparison of the order-eight bare series from the low- and high-field expansion with the variational energy (dashed line) and the QMC data (black circles) from Ref.~[\onlinecite{Devakul_2018}]. The vertical solid line in both panels indicates the phase transition point  $\lambda_{x,{\rm c}}^{\rm pCUT}=0.9196 \pm 0.0012$ according to the Pad\'{e} analysis of the pCUT results. Insets represent zooms close to $\lambda_{x,{\rm c}}$. \label{fig::xcube_x_pd}}
 \end{figure}

Using the pCUT approach, we reached order eight in the low- and high-field expansion for the ground-state energy per link $\epsilon$. The explicit expression for the low-field expansion reads
\begin{equation}
  \frac{\epsilon_{\rm lf}+J/3}{J} = - 1 -\frac{\lambda^2_x}{8}-\frac{11 \lambda^4_x}{1536}-\frac{349 \lambda^6_x}{221184}-\frac{902473 \lambda^8_x}{1698693120}\, ,
\end{equation}
while the high-field expansion in terms of \mbox{$\tilde{\lambda}_x\equiv 1/\lambda_x$} is given by 
\begin{equation}
   \begin{aligned}
    \frac{\epsilon_{\rm hf }+J/3 }{h_x} =-1-\frac{1}{8}\tilde{\lambda}_x^{2}-\frac{1}{32} \tilde{\lambda}_x^{3} -\frac{11}{1536}\tilde{\lambda}_x^{4}-\frac{55}{18432}\tilde{\lambda}^5_x  \\ -\frac{733}{442368}\tilde{\lambda}^6_x -\frac{9403}{10616832}\tilde{\lambda}^7_x -\frac{2605817}{5096079360 }\tilde{\lambda}^8_x  \, .
   \end{aligned}
   \end{equation}
 Note that in this series also odd orders appear due to the local constraint $\prod_{\kappa} B_{s}^{(\kappa)} = \mathds{1}$, e.g., the action of three distinct $B_{s}^{(\kappa)}$ at vertex $s$ gives a finite contribution in perturbation theory of order three for the ground-state energy. The order-by-order comparison of these two series is shown in the upper panel of Fig.~\ref{fig::xcube_x_pd}. Obviously, the bare series are well converged up to the crossing at $\lambda_{x,{\rm c}}^{\rm pCUT bare}=0.9226$  in order eight. A Pad\'{e} analysis yields almost the same value $\lambda_{x,{\rm c}}^{\rm pCUT}=0.9196 \pm 0.0012$. 
This value quantitatively agrees with the result $\lambda_{x,{\rm c}}^{\rm QMC}\approx 0.922$ from QMC simulations \cite{Devakul_2018} (see also lower panel of Fig.~\ref{fig::xcube_x_pd}). In contrast, the variational calculation yields a lower critical value $\lambda_{x,{\rm c}}^{\rm var}=0.844$ in a similar fashion as our variational results for Haah's code in a field. All approaches result in a phase transition which is strongly first order.
 
\begin{figure}
  \includegraphics[width= 0.9\columnwidth]{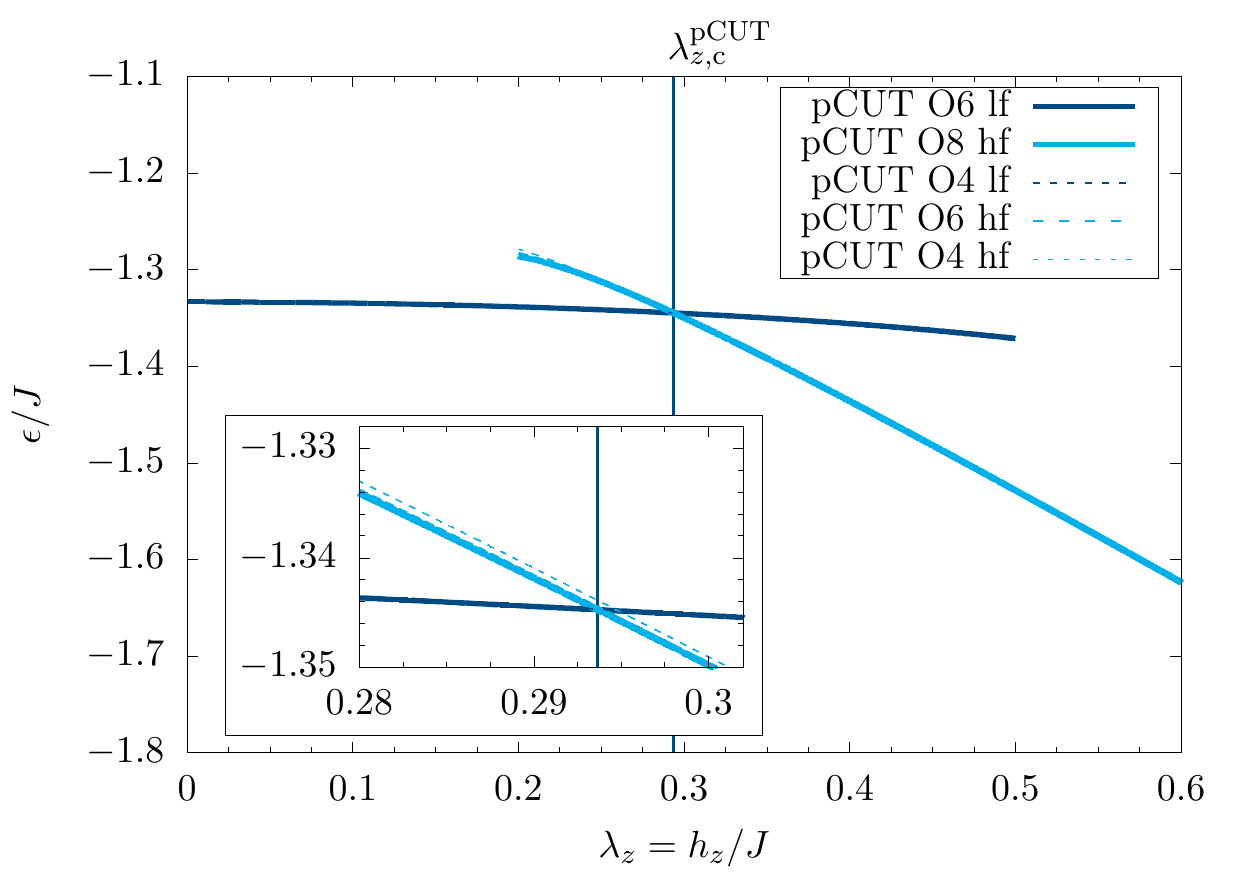}
\includegraphics[width= 0.9\columnwidth]{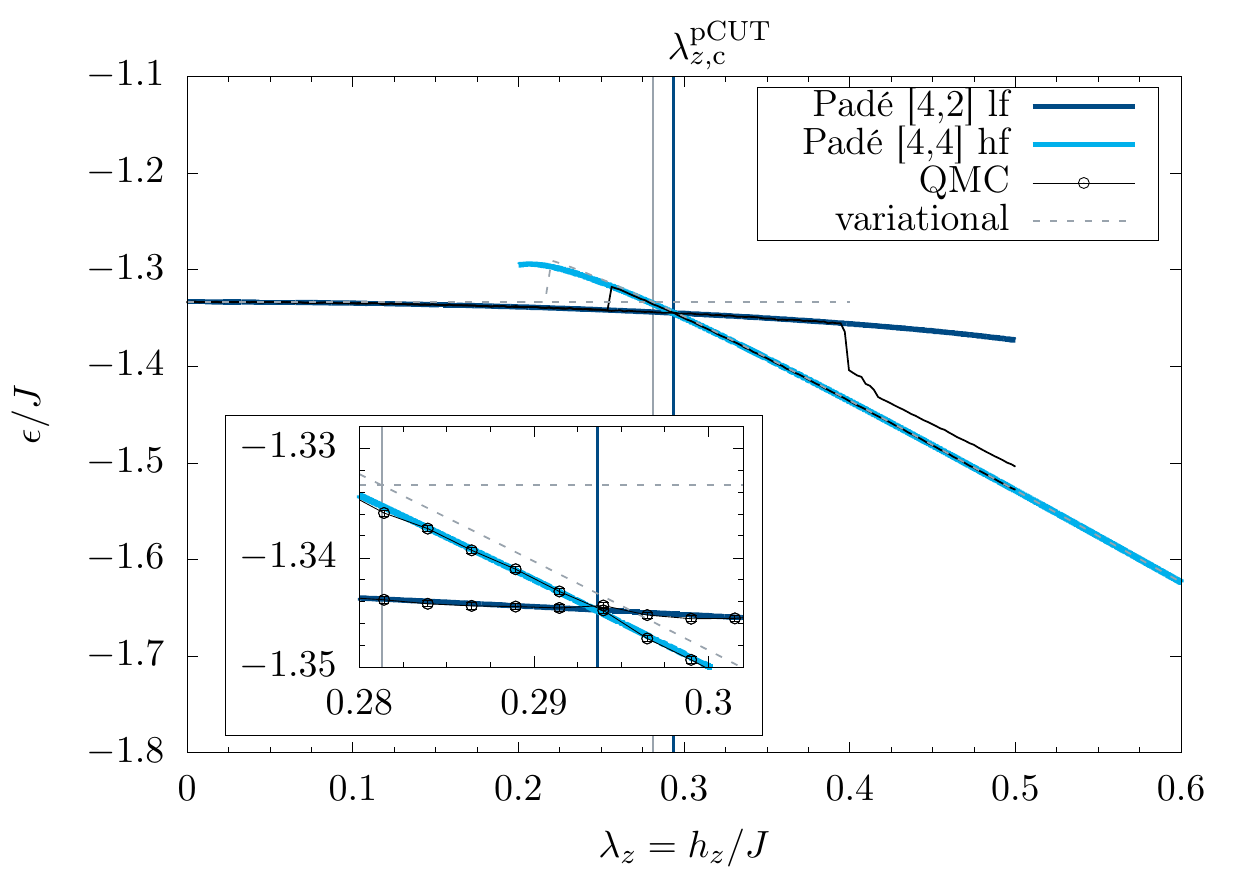} 
\caption{The ground-state energy $\epsilon/J$ of the X-Cube model per link as a function of $\lambda_z$. {\it Upper panel}: Different bare series of the low-field (darker lines) and high-field (lighter lines) expansion in order 4, 6, and 8. {\it Lower panel}: Comparison of the order six (eight) bare series from the low-field (high-field) expansion with the variational energy (dashed line) and the QMC data (black circles) from Ref.~[\onlinecite{Devakul_2018}]. The vertical solid line in both panels indicates the phase transition point $\lambda_{z,{\rm c}}^{\rm pCUT}=0.29364\pm 0.00017$ according to the Pad\'{e} analysis of the pCUT results. Insets represent zooms close to $\lambda_{z,{\rm c}}$.\label{fig::xcube_z_pd}}
\end{figure}

\subsubsection*{Single $z$-field}
Next we consider \mbox{$\boldsymbol{h}=(0,0,h_z)$} and use the parameter $\lambda_z=h_z/J$. In this case the eigenvalues $b_s^{(\kappa)}$ of the $\hat{B}_s^{(\kappa)}$ operators are conserved and the low-energy physics takes place in the sector with $b_s^{(\kappa)}=+1$ for all $s$ and $\kappa$. In this sector the model is dual to the transverse-field plaquette Ising model of Eq.~\eqref{eq:xcube_z_duality} as described in Sect.~\ref{sect::FF}.  

We reached order six (eight) in the low-field (high-field) expansion for the ground-state energy per link $\epsilon$  using the pCUT approach. The series from the low-field expansion is
   \begin{equation}
    \frac{\epsilon_{\rm lf}+J}{J} = -\frac{1}{3}-\frac{1}{8}\lambda_z^2-\frac{113 }{1536}\lambda_z^4-\frac{21427 }{163840}\lambda_z^6 \, ,
    \end{equation}
whereas the high-field expansion reads
    \begin{equation}
    \begin{aligned}
     \frac{\epsilon_{\rm hf} + J }{h_z}=& -1-\frac{1}{72} \tilde{\lambda}_z^{2} -\frac{17}{228096}\tilde{\lambda}_z^{4}-\frac{2307925}{1653327396864}\tilde{\lambda}_z^{6} \\ 
     &-\frac{3488211785451004153}{88061074183163574484992000}\tilde{\lambda}_z^{8}\, .
      \end{aligned}
    \end{equation}
 
The order-by-order comparison of these two series is shown in the upper panel of Fig.~\ref{fig::xcube_z_pd}. The bare series are again well converged and a Pad\'{e} analysis yields the value $\lambda_{z,{\rm c}}^{\rm pCUT}=0.29364\pm 0.00017$ with a very tiny uncertainty. This value is also in quantitative agreement with the result from QMC simulations \cite{Devakul_2018} $\lambda_{z,{\rm c}}^{\rm QMC}\approx 0.293$ (see also lower panel of Fig.~\ref{fig::xcube_z_pd}). The variational calculation yields a slightly lower critical value  $\lambda_{z,{\rm c}}^{\rm var}=0.281$ which can be explained consistently to all the cases before. Again, all approaches signal a strong first-order phase transition.

\section{Fracton quasi-particles}
\label{sect::fractons}

In the last section we have seen that both fracton phases break down via first-order phase transitions in all considered field directions. In a next step we study the energetic properties of the excitations in the fracton phases at finite fields. We find that all considered excitations including single- and multi-fracton excitations remain gapped until the phase transition confirming the first-order nature of the phase transition.   

\subsection{Haah's code in a field}
\label{ssect::haah_fractons}
In the following we discuss the excitation energies in the single-, two-, and four-fracton sector using the pCUT method. We start with the single-type case and turn to the two-type parallel case afterwards. The two-type orthogonal case is not discussed separately, since it is equivalent to two independent single-type cases. 

\subsubsection*{Single-type case}
The single-type case corresponds to the field configuration \mbox{$\boldsymbol{h}^\sigma=(h_x^\sigma,0,0)$} and $\boldsymbol{h}^\mu=0$. As already demonstrated and exploited before, the $\hat{B}_c$ operators still commute with the Hamiltonian and excitations with $b_c=-1$ do not play a role for the low-energy physics. Furthermore, the phase transition point $\lambda_{x,{\rm c}}\equiv (h_x^\sigma/J)_{\rm c}=1$ is known exactly due to the self-duality.

\begin{figure}
	\centering
	\includegraphics[width = \columnwidth]{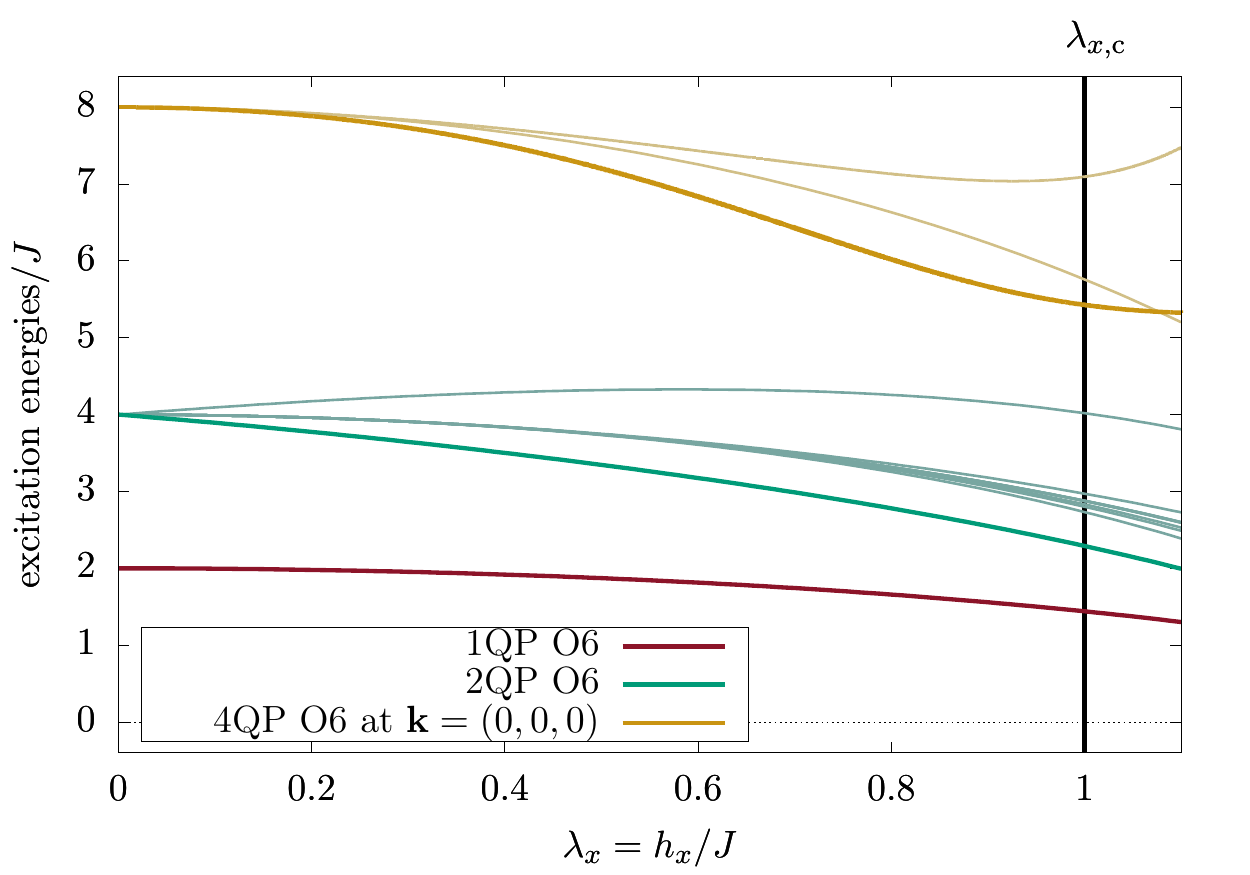}
	\caption{Relevant excitation energies of the one-, two-, and four-fracton sector in the fracton phase of Haah's code for the single-type case as a function of $\lambda_x = \frac{h_x}{J}$. Shown are bare series in order six in $\lambda_x$ for all sectors.}
	\label{fig::haah_fractons::SingleField}		
\end{figure}

Consequently, the relevant low-energy excitations in the fracton phase relate to eigenvalues $a_c=-1$ of the $\hat{A}_c$ operators and we focus on the one-, two-, and four-fracton sector, which we analyzed up to order six in \mbox{$\lambda_x=h_x/J$}. The corresponding bare series of these sectors are illustrated in Fig.~\ref{fig::haah_fractons::SingleField}. Clearly, none of these bare series becomes zero for $\lambda\leq \lambda_{x,{\rm c}}$, analyzed in detail below. Note that the corresponding bare series of spin-flip excitations of the high-field polarized phase can be obtained due to the self-duality by interchanging $J$ and $h_x$ in the series of the fracton phase. 

\begin{figure}
	\centering
	\includegraphics[width = \columnwidth]{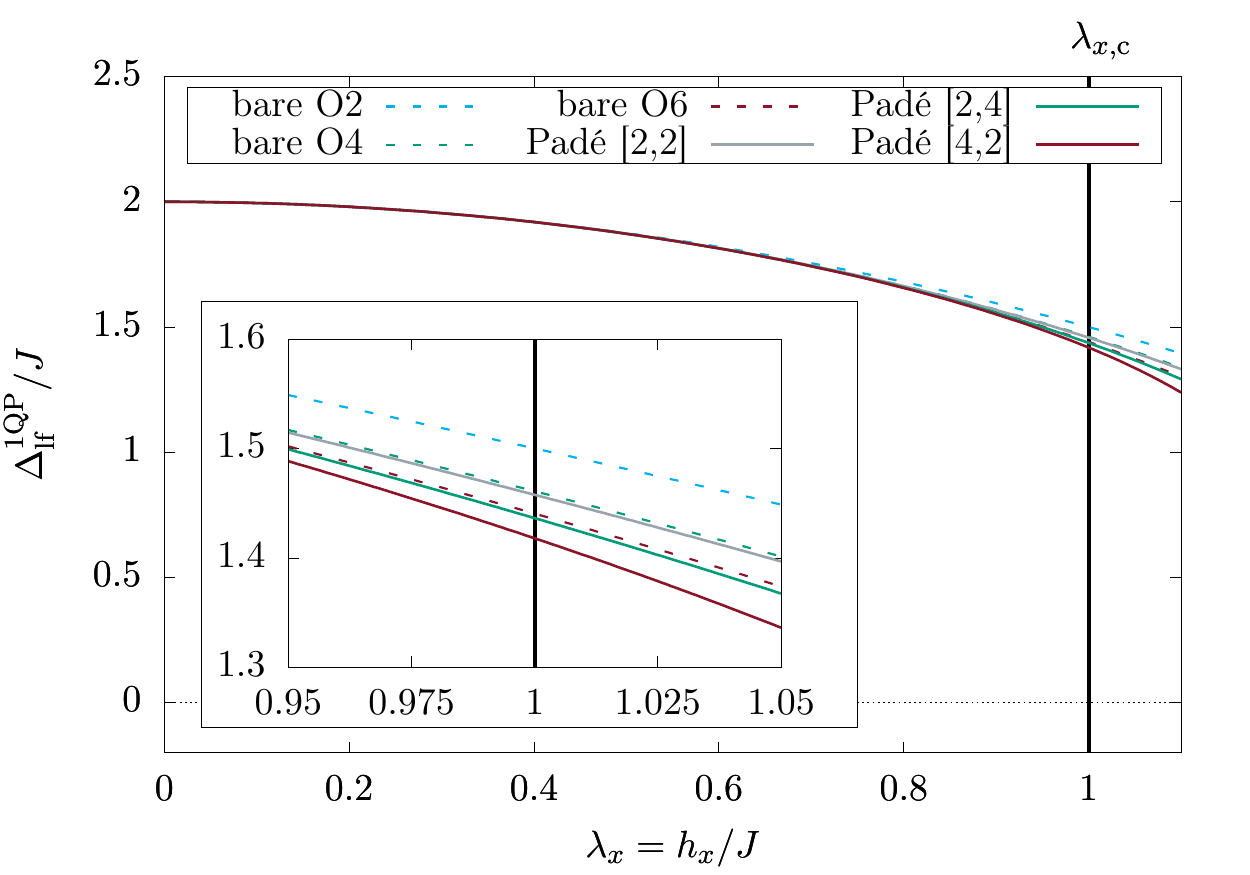}
	\caption{Bare series as well as Pad\'{e} extrapolants of the one-fracton gap $\Delta_{\rm lf}^{\rm 1qp}/J$ as a function of $\lambda_x$ for the single-type case of Haah's Code. The vertical solid line indicates the exactly known phase transition point $\lambda_{x,{\rm c}}=1$. }
	\label{fig::haah_fractons::SingleField::singleParticlesPade}		
\end{figure}

A single fracton is not allowed to hop due to the magnetic field, but it can reduce its energy by vacuum fluctuations. Its dispersion realation is then obviously flat and the momentum-independent one-fracton gap is given by
\begin{equation}    
\frac{\Delta_{\rm lf}^{\text{1qp}}}{J} = 2 - \frac{1}{2}\lambda_x^2 - \frac{5}{128}\lambda_x^4 - \frac{775}{73728} \lambda_x^6\, ,
\end{equation}
which is displayed in  Fig.~\ref{fig::haah_fractons::SingleField::singleParticlesPade}. It can be clearly seen that the one-fracton gap is very well converged up to the phase transition point.

\begin{figure}
	\centering
	\begin{minipage}{0.31\columnwidth}
		\adjincludegraphics[width = \textwidth]{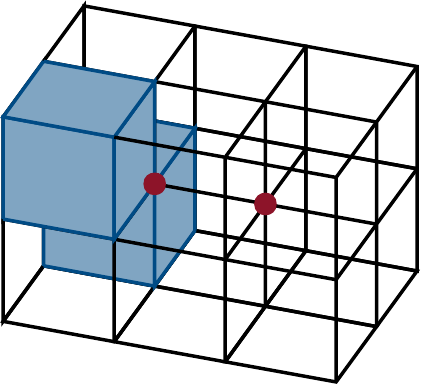}
	\end{minipage}
	\hfill
	\begin{minipage}{0.31\columnwidth}
		\adjincludegraphics[width = \textwidth]{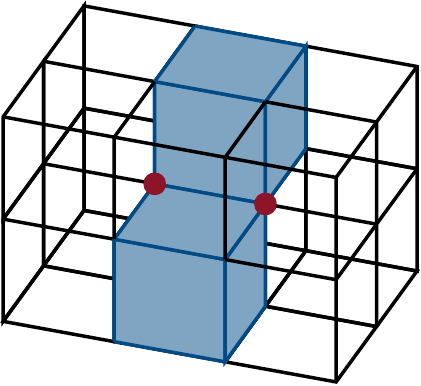}
	\end{minipage}
	\hfill
	\begin{minipage}{0.31\columnwidth}
		\adjincludegraphics[width = \textwidth]{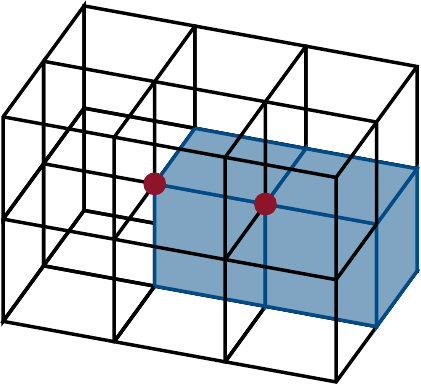}
	\end{minipage}
	\caption{Illustration of the most relevant unperturbed two-fracton states at small fields for the low-energy physics in the two-fracton sector of Haah's code. A cube $c$ colored in blue indicates that its eigenvalue is $a_c = +1$ and hence represents a fracton excitation. For the single-type case, an $\sigma^x$ operator acting on the spin at the right red vertex changes the state shown in the center into the state shown at the right. Hence, we have a hopping element between these states in first order perturbation. Additionally, for the two-type parallel case, a $\mu^x$ acting on the left red spin changes the state depicted in the center to the one depicted on the left.}
	\label{fig::HaahsCode::TwoFractonStates}		
\end{figure}

The effective Hamiltonian of a two-particle sector is typically harder to diagonalize, since the relative distance of the two particles (here fractons corresponding to two cubes with $a_c=-1$) can be arbitrary so that an Hamiltonian acting on an infinite-dimensional Hilbert space remains to be treated even if the total momentum is conserved. However, the situation is different for Haah's code in a field, since the individual fractons are not allowed to hop at all. As a consequence, most two-fracton configurations remain decoupled from all the other two-fracton states in a given perturbative order and the pCUT method directly gives a single series for the excitation energy of these two-fracton states. Only certain two-fracton configurations having fractons close to each other can be coupled by the magnetic field yielding still a finite-dimensional problem in the form of a finite Hamiltonian matrix (see Fig.~\ref{fig::HaahsCode::TwoFractonStates} for an illustration of the most relevant unperturbed two-fracton states at small fields for the low-energy physics in the two-fracton sector.). The diagonalization of these matrices gives binding and anti-binding eigenstates with respect to the energy of two independent fractons. The series for the two-fracton configuration with the highest binding energy representing the two-fracton gap reads   
\begin{eqnarray}    
\frac{\Delta_{\rm lf}^{\rm 2qp}}{J} &=& 4-\lambda_x-\frac{5}{8} \lambda_x^2 + \frac{3}{32} \lambda_x^3-\frac{241}{1536}\lambda_x^4 \nonumber \\
&& + \frac{1553}{36864}\lambda_x^5-\frac{558043}{8847360}\lambda_x^6\, .
\end{eqnarray}

\begin{figure}
	\centering
	\includegraphics[width = \columnwidth]{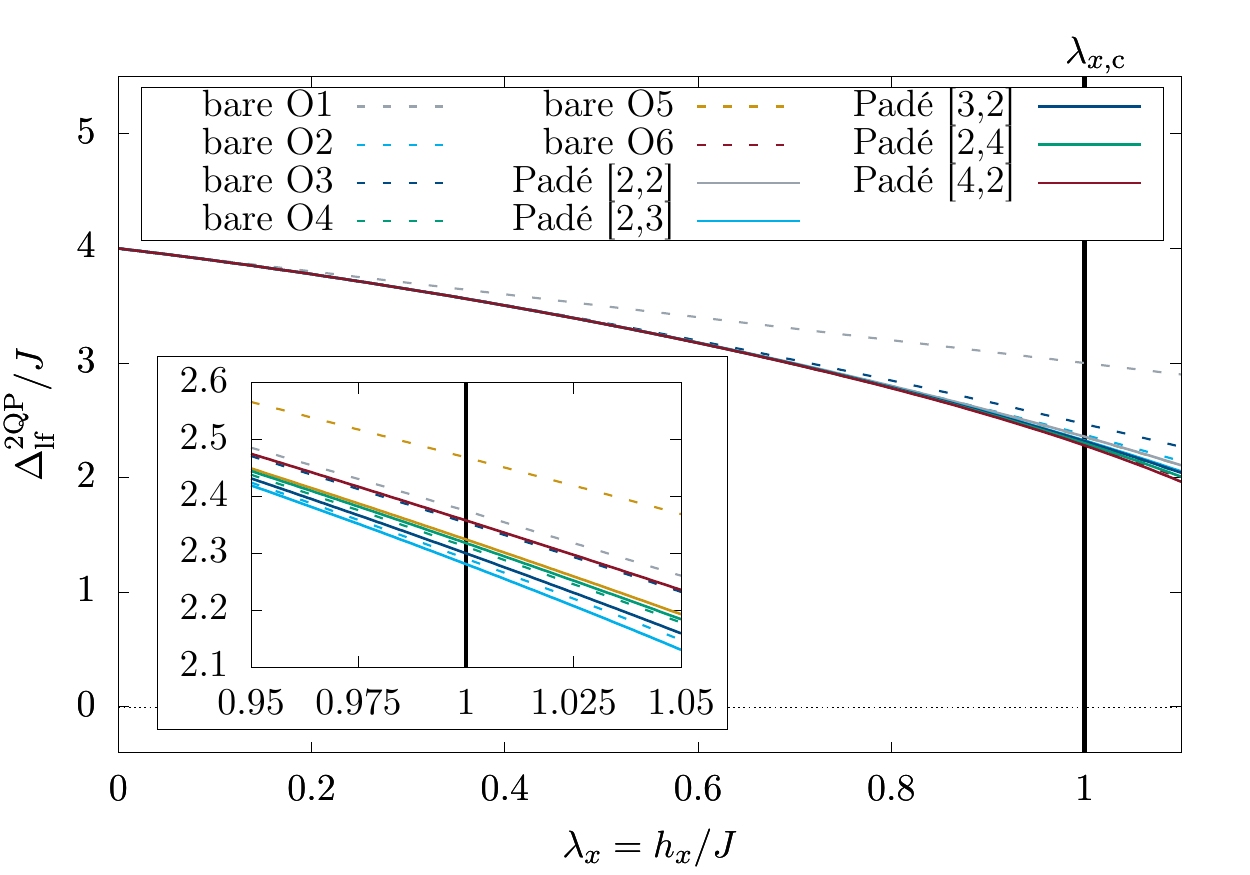}
	\caption{Bare series as well as Pad\'{e} extrapolants of the two-fracton gap $\Delta_{\rm lf}^{\rm 2qp}/J$ as a function of $\lambda_x$ for the single-type case of Haah's code. The vertical solid line indicates the exactly known phase transition point $\lambda_{x,{\rm c}}=1$.}
	\label{fig::haah_fractons::SingleField::twoParticlesPade}		
\end{figure}

Bare series as well as Pad\'{e} extrapolants of the two-fracton gap $\Delta_{\rm lf}^{\rm 2qp}$ are shown in Fig.~\ref{fig::haah_fractons::SingleField::twoParticlesPade}. Clearly, this quantity is well converged yielding a finite two-fracton gap $\Delta_{\rm lf,c}^{\rm 2qp}/J\approx 2.3$ at the phase transition point $\lambda_{x,{\rm c}}=1$. 

\begin{figure}
	\centering
	\begin{minipage}{0.45\columnwidth}
		\adjincludegraphics[width = \textwidth]{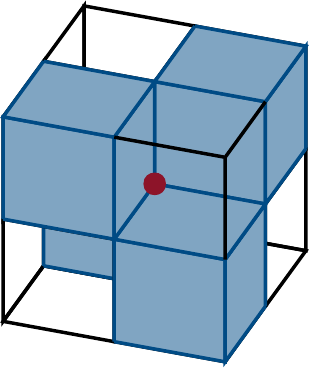}
	\end{minipage}
	\hfill
	\begin{minipage}{0.45\columnwidth}
		\adjincludegraphics[width = \textwidth]{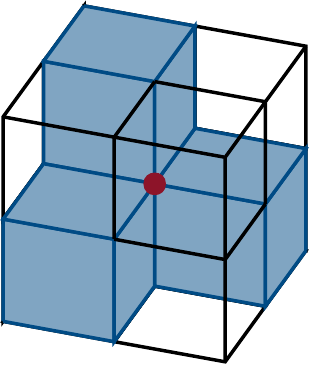}
	\end{minipage}
	\caption{Illustration of the most relevant unperturbed four-fracton states at small fields for the low-energy physics in the four-fracton sector of Haah's code. A cube $c$ colored in blue indicates that its eigenvalue $a_c = -1$ and hence represents a fracton excitation. {\it Left}: The action of a $\sigma^x$ operator at the red vertex on the bare fracton ground state creates the four-fracton state depicted on the left. This four-fracton configuration can hop in second-order perturbation theory and is therefore most relevant for the low-energy physics in the four-fracton sector. {\it Right}: An operator $\mu^x$ acting on the spin at the red vertex, creates the state shown on the right. Again this state is mobile. Furthermore,  the left and the right configuration can be interact with each other in the two-type parallel case.}
	\label{fig::HaahsCode::FourFractonStates}		
\end{figure}

\begin{figure}
	\centering
	\includegraphics[width = \columnwidth]{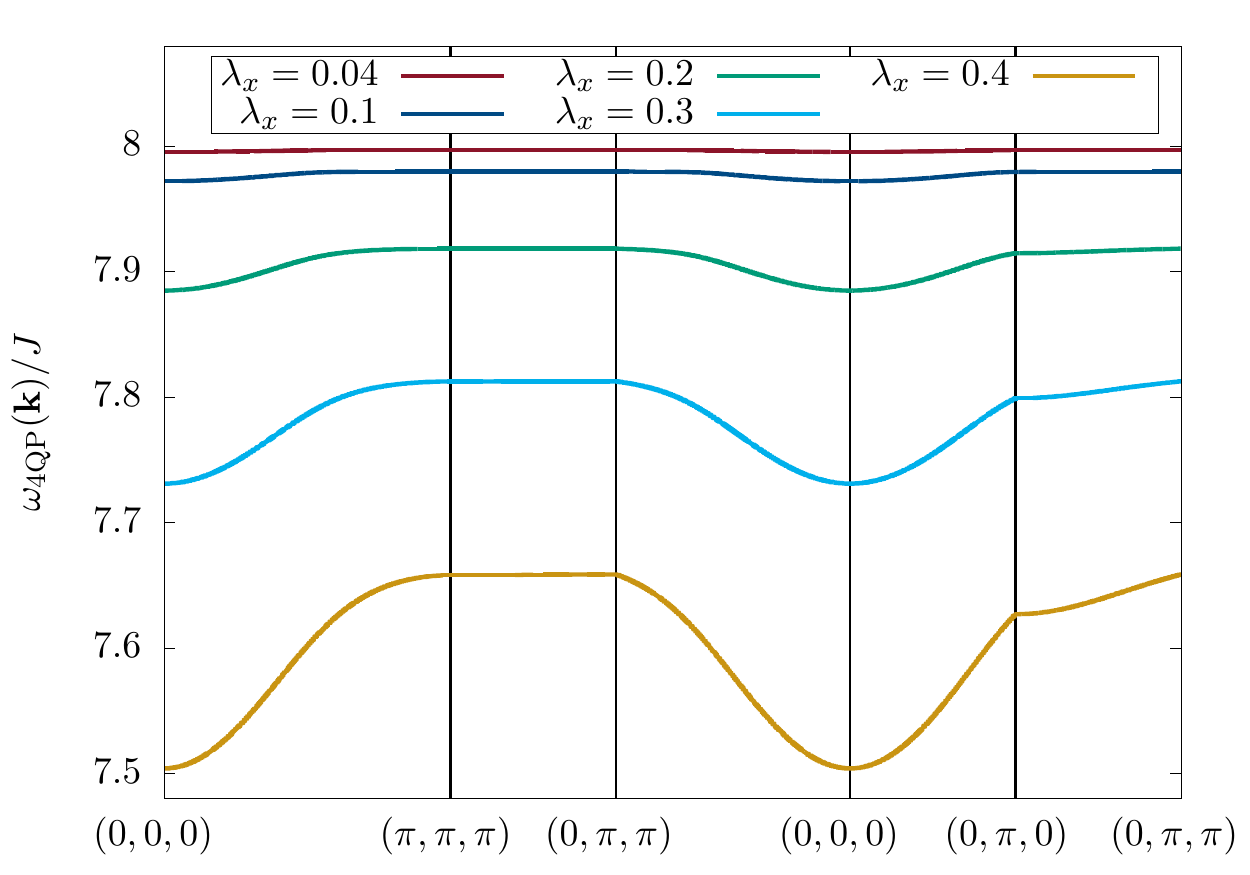}
	\caption{Dispersion $\omega^{\rm 4qp}(\boldsymbol{k})/J$ of the lowest mode of the four fracton sector of Haah's code in the single-type case for different $\lambda_x$.}
	\label{fig::haah_fractons::SingleField::fourParticlesDispersion}
\end{figure}

\begin{figure}
	\centering
	\includegraphics[width = \columnwidth]{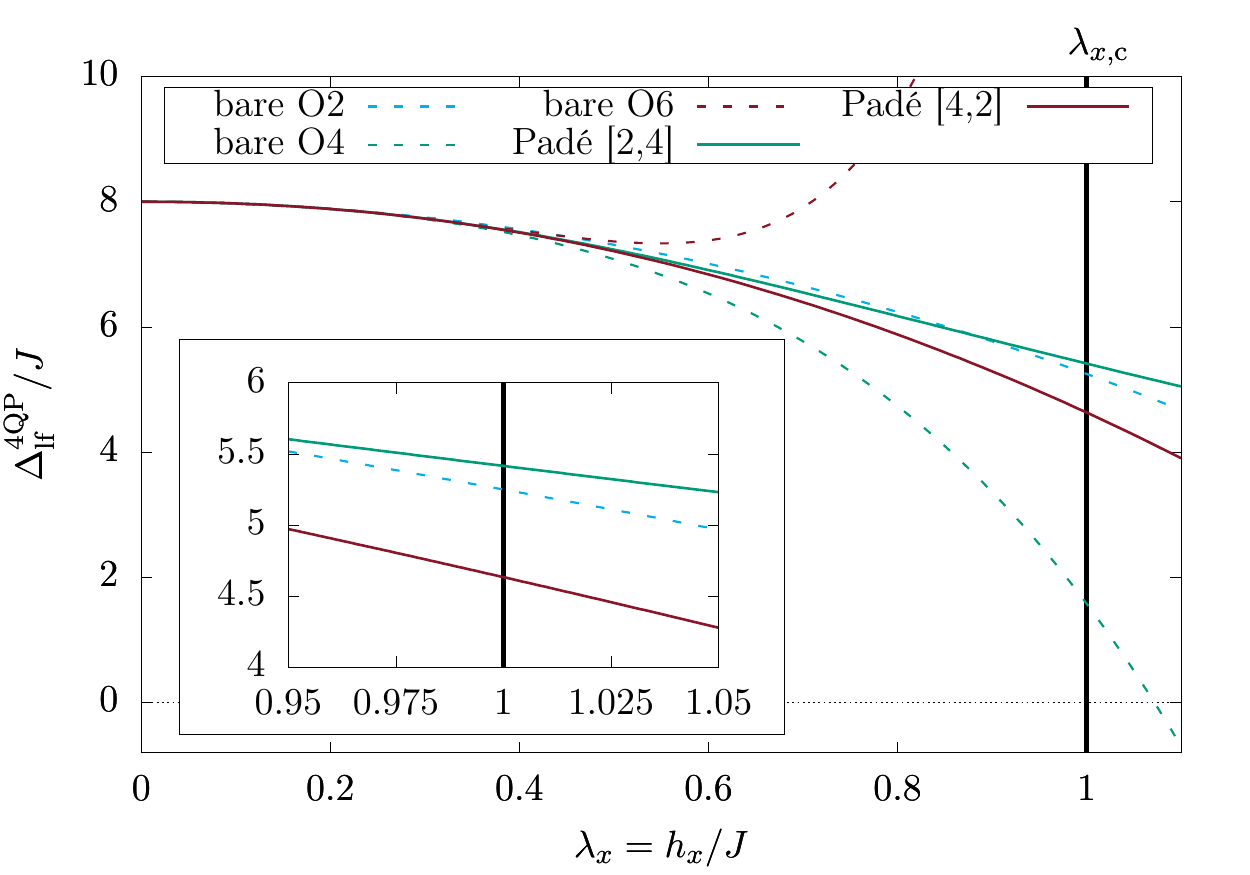}
	\caption{Bare series as well as Pad\'{e} extrapolants of the four-fracton gap $\Delta_{\rm lf}^{\rm 4qp}/J$ at $\boldsymbol{k} = (0,0,0)$ as a function of $\lambda_x$ for the single-type case of Haah's code. The vertical solid line indicates the exactly known phase transition point \mbox{$\lambda_{x,{\rm c}}=1$.}}
	\label{fig::haah_fractons::SingleField::fourParticlesPade}		
\end{figure}

In contrast to the one- and two-fracton sector, there exist certain topologically trivial fracton configurations in the four-fracton sector which are able to hop. The largest four-fracton hopping element at small fields starts in second-order perturbation theory in $\lambda_x$ for a local four-fracton configuration (see Fig.~\ref{fig::HaahsCode::FourFractonStates}). Here we have calculated the four-fracton contribution up to order six in $\lambda_x$, which is connected to this four-fracton configuration. Interestingly, there is only a single four-fracton configuration which has a finite transition amplitude to this configuration in sixth-order perturbation theory (see also Subsubsect.~\ref{sssect::pCUT_haah}). The resulting Hamiltonian matrix is therefore a $2 \times 2$ matrix in momentum space. The resulting low-energy branch $\omega^{\rm 4qp} (\boldsymbol{k})$ of the four-fracton excitation is shown for different values of $\lambda_x$ in Fig.~\ref{fig::haah_fractons::SingleField::fourParticlesDispersion} along a high-symmetry path in the three-dimensional Brillouin zone using the bare order-six series. Overall, the dispersion develops only slowly a finite bandwidth as a function of $\lambda_x$, which is a consequence of the restricted mobility of individual fracton excitations. However, the gap $\Delta_{\rm lf}^{\rm 4qp}$ in this sector is clearly located at zero wave vector. 
Here we can obtain an exact series for the four-fracton gap due to the analytic expressions for the eigenvalues of the $2\times2$ Hamiltonian matrix.
The resulting series reads 
	\begin{equation}    
	\frac{\Delta_{\rm lf}^{\text{4qp}}}{J} = 8 -\frac{11}{4}\lambda_x^2 -\frac{939}{256}\lambda_x^4 + \frac{5357083}{294912} \lambda_x^6.
	\end{equation}
The gap is plotted in Fig.~\ref{fig::haah_fractons::SingleField::fourParticlesPade}. The bare series for this quantity is alternating so that the convergence is limited to rather small values of $\lambda_x$. Additionally, the only two available Pad\'e extrapolants are not very well converged close to the phase transition point. Nevertheless, all extrapolants yield a finite gap $\Delta_{\rm lf}^{\rm 4qp}> 2$ at $\lambda_{x,{\rm c}}=1$.


\subsubsection*{Two-type parallel case}
The two-type parallel case corresponds to \mbox{$\boldsymbol{h}^\sigma=\boldsymbol{h}^\mu=(h_x,0,0)$} and we analyze the energetic properties of excitations in the fracton phase as a function of the parameter $\lambda_x\equiv h_x/J$. As in the single-type case, the eigenvalues $b_c$ of $\hat{B}_c$ operators are conserved quantities and the associated excitations with $b_c=-1$ do not play a role for the low-energy physics. The phase transition is located at $\lambda_{x,{\rm c}}^{\rm pCUT}\equiv (h_x/J)_{\rm c}=0.456 \pm 0.006$  according to the pCUT analysis of the ground-state energy in Subsect.~\ref{ssect::haah_pd}.

\begin{figure}
	\centering
	\includegraphics[width = \columnwidth]{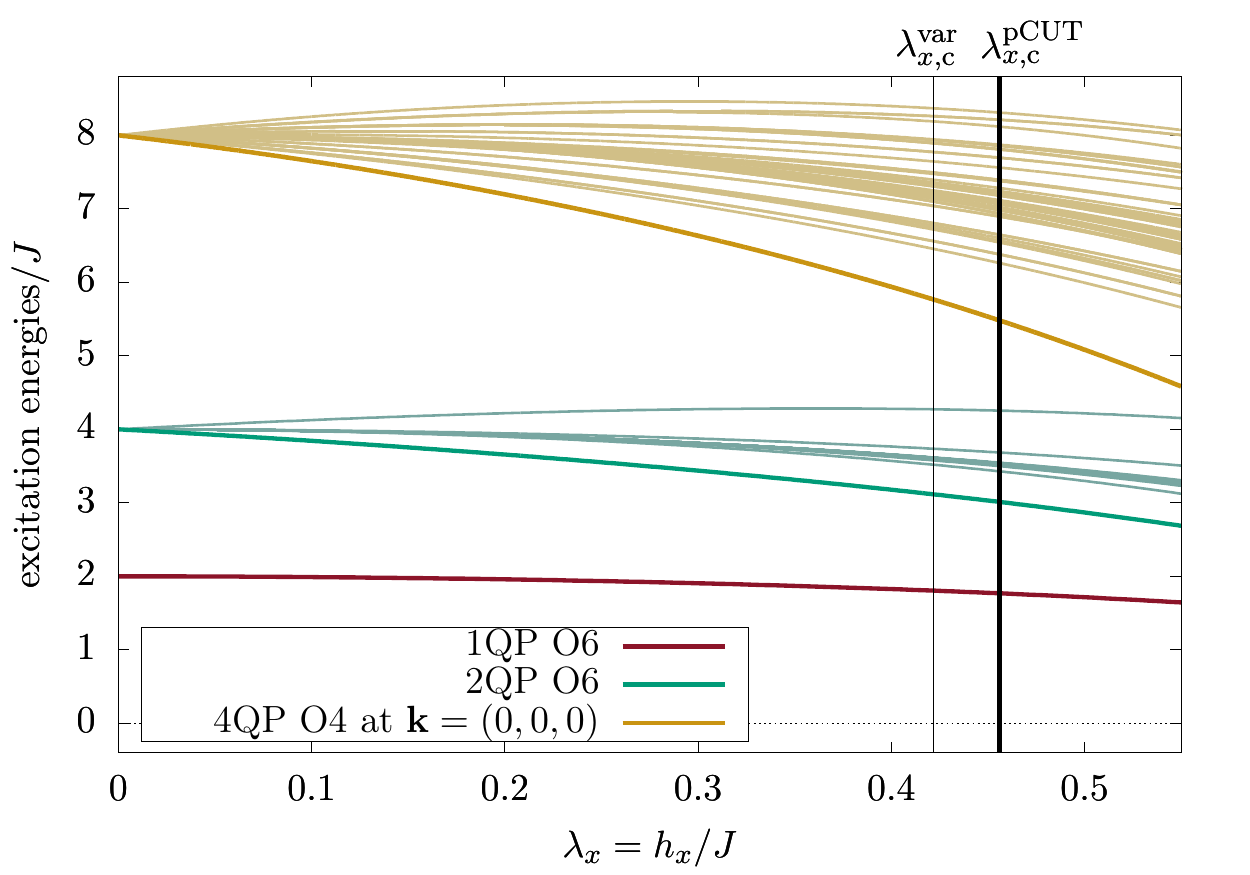}
	\caption{Relevant excitation energies of the one-, two-, and four-fracton sector in the fracton phase of Haah's code for the two-type parallel case as a function of $\lambda_x = h_x/J$. Shown are bare series in order six in $\lambda_x$ for the one- and two-fracton sectors while order four is displayed for the four-fracton sector.}
	\label{fig::haah_fractons::TwoField}		
\end{figure}

The relevant low-energy excitations in the fracton phase are therefore again eigenvalues $a_c=-1$ of the $\hat{A}_c$ operators. We calculated the series up to order six in $\lambda_x$ in the one- and two-fracton sector while we reached order four in the four-fracton sector. The corresponding bare series of these sectors are illustrated in Fig.~\ref{fig::haah_fractons::TwoField}. Clearly, none of these bare series becomes zero for $\lambda\leq \lambda_{x,{\rm c}}^{\rm pCUT}$, which we analyze next. 

\begin{figure}
	\centering
	\includegraphics[width = \columnwidth]{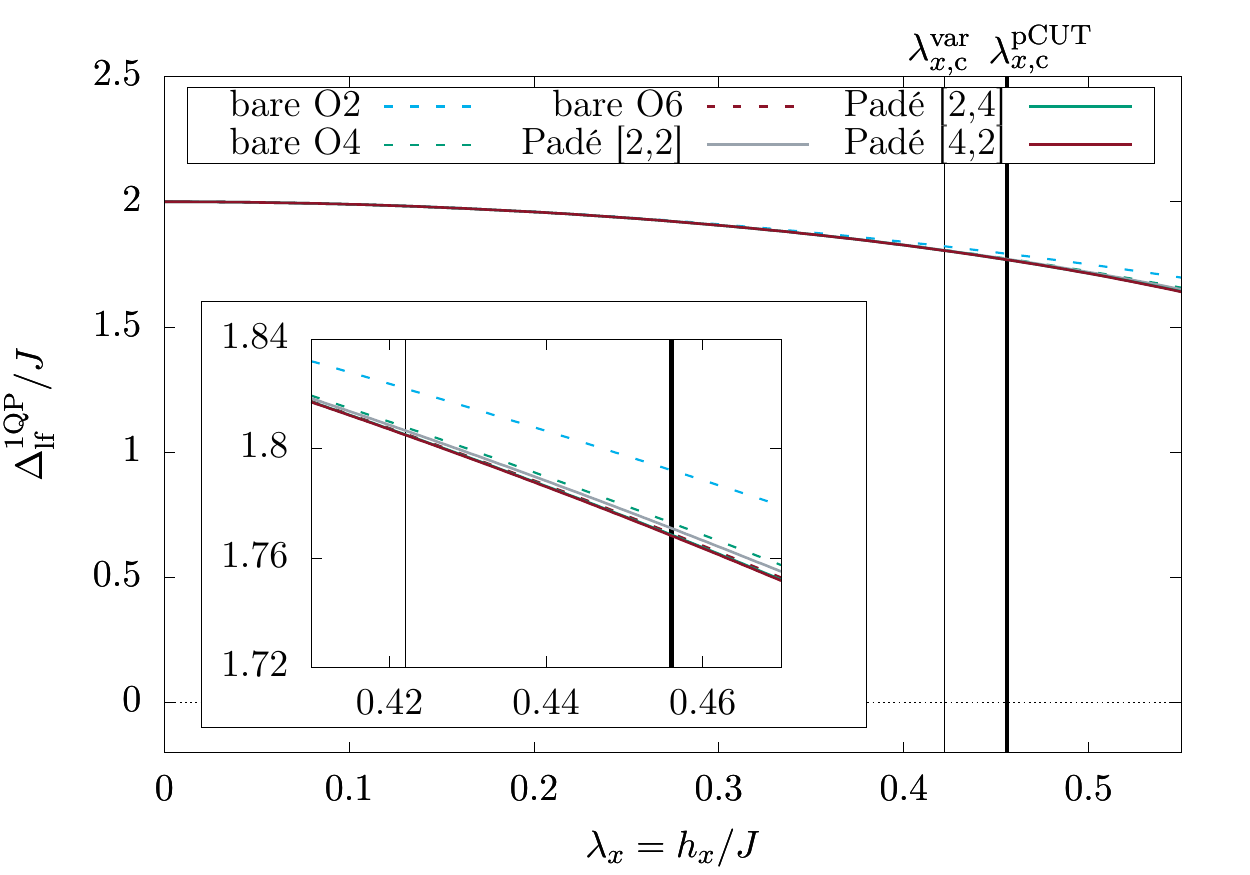}
	\caption{Bare series as well as Pad\'{e} extrapolants of the one-fracton gap $\Delta_{\rm lf}^{\rm 1qp}/J$ as a function of $\lambda_x$ for the two-type parallel case of Haah's code. The vertical solid line indicates the phase transition point \mbox{$\lambda_{x,{\rm c}}^{\rm pCUT}=0.456$.}}
	\label{fig::haah_fractons::TwoField::singleParticlesPade}
\end{figure}

A single fracton is strictly immobile also in the two-type parallel case, but it can again reduce its energy by vacuum fluctuations. The one-fracton gap reads
\begin{equation}    
\frac{\Delta_{\rm lf}^{\rm 1qp}}{J} = 2-\lambda_x^2-\frac{57}{128} \lambda_x^4 -\frac{10297}{24576}\lambda_x^6\, .
\end{equation}
Bare series as well as Pad\'{e} extrapolants of $\Delta_{\rm lf}^{\rm 1qp}$ are shown in  Fig.~\ref{fig::haah_fractons::TwoField::singleParticlesPade}, which look similar to the one-fracton gap in the single-type case.

Similarly to the single-type case, the two-fracton sector decouples into sets of effective Hamiltonians in finite-dimensional Hilbert spaces which can be solved easily by matrix diagonalization. The diagonalization of these Hamiltonian matrices gives binding and anti-binding eigenstates with respect to the energy of two independent fractons. The series for the two-fracton configuration with the highest binding energy representing the two-fracton gap reads   
\begin{eqnarray}    
\frac{\Delta_{\rm lf}^{\rm 2qp}}{J} &=& 4 -1.414213562373095 \ \lambda_x -1.5 \ \lambda_x^2  \\
&-& 0.198873782208717 \ \lambda_x^3 -1.005533854166667 \ \lambda_x^4 \nonumber \\
&-& 0.279436078242340 \ \lambda_x^5 -1.185076113100405 \ \lambda_x^6\, . \nonumber
\end{eqnarray}
\begin{figure}
	\centering
	\includegraphics[width = \columnwidth]{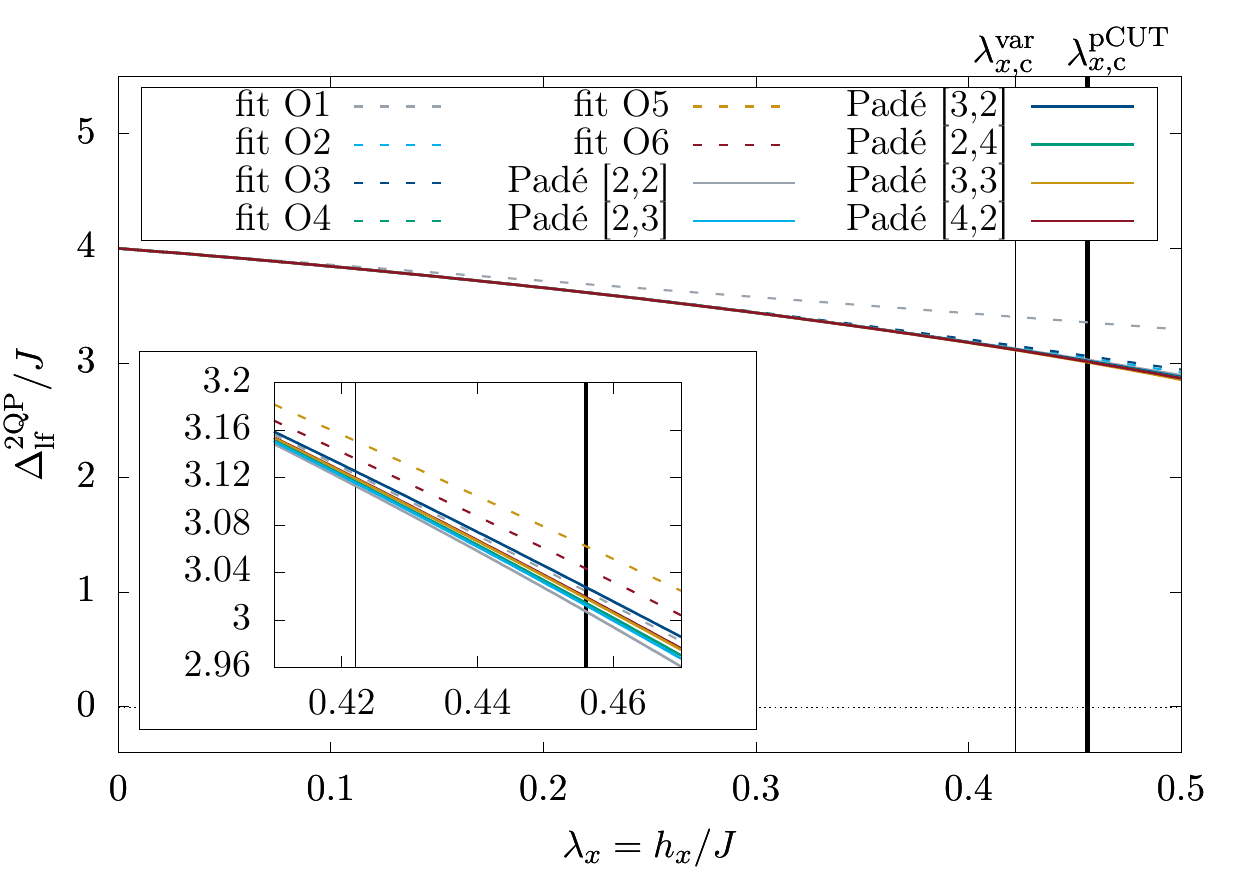}
	\caption{Bare series as well as Pad\'{e} extrapolants of the two-fracton gap $\Delta_{\rm lf}^{\rm 2qp}/J$ as a function of $\lambda_x$ for the two-type parallel case of Haah's code. The vertical solid line indicates the phase transition point \mbox{$\lambda_{x,{\rm c}}^{\rm pCUT}=0.456$.}}
	\label{fig::haah_fractons::TwoField::twoParticlesPade}		
\end{figure}
Bare series as well as Pad\'{e} extrapolants of $\Delta_{\rm lf}^{\rm 2qp}$ are shown in Fig.~\ref{fig::haah_fractons::TwoField::twoParticlesPade} yielding a well-converged finite two-fracton gap $\Delta_{\rm lf,c}^{\rm 2qp}/J\approx 3 $ at the phase transition $\lambda_{x,{\rm c}}^{\rm pCUT}=0.456$. 

\begin{figure}
	\centering
	\includegraphics[width = \columnwidth]{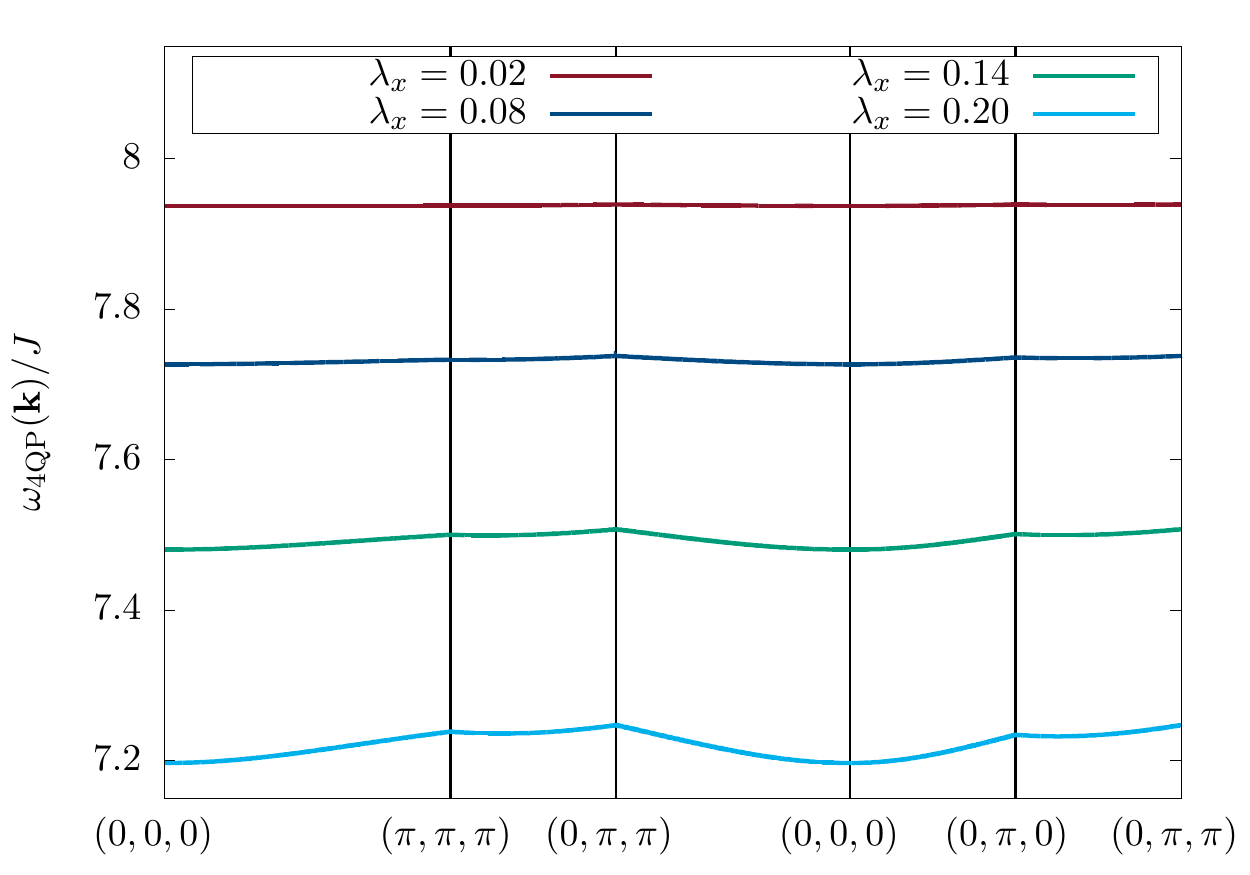}
	\caption{Dispersion $\omega^{\rm 4qp}(\boldsymbol{k})/J$ of the lowest mode in the four-fracton sector for the two-type parallel case of Haah's code along a high-symmetry path in the three-dimensional Brillouin zone for different fixed $\lambda_x$ using the bare order-six series.}
	\label{fig::haah_fractons::TwoField::fourParticlesDispersion}		
\end{figure}

The resulting dispersion $\omega^{\rm 4qp} (\boldsymbol{k})$ of the lowest four-fracton excitation is shown for different values of $\lambda_x$ in Fig.~\ref{fig::haah_fractons::TwoField::fourParticlesDispersion} along a high-symmetry path in the three-dimensional Brillouin zone using the bare order-six series. The dominant effect of $\lambda_x$ is to lower the overall energy for all $\boldsymbol{k}$, which mainly originates from the local quantum fluctuations of the involved individual fractons. Moreover, the gap $\Delta_{\rm lf}^{\rm 4qp}$ in this sector is located again at zero wave vector. 
The Hamiltonian matrix for all connected states is a $77\times77$ matrix that we diagonalized numerically. To regain a series we made a sequence of fits for each order as before yielding
	\begin{eqnarray}    
	\frac{\Delta_{\rm lf}^{\text{4qp}}}{J} &=& 8 - 3.026151005915648 \lambda_x \nonumber \\
	&-&4.805118904820233 \lambda_x^2 -0.127432688836401\lambda_x^3 \nonumber \\
	&-& 2.546990725837421 \lambda_x^4\, .
	\end{eqnarray}
	The fit errors were significantly lower than the machine precision, but we restrict ourself to the displayed accuracy for practical reasons.

\begin{figure}
	\centering
	\includegraphics[width = \columnwidth]{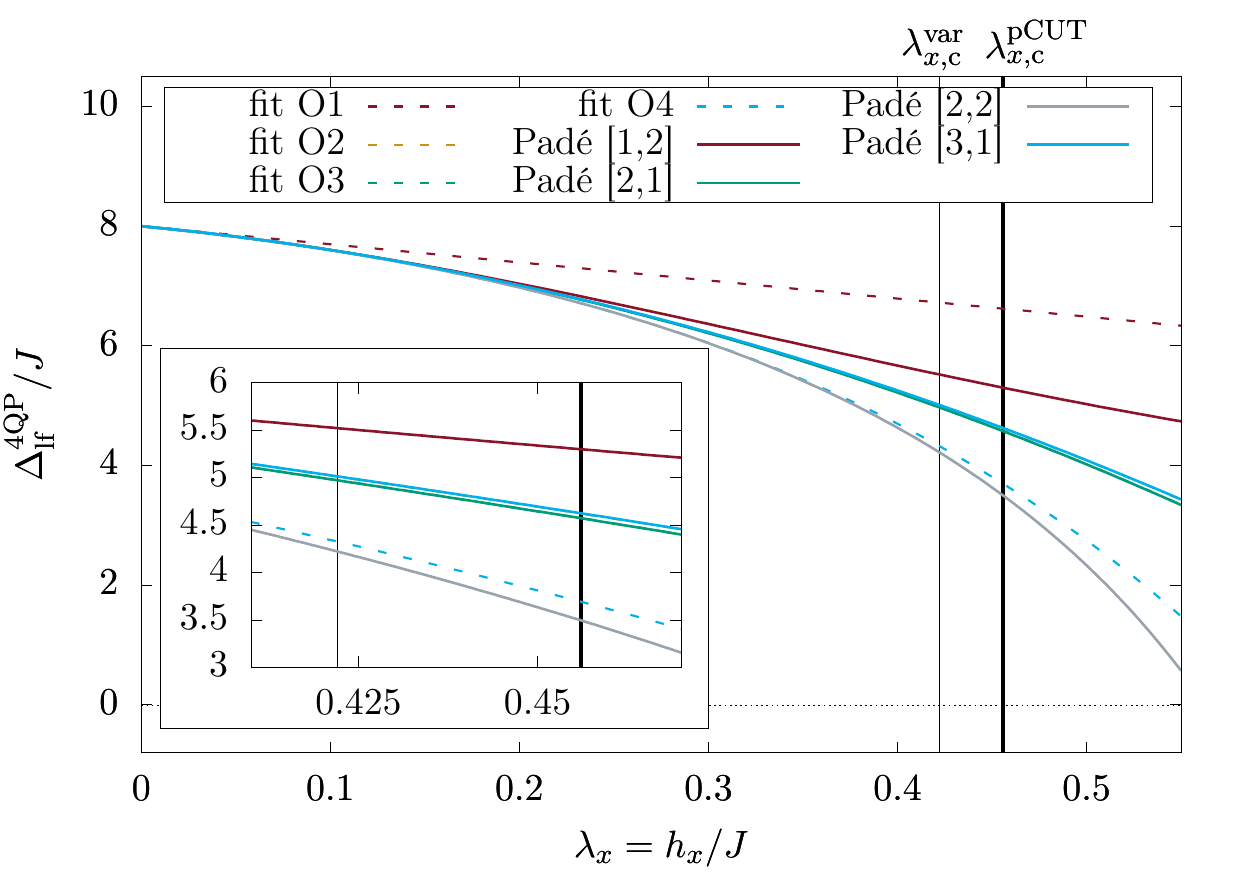}
	\caption{Bare series as well as Pad\'{e} extrapolants of the four-fracton gap $\Delta_{\rm lf}^{\rm 4qp}/J$ at $\boldsymbol{k} = (0,0,0)$ as a function of $\lambda_x$ for the two-type parallel case of Haah's code. The vertical solid line indicates the phase transition point $\lambda_{x,{\rm c}}^{\rm pCUT}=0.456$.}
	\label{fig::haah_fractons::TwoField::fourParticlesPade}
\end{figure}
An analysis of this gap is plotted in Fig.~\ref{fig::haah_fractons::TwoField::fourParticlesPade}. In contrast to the single-type case, the convergence of the bare series as well as of the Pad\'e extrapolants is better although the maximum order is only four. Clearly, the four-fracton gap remains finite up to the phase transition point. 

\subsection{X-Cube in a field}
\label{ssect::xcube_fractons}
In the last subsection for Haah's code in a field we have seen that no gap of single- or multi-fracton excitations in the fracton phase closes before the detected ground-state level crossings in Subsect.~\ref{ssect::haah_pd}. The first-order nature of the phase transitions is therefore confirmed for all considered field configurations. Physically, this can be traced back to the immobility of all topologically non-trivial excitations of type-II fracton phases like Haah's code. Now we investigate the relevant excitation gaps for the X-Cube model in a field, which is qualitatively different, since in type-I fracton phases composites of fractons are allowed to move in lower dimensional submanifolds of the whole system. 

\subsubsection*{Single $x$-field}

We start by considering the effect of a field in $x$-direction parametrized by \mbox{$\boldsymbol{h}=(h_x,0,0)$} on the excitations of the fracton phase. Since the eigenvalues $a_c$ of the $\hat{A}_c$ operators remain conserved, the relevant low-energy excitations correspond to fracton excitations with \mbox{$b_s^{(\kappa)}=-1$}. Keeping in mind the local constraint $\prod_\kappa \hat{B}_{s}^{(\kappa)} = \mathds{1}$, a single $b_{s}^{(\kappa)}=-1$ does not exist and the elementary excitation is given by a lineon excitation. This is also reflected by dual description Eq.~\eqref{eq:xcube_x_duality} given in Sect.~\ref{sect::FF}.  The lineon corresponds to a pair of \mbox{$b_{s}^{(\kappa)}=b_{s}^{(\kappa')}=-1$} at vertex $s$ with $\kappa\neq\kappa'$ and can only move along the cartesian direction corresponding to the cut of the planes $\kappa$ and $\kappa'$.
\begin{figure}
	\centering
	\includegraphics[width = \columnwidth]{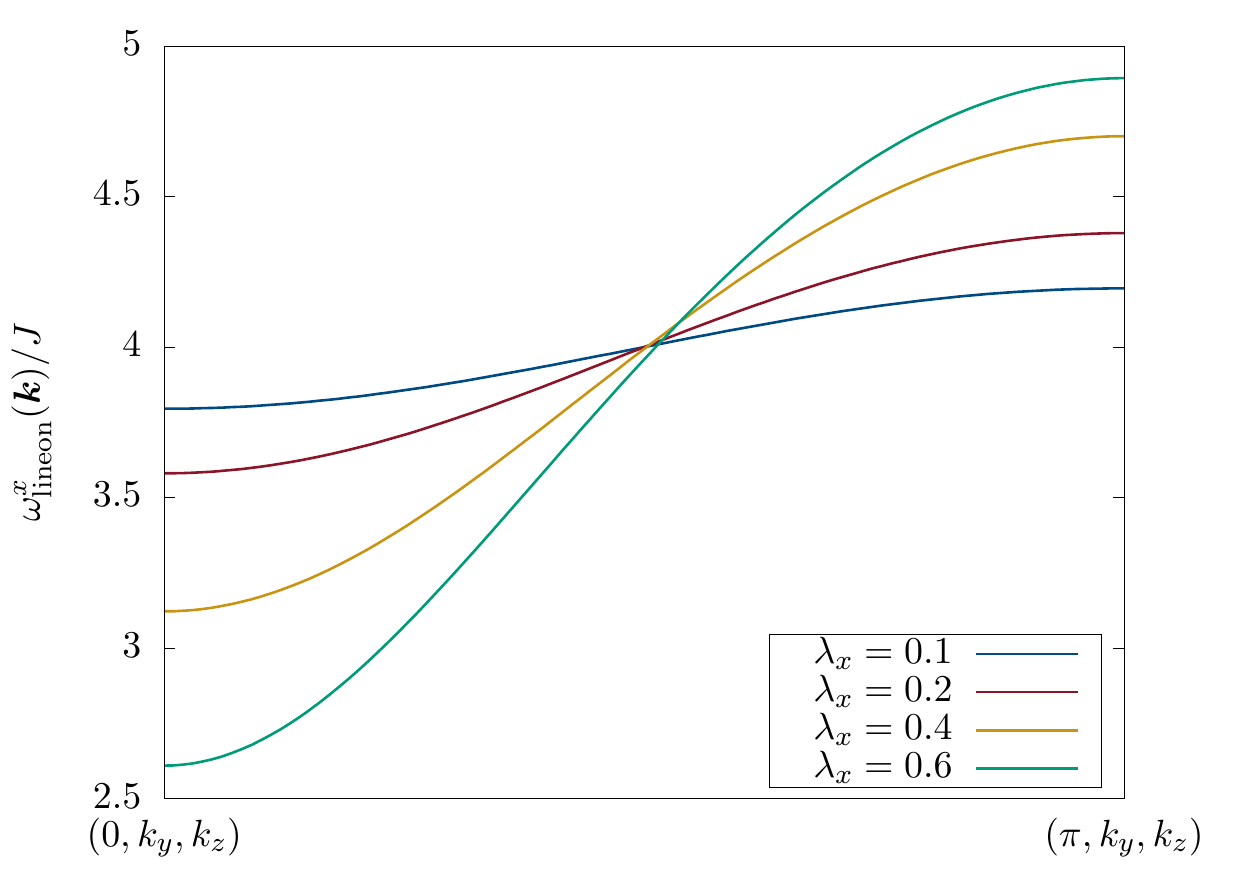}
	\caption{Dispersion $\omega^{x}_{\rm lineon}(\boldsymbol{k})/J$ of the $x$-type lineon excitation of the X-Cube model in a single $x$-field as a function of $k_x$ for different values of $\lambda_x$ using the bare pCUT series of order ten.}
	\label{fig::xcube_fractons::xfield::lineon_disp}
\end{figure}

All three types of lineons are energetically degenerate; in the following we consider a lineon able to move in $x$-direction. We calculated all hopping elements in $x$-direction up to order ten in $\lambda_x$ using the pCUT approach. Applying a Fourier transformation yields the lineon dispersion $\omega^{x}_{\rm lineon}(\boldsymbol{k})$, which is only dependent on $k_x$ but flat in the $k_y$-$k_z$-planes. This dispersion is plotted in Fig.~\ref{fig::xcube_fractons::xfield::lineon_disp} for different values of $\lambda_x$ using the bare pCUT series. 
\begin{figure}
	\centering
	\includegraphics[width = \columnwidth]{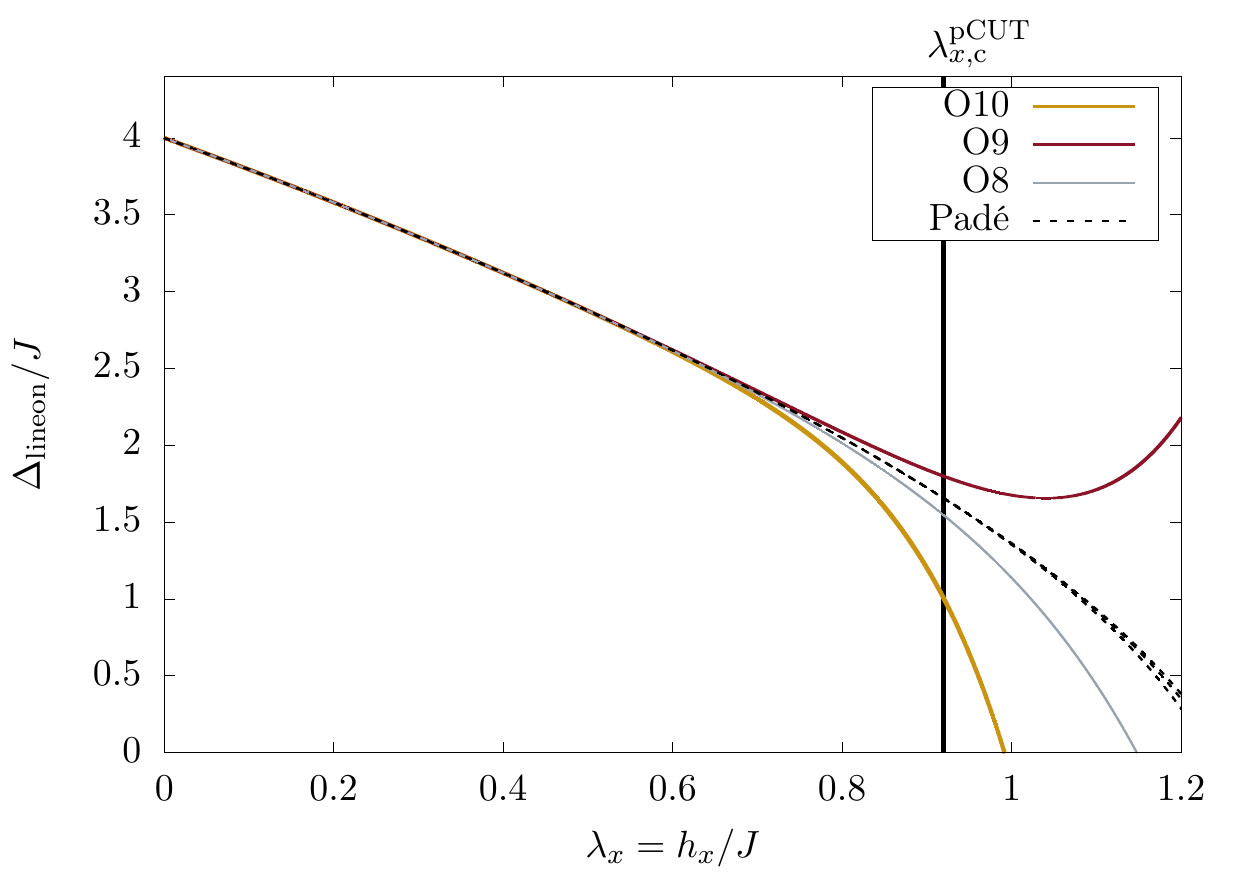}
	\caption{Bare series (solid lines) as well as Pad\'{e} extrapolants (dashed lines) of the lineon gap $\Delta_{\rm lineon}/J$ of the X-Cube model in a $x$-field as a function of $\lambda_x$. The vertical solid line indicates the phase transition point at \mbox{$\lambda_{x,{\rm c}}^{\rm pCUT}=0.9196 \pm 0.0012$} obtained from the pCUT analysis of the ground-state energy.}
	\label{fig::xcube_fractons::xfield::lineon_gap}
\end{figure}
The lineon gap $\Delta_{\rm lineon}$ is located at the momenta $\boldsymbol{k}_{\rm gap}=(0,k_y,k_z)$ and its pCUT series reads
 \begin{equation}
    \begin{aligned} 
     \frac{\Delta_{\mathrm{lineon}}}{J}  &= 4-2 \lambda_x-\frac{1}{2}\lambda_x^2 +\frac{1}{8}\lambda_x^3-\frac{53}{192}\lambda_x^4\\ 
     \phantom{\frac{\Delta_{\mathrm{lineon}}}{J}}&+\frac{973 }{4608}\lambda_x^5-\frac{34273 }{110592}\lambda_x^6  +\frac{1177037}{3538944} \lambda_x^7\\
     \phantom{\frac{\Delta_{\mathrm{lineon}}}{J}=}&-\frac{1133315183}{2548039680} \lambda_x^8 +\frac{368493295324181 }{684913065984000}\lambda_x^9 \\
     \phantom{\frac{\Delta_{\mathrm{lineon}}}{J}=}&-\frac{263340353497788689 }{143831743856640000} \lambda_x^{10}\, .
   \end{aligned}
   \end{equation}
Bare series as well as Pad\'{e} extrapolants of the lineon gap are shown in Fig.~\ref{fig::xcube_fractons::xfield::lineon_gap}. Although the bare series is alternating, the Pad\'{e} extrapolants are well converged up to the phase transition at $\lambda_{x,{\rm c}}^{\rm pCUT}=0.9196 \pm 0.0012$ implying a finite gap at the phase transition in agreement with the first-order phase transition found in Subsect.~\ref{ssect::xcube_pd} and the QMC simulations in Ref.~[\onlinecite{Devakul_2018}]. 

\subsubsection*{Single $z$-field}
Next we focus on how a field in $z$-direction parametrized by \mbox{$\boldsymbol{h}=(0,0,h_z)$} affects the excitations of the fracton phase. Since in this case the eigenvalues $b_s^{(\kappa)}$ of the $\hat{B}_s^{(\kappa)}$ operators are conserved and lineon excitations remain static at finite fields, the relevant low-energy excitations are fracton excitations with \mbox{$a_c=-1$}.

\begin{figure}
 \includegraphics[width=\columnwidth]{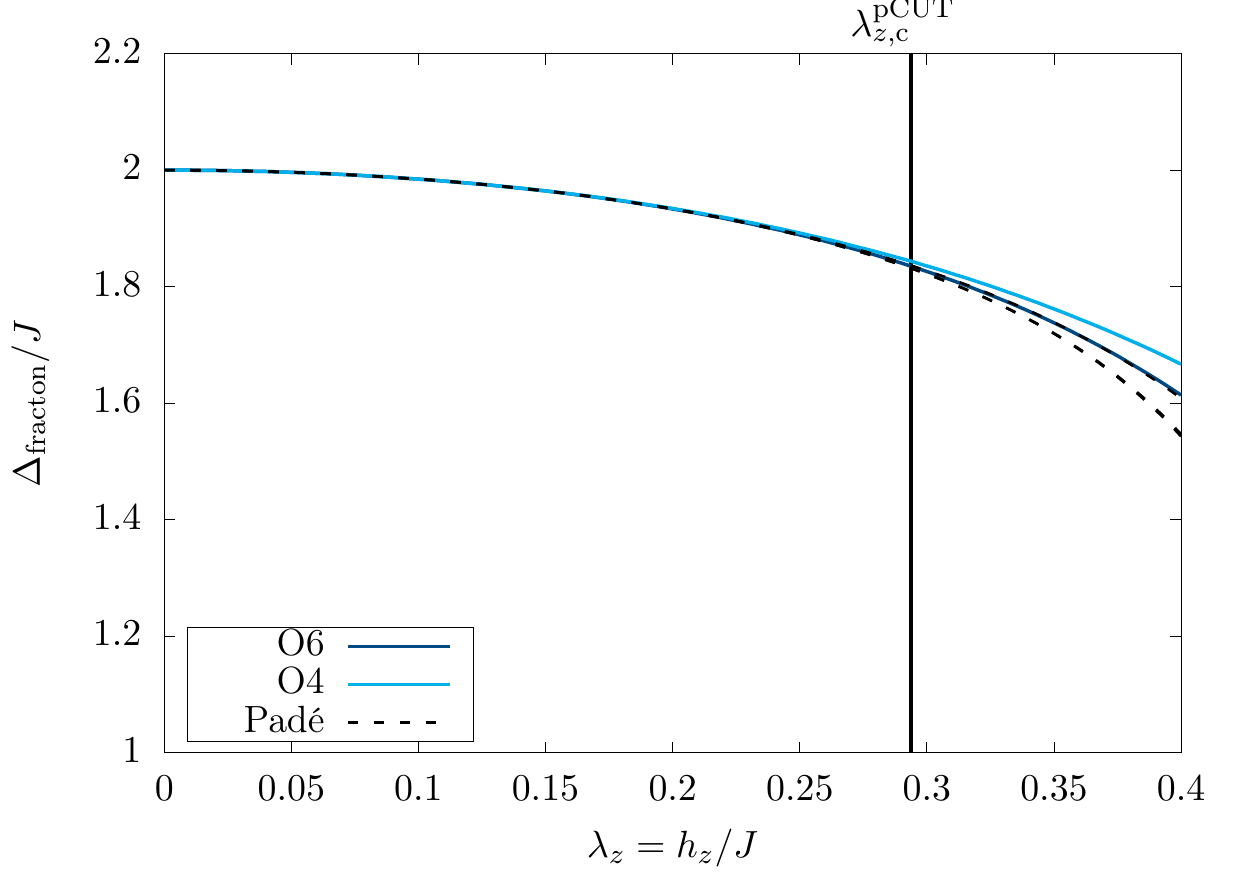}
 \caption{Bare series (solid lines) as well as Pad\'{e} extrapolants (dashed lines) of the one-fracton gap $\Delta_{\text{fracton}}/J$ of the X-Cube model in a $z$-field. The vertical solid line indicates the phase transition at \mbox{$\lambda_{z,{\rm c}}^{\rm pCUT}=0.29364 \pm 0.00017$} obtained from the pCUT analysis of the ground-state energy.}
 \label{fig::xcube_fractons::zfield::fracton_disp}
\end{figure}

A single fracton with \mbox{$a_c=-1$} does not hop despite of the magnetic field, but it can reduce its energy by vacuum fluctuations like single fractons in the perturbed Haah's code. The momentum-independent one-fracton gap is then given by
\begin{equation}    
  \frac{\Delta_{\rm fracton}}{J} = 2-\frac{3}{2}\lambda_z^2-\frac{465}{128}\lambda_z^4-\frac{478181}{36864}\lambda_z^6 \quad,
\end{equation}
which is illustrated in Fig. \ref{fig::xcube_fractons::zfield::fracton_disp} including Pad\'e extrapolants. 
\begin{figure}[t!]
	\centering
	\includegraphics[width = \columnwidth]{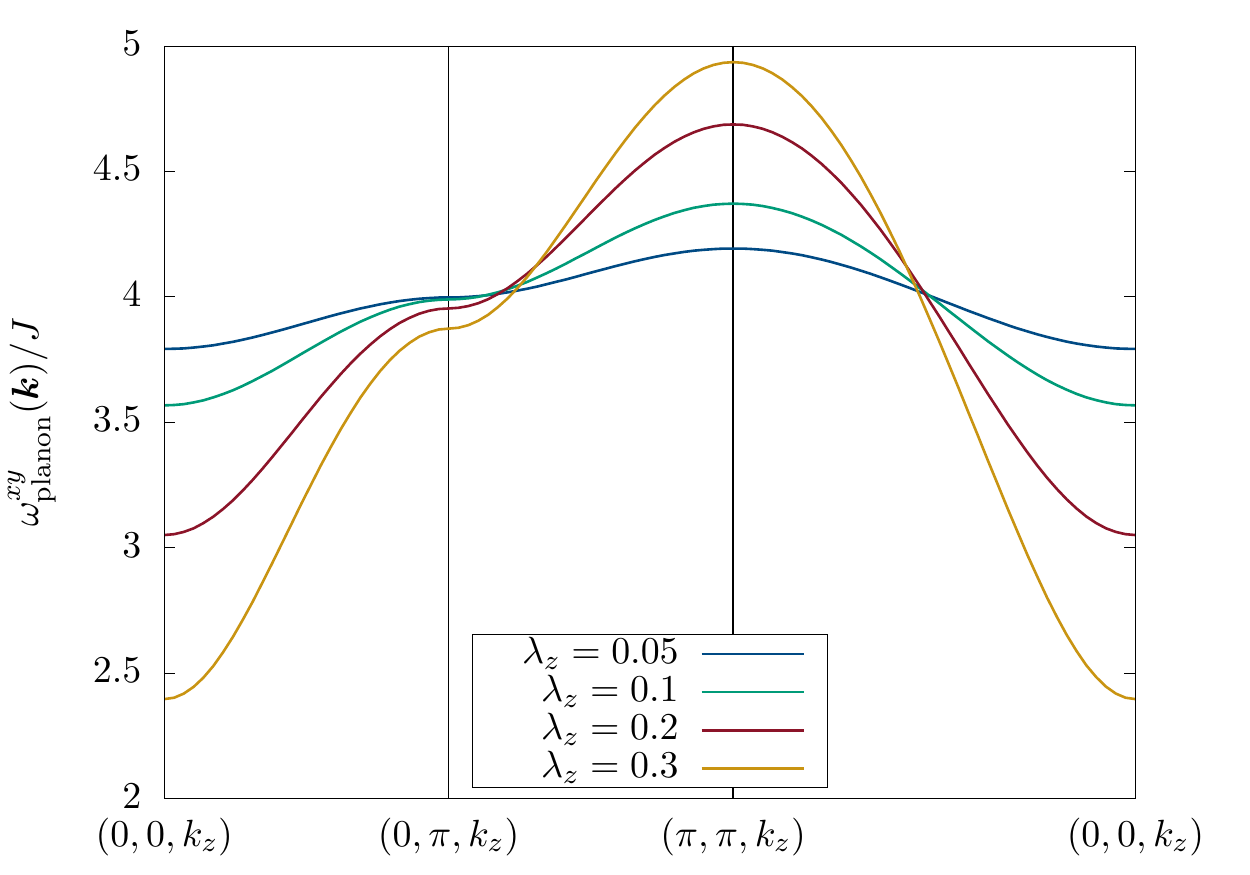}
	\caption{Dispersion of the planon excitation $\omega^{xy}_{\rm planon}(\boldsymbol{k})/J$ of the X-Cube model in a $z$-field along a high-symmetry path in the Brillouin zone for different values of $\lambda_z$ using the bare order-seven pCUT series.}
	\label{fig::xcube_fractons::zfield::planon_disp}
\end{figure}
\begin{figure}[t!]
	\centering
	\includegraphics[width = \columnwidth]{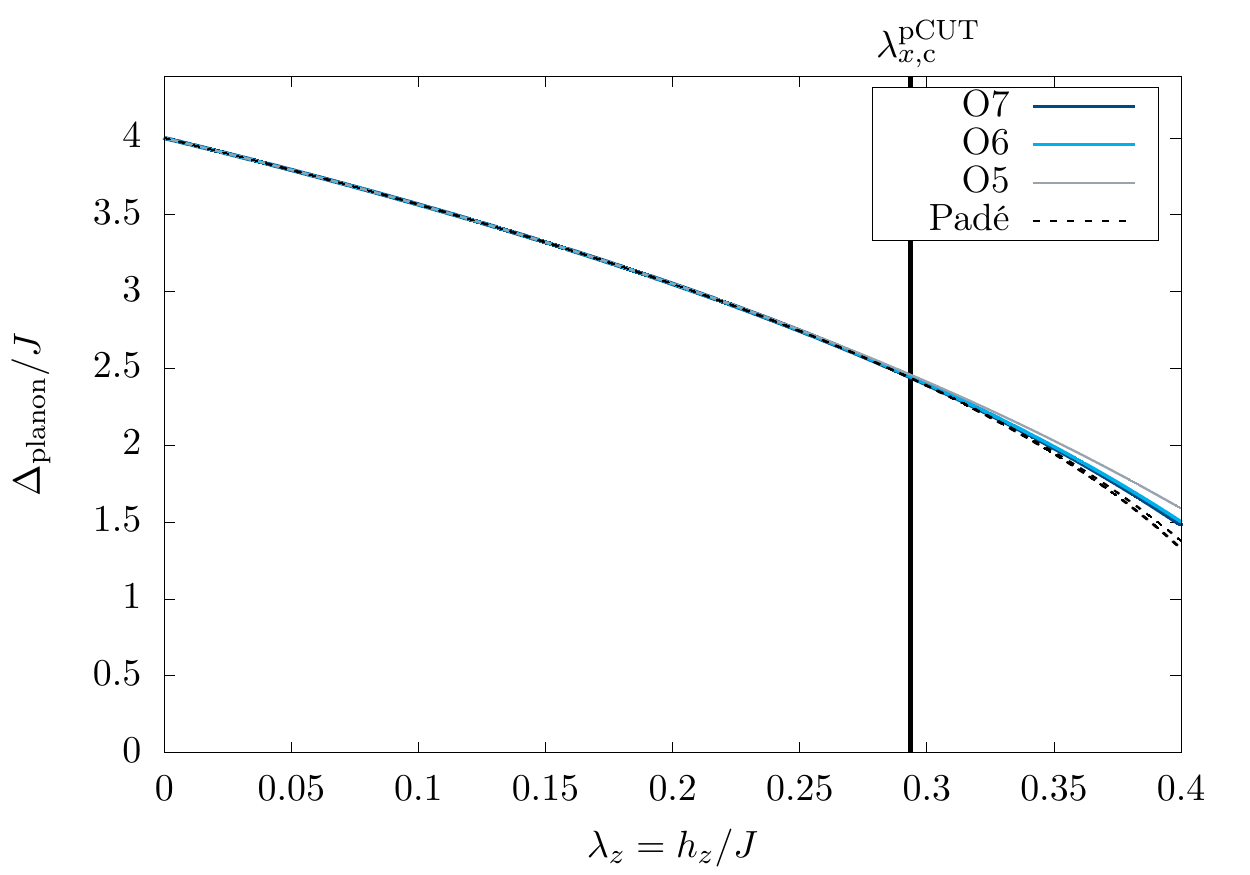}
	\caption{Bare series (solid lines) as well as Pad\'{e} extrapolants (dashed lines) of the planon gap $\Delta_{\rm planon}/J$ as a function of $\lambda_z$ for the $z$-field case of the X-Cube model. As before the vertical solid line indicates the phase transition point $\lambda_{z,{\rm c}}^{\rm pCUT}=0.29364 \pm 0.00017$ obtained from the pCUT analysis of the ground-state energy.}
	\label{fig::xcube_fractons::zfield::planon_gap}
\end{figure}
The relevant mobile low-energy excitation is the planon which corresponds to two cubes with 
\mbox{$a_c=-1$} sharing a cartesian coordinate so that the planon can move in the two-dimensional plane orthogonal to this direction. Planons in the three different kinds of planes $xy$, $xz$, and $yz$ have identical physical properties. In the following we focus on the $xy$-planon whose dispersion relation $\omega^{xy}_{\rm planon}(\boldsymbol{k})/J$ for different values of $\lambda_z$ using the bare order-seven series is displayed in Fig \ref{fig::xcube_fractons::zfield::planon_disp}. The gap $\Delta_{\mathrm{planon}}$ of the $xy$-planon is located at momenta $\boldsymbol{k}=(0,0,k_z)$ and its series expansion up to order seven reads
\begin{eqnarray}
 \label{eq::xcube_z_planon_gap}
    \frac{\Delta_{\mathrm{planon}}}{J} &=&4-4 \lambda_z-3 \lambda_z^2-\frac{17 }{8}\lambda_z^3-\frac{1151}{192} \lambda_z^4\\ 
  \phantom{\Delta_{\mathrm{planon}}(\lambda_z)}&-& \frac{37165}{9216} \lambda_z^5 -\frac{2591423 }{122880}\lambda_z^6-\frac{6264944713 }{530841600}\lambda_z^7\, . \nonumber
\end{eqnarray}
Bare series as well as Pad\'{e} extrapolants of the planon gap are shown in Fig.~\ref{fig::xcube_fractons::zfield::planon_gap}. Both the bare monotonic series \eqref{eq::xcube_z_planon_gap} as well as the Pad\'{e} extrapolants are well converged until the phase transition point \mbox{$\lambda_{z,{\rm c}}^{\rm pCUT}=0.29364 \pm 0.00017$}. So although the planon is allowed to move in a two-dimensional subspace (in contrast to the lineon), the planon gap remains finite at the phase transition in agreement with the first-order phase transition found in Subsect.~\ref{ssect::xcube_pd} and the QMC simulations in Ref.~[\onlinecite{Devakul_2018}].

\section{Conclusions}
\label{sect::conclusion}
In this work we have investigated the quantum robustness of type-I and type-II fracton phases by considering the exactly solvable Haah's code and the X-Cube model in magnetic fields. The latter introduce quantum fluctuations leading to the breakdown of the fracton phases at certain field strengths via quantum phase transitions to the topologically trivial polarized phase. 

We have located the zero-temperature phase transitions quantitatively in all considered cases by performing low- and high-field series expansions for the ground-state energy. For the X-Cube model in a field, our results agree with QMC simulations \cite{Devakul_2018}. Furthermore, all phase transitions are classified as strongly first order, which can be deduced from the kink in the ground-state energy as well as from the excitation energies of fracton quasi-particles. The excitation energies for single and composites of fracton quasi-particles have been calculated as high-order series expansions showing that all energy gaps remain finite in the whole fracton phases. Physically, therefore the (partial) immobility of fracton quasi-particles hinders the occurence of second-order phase transitions out of fracton phases. In conclusion it should be investigated whether the defining properties of fracton codes exclude quantum critical behavior in general.

\begin{acknowledgments}
  We thank T.~Devakul for providing us with the QMC data from Ref.~[\onlinecite{Devakul_2018}]. KPS acknowledges financial support by the German Science Foundation (DFG) through the grant SCHM 2511/11-1. 
\end{acknowledgments}

\bibliography{./MT_paper_bib_10_01_19}

\begin{thebibliography}{81}%
\makeatletter
\providecommand \@ifxundefined [1]{%
 \@ifx{#1\undefined}
}%
\providecommand \@ifnum [1]{%
 \ifnum #1\expandafter \@firstoftwo
 \else \expandafter \@secondoftwo
 \fi
}%
\providecommand \@ifx [1]{%
 \ifx #1\expandafter \@firstoftwo
 \else \expandafter \@secondoftwo
 \fi
}%
\providecommand \natexlab [1]{#1}%
\providecommand \enquote  [1]{``#1''}%
\providecommand \bibnamefont  [1]{#1}%
\providecommand \bibfnamefont [1]{#1}%
\providecommand \citenamefont [1]{#1}%
\providecommand \href@noop [0]{\@secondoftwo}%
\providecommand \href [0]{\begingroup \@sanitize@url \@href}%
\providecommand \@href[1]{\@@startlink{#1}\@@href}%
\providecommand \@@href[1]{\endgroup#1\@@endlink}%
\providecommand \@sanitize@url [0]{\catcode `\\12\catcode `\$12\catcode
  `\&12\catcode `\#12\catcode `\^12\catcode `\_12\catcode `\%12\relax}%
\providecommand \@@startlink[1]{}%
\providecommand \@@endlink[0]{}%
\providecommand \url  [0]{\begingroup\@sanitize@url \@url }%
\providecommand \@url [1]{\endgroup\@href {#1}{\urlprefix }}%
\providecommand \urlprefix  [0]{URL }%
\providecommand \Eprint [0]{\href }%
\providecommand \doibase [0]{http://dx.doi.org/}%
\providecommand \selectlanguage [0]{\@gobble}%
\providecommand \bibinfo  [0]{\@secondoftwo}%
\providecommand \bibfield  [0]{\@secondoftwo}%
\providecommand \translation [1]{[#1]}%
\providecommand \BibitemOpen [0]{}%
\providecommand \bibitemStop [0]{}%
\providecommand \bibitemNoStop [0]{.\EOS\space}%
\providecommand \EOS [0]{\spacefactor3000\relax}%
\providecommand \BibitemShut  [1]{\csname bibitem#1\endcsname}%
\let\auto@bib@innerbib\@empty
\bibitem [{\citenamefont {Wen}(1989)}]{Wen_1989}%
  \BibitemOpen
  \bibfield  {author} {\bibinfo {author} {\bibfnamefont {X.-G.}\ \bibnamefont
  {Wen}},\ }\href {\doibase 10.1103/PhysRevB.40.7387} {\bibfield  {journal}
  {\bibinfo  {journal} {Phys. Rev. B}\ }\textbf {\bibinfo {volume} {40}},\
  \bibinfo {pages} {7387} (\bibinfo {year} {1989})}\BibitemShut {NoStop}%
\bibitem [{\citenamefont {Wen}(1990)}]{Wen_1990}%
  \BibitemOpen
  \bibfield  {author} {\bibinfo {author} {\bibfnamefont {X.-G.}\ \bibnamefont
  {Wen}},\ }\href {\doibase 10.1142/S0217979290000139} {\bibfield  {journal}
  {\bibinfo  {journal} {Int. J. Mod. Phys. B}\ }\textbf {\bibinfo {volume}
  {4}},\ \bibinfo {pages} {239} (\bibinfo {year} {1990})}\BibitemShut {NoStop}%
\bibitem [{\citenamefont {Wen}(2004)}]{Wen_2004}%
  \BibitemOpen
  \bibfield  {author} {\bibinfo {author} {\bibfnamefont {X.-G.}\ \bibnamefont
  {Wen}},\ }\href@noop {} {\enquote {\bibinfo {title} {Quantum field theory of
  many-body systems: From the origin of sound to an origin of light and
  electrons},}\ }\bibinfo {howpublished} {Oxford University Press} (\bibinfo
  {year} {2004})\BibitemShut {NoStop}%
\bibitem [{\citenamefont {Leinaas}\ and\ \citenamefont
  {Myrheim}(1977)}]{Leinaas_1977}%
  \BibitemOpen
  \bibfield  {author} {\bibinfo {author} {\bibfnamefont {J.~M.}\ \bibnamefont
  {Leinaas}}\ and\ \bibinfo {author} {\bibfnamefont {J.}~\bibnamefont
  {Myrheim}},\ }\href {\doibase 10.1007/BF02727953} {\bibfield  {journal}
  {\bibinfo  {journal} {Il Nuovo Cimento B}\ }\textbf {\bibinfo {volume}
  {37}},\ \bibinfo {pages} {1} (\bibinfo {year} {1977})}\BibitemShut {NoStop}%
\bibitem [{\citenamefont {Wilczek}(1982)}]{Wilczek_1982}%
  \BibitemOpen
  \bibfield  {author} {\bibinfo {author} {\bibfnamefont {F.}~\bibnamefont
  {Wilczek}},\ }\href {\doibase 10.1103/PhysRevLett.48.1144} {\bibfield
  {journal} {\bibinfo  {journal} {Phys. Rev. Lett.}\ }\textbf {\bibinfo
  {volume} {48}},\ \bibinfo {pages} {1144} (\bibinfo {year}
  {1982})}\BibitemShut {NoStop}%
\bibitem [{\citenamefont {Hamma}\ \emph {et~al.}(2005)\citenamefont {Hamma},
  \citenamefont {Zanardi},\ and\ \citenamefont {Wen}}]{Hamma_2005}%
  \BibitemOpen
  \bibfield  {author} {\bibinfo {author} {\bibfnamefont {A.}~\bibnamefont
  {Hamma}}, \bibinfo {author} {\bibfnamefont {P.}~\bibnamefont {Zanardi}}, \
  and\ \bibinfo {author} {\bibfnamefont {X.-G.}\ \bibnamefont {Wen}},\ }\href
  {\doibase 10.1103/PhysRevB.72.035307} {\bibfield  {journal} {\bibinfo
  {journal} {Phys. Rev. B}\ }\textbf {\bibinfo {volume} {72}},\ \bibinfo
  {pages} {035307} (\bibinfo {year} {2005})}\BibitemShut {NoStop}%
\bibitem [{\citenamefont {Nussinov}\ and\ \citenamefont
  {Ortiz}(2008)}]{Nussinov_2008}%
  \BibitemOpen
  \bibfield  {author} {\bibinfo {author} {\bibfnamefont {Z.}~\bibnamefont
  {Nussinov}}\ and\ \bibinfo {author} {\bibfnamefont {G.}~\bibnamefont
  {Ortiz}},\ }\href {\doibase 10.1103/PhysRevB.77.064302} {\bibfield  {journal}
  {\bibinfo  {journal} {Phys. Rev. B}\ }\textbf {\bibinfo {volume} {77}},\
  \bibinfo {pages} {064302} (\bibinfo {year} {2008})}\BibitemShut {NoStop}%
\bibitem [{\citenamefont {Reiss}\ and\ \citenamefont
  {Schmidt}(2019)}]{Reiss_2019}%
  \BibitemOpen
  \bibfield  {author} {\bibinfo {author} {\bibfnamefont {D.~A.}\ \bibnamefont
  {Reiss}}\ and\ \bibinfo {author} {\bibfnamefont {K.~P.}\ \bibnamefont
  {Schmidt}},\ }\href {\doibase 10.21468/SciPostPhys.6.6.078} {\bibfield
  {journal} {\bibinfo  {journal} {SciPost Phys.}\ }\textbf {\bibinfo {volume}
  {6}},\ \bibinfo {pages} {78} (\bibinfo {year} {2019})}\BibitemShut {NoStop}%
\bibitem [{\citenamefont {Kitaev}(2003)}]{Kitaev_2003}%
  \BibitemOpen
  \bibfield  {author} {\bibinfo {author} {\bibfnamefont {A.}~\bibnamefont
  {Kitaev}},\ }\href {\doibase 10.1016/S0003-4916(02)00018-0} {\bibfield
  {journal} {\bibinfo  {journal} {Ann. Phys.}\ }\textbf {\bibinfo {volume}
  {303}},\ \bibinfo {pages} {2} (\bibinfo {year} {2003})}\BibitemShut {NoStop}%
\bibitem [{\citenamefont {Nayak}\ \emph {et~al.}(2008)\citenamefont {Nayak},
  \citenamefont {Simon}, \citenamefont {Stern}, \citenamefont {Freedman},\ and\
  \citenamefont {Sarma}}]{Nayak_2008}%
  \BibitemOpen
  \bibfield  {author} {\bibinfo {author} {\bibfnamefont {C.}~\bibnamefont
  {Nayak}}, \bibinfo {author} {\bibfnamefont {S.~H.}\ \bibnamefont {Simon}},
  \bibinfo {author} {\bibfnamefont {A.}~\bibnamefont {Stern}}, \bibinfo
  {author} {\bibfnamefont {M.}~\bibnamefont {Freedman}}, \ and\ \bibinfo
  {author} {\bibfnamefont {S.~D.}\ \bibnamefont {Sarma}},\ }\href {\doibase
  10.1103/RevModPhys.80.1083} {\bibfield  {journal} {\bibinfo  {journal} {Rev.
  Mod. Phys.}\ }\textbf {\bibinfo {volume} {80}},\ \bibinfo {pages} {1083}
  (\bibinfo {year} {2008})}\BibitemShut {NoStop}%
\bibitem [{\citenamefont {Alicki}\ \emph {et~al.}(2009)\citenamefont {Alicki},
  \citenamefont {Fannes},\ and\ \citenamefont {Horodecki}}]{Alicki_2009}%
  \BibitemOpen
  \bibfield  {author} {\bibinfo {author} {\bibfnamefont {R.}~\bibnamefont
  {Alicki}}, \bibinfo {author} {\bibfnamefont {M.}~\bibnamefont {Fannes}}, \
  and\ \bibinfo {author} {\bibfnamefont {M.}~\bibnamefont {Horodecki}},\ }\href
  {\doibase 10.1088/1751-8113/42/6/065303} {\bibfield  {journal} {\bibinfo
  {journal} {J. Phys. A: Math. Theor.}\ }\textbf {\bibinfo {volume} {42}},\
  \bibinfo {pages} {065303} (\bibinfo {year} {2009})}\BibitemShut {NoStop}%
\bibitem [{\citenamefont {Castelnovo}\ and\ \citenamefont
  {Chamon}(2007)}]{Castelnovo_2007}%
  \BibitemOpen
  \bibfield  {author} {\bibinfo {author} {\bibfnamefont {C.}~\bibnamefont
  {Castelnovo}}\ and\ \bibinfo {author} {\bibfnamefont {C.}~\bibnamefont
  {Chamon}},\ }\href {\doibase 10.1103/PhysRevB.76.184442} {\bibfield
  {journal} {\bibinfo  {journal} {Phys. Rev. B}\ }\textbf {\bibinfo {volume}
  {76}},\ \bibinfo {pages} {184442} (\bibinfo {year} {2007})}\BibitemShut
  {NoStop}%
\bibitem [{\citenamefont {Nussinov}\ and\ \citenamefont
  {Ortiz}(2009)}]{Nussinov_2009}%
  \BibitemOpen
  \bibfield  {author} {\bibinfo {author} {\bibfnamefont {Z.}~\bibnamefont
  {Nussinov}}\ and\ \bibinfo {author} {\bibfnamefont {G.}~\bibnamefont
  {Ortiz}},\ }\href@noop {} {\bibfield  {journal} {\bibinfo  {journal} {Ann.
  Phys.}\ }\textbf {\bibinfo {volume} {324}},\ \bibinfo {pages} {977} (\bibinfo
  {year} {2009})}\BibitemShut {NoStop}%
\bibitem [{\citenamefont {Bravyi}\ and\ \citenamefont
  {Terhal}(2009)}]{Bravyi_2009}%
  \BibitemOpen
  \bibfield  {author} {\bibinfo {author} {\bibfnamefont {S.}~\bibnamefont
  {Bravyi}}\ and\ \bibinfo {author} {\bibfnamefont {B.~M.}\ \bibnamefont
  {Terhal}},\ }\href@noop {} {\bibfield  {journal} {\bibinfo  {journal} {New J.
  Phys.}\ }\textbf {\bibinfo {volume} {11}} (\bibinfo {year}
  {2009})}\BibitemShut {NoStop}%
\bibitem [{\citenamefont {Chamon}(2005)}]{Chamon_2005}%
  \BibitemOpen
  \bibfield  {author} {\bibinfo {author} {\bibfnamefont {C.}~\bibnamefont
  {Chamon}},\ }\href {\doibase 10.1103/PhysRevLett.94.040402} {\bibfield
  {journal} {\bibinfo  {journal} {Phys. Rev. Lett.}\ }\textbf {\bibinfo
  {volume} {94}},\ \bibinfo {pages} {040402} (\bibinfo {year}
  {2005})}\BibitemShut {NoStop}%
\bibitem [{\citenamefont {Bravyi}\ \emph {et~al.}(2011)\citenamefont {Bravyi},
  \citenamefont {Leemhuis},\ and\ \citenamefont {Terhal}}]{Bravyi_2011}%
  \BibitemOpen
  \bibfield  {author} {\bibinfo {author} {\bibfnamefont {S.}~\bibnamefont
  {Bravyi}}, \bibinfo {author} {\bibfnamefont {B.}~\bibnamefont {Leemhuis}}, \
  and\ \bibinfo {author} {\bibfnamefont {B.~M.}\ \bibnamefont {Terhal}},\
  }\href {\doibase 10.1016/j.aop.2010.11.002} {\bibfield  {journal} {\bibinfo
  {journal} {Ann. Phys.}\ }\textbf {\bibinfo {volume} {326}},\ \bibinfo {pages}
  {839} (\bibinfo {year} {2011})}\BibitemShut {NoStop}%
\bibitem [{\citenamefont {Haah}(2011)}]{Haah_2011}%
  \BibitemOpen
  \bibfield  {author} {\bibinfo {author} {\bibfnamefont {J.}~\bibnamefont
  {Haah}},\ }\href {\doibase 10.1103/PhysRevA.83.042330} {\bibfield  {journal}
  {\bibinfo  {journal} {Phys. Rev. A}\ }\textbf {\bibinfo {volume} {83}},\
  \bibinfo {pages} {042330} (\bibinfo {year} {2011})}\BibitemShut {NoStop}%
\bibitem [{\citenamefont {Bravyi}\ and\ \citenamefont
  {Haah}(2013)}]{Bravyi_2013}%
  \BibitemOpen
  \bibfield  {author} {\bibinfo {author} {\bibfnamefont {S.}~\bibnamefont
  {Bravyi}}\ and\ \bibinfo {author} {\bibfnamefont {J.}~\bibnamefont {Haah}},\
  }\href {\doibase 10.1103/PhysRevLett.111.200501} {\bibfield  {journal}
  {\bibinfo  {journal} {Phys. Rev. Lett.}\ }\textbf {\bibinfo {volume} {111}},\
  \bibinfo {pages} {200501} (\bibinfo {year} {2013})}\BibitemShut {NoStop}%
\bibitem [{\citenamefont {Yoshida}(2013)}]{Yoshida_2013}%
  \BibitemOpen
  \bibfield  {author} {\bibinfo {author} {\bibfnamefont {B.}~\bibnamefont
  {Yoshida}},\ }\href {\doibase 10.1103/PhysRevB.88.125122} {\bibfield
  {journal} {\bibinfo  {journal} {Phys. Rev. B}\ }\textbf {\bibinfo {volume}
  {88}},\ \bibinfo {pages} {125122} (\bibinfo {year} {2013})}\BibitemShut
  {NoStop}%
\bibitem [{\citenamefont {Vijay}\ \emph {et~al.}(2015)\citenamefont {Vijay},
  \citenamefont {Haah},\ and\ \citenamefont {Fu}}]{Vijay_2015}%
  \BibitemOpen
  \bibfield  {author} {\bibinfo {author} {\bibfnamefont {S.}~\bibnamefont
  {Vijay}}, \bibinfo {author} {\bibfnamefont {J.}~\bibnamefont {Haah}}, \ and\
  \bibinfo {author} {\bibfnamefont {L.}~\bibnamefont {Fu}},\ }\href {\doibase
  10.1103/PhysRevB.92.235136} {\bibfield  {journal} {\bibinfo  {journal} {Phys.
  Rev. B}\ }\textbf {\bibinfo {volume} {92}},\ \bibinfo {pages} {235136}
  (\bibinfo {year} {2015})}\BibitemShut {NoStop}%
\bibitem [{\citenamefont {Vijay}\ \emph
  {et~al.}(2016{\natexlab{a}})\citenamefont {Vijay}, \citenamefont {Haah},\
  and\ \citenamefont {Fu}}]{Vijay_2016}%
  \BibitemOpen
  \bibfield  {author} {\bibinfo {author} {\bibfnamefont {S.}~\bibnamefont
  {Vijay}}, \bibinfo {author} {\bibfnamefont {J.}~\bibnamefont {Haah}}, \ and\
  \bibinfo {author} {\bibfnamefont {L.}~\bibnamefont {Fu}},\ }\href {\doibase
  10.1103/PhysRevB.94.235157} {\bibfield  {journal} {\bibinfo  {journal} {Phys.
  Rev. B}\ }\textbf {\bibinfo {volume} {94}},\ \bibinfo {pages} {235157}
  (\bibinfo {year} {2016}{\natexlab{a}})}\BibitemShut {NoStop}%
\bibitem [{\citenamefont {Slagle}\ and\ \citenamefont
  {Kim}(2017{\natexlab{a}})}]{Slagle_2017}%
  \BibitemOpen
  \bibfield  {author} {\bibinfo {author} {\bibfnamefont {K.}~\bibnamefont
  {Slagle}}\ and\ \bibinfo {author} {\bibfnamefont {Y.~B.}\ \bibnamefont
  {Kim}},\ }\href {\doibase 10.1103/PhysRevB.96.165106} {\bibfield  {journal}
  {\bibinfo  {journal} {Phys. Rev. B}\ }\textbf {\bibinfo {volume} {96}},\
  \bibinfo {pages} {165106} (\bibinfo {year} {2017}{\natexlab{a}})}\BibitemShut
  {NoStop}%
\bibitem [{\citenamefont {Slagle}\ and\ \citenamefont
  {Kim}(2017{\natexlab{b}})}]{Slagle17}%
  \BibitemOpen
  \bibfield  {author} {\bibinfo {author} {\bibfnamefont {K.}~\bibnamefont
  {Slagle}}\ and\ \bibinfo {author} {\bibfnamefont {Y.~B.}\ \bibnamefont
  {Kim}},\ }\href@noop {} {\bibfield  {journal} {\bibinfo  {journal} {Physical
  Review B}\ }\textbf {\bibinfo {volume} {96}},\ \bibinfo {pages} {195139}
  (\bibinfo {year} {2017}{\natexlab{b}})}\BibitemShut {NoStop}%
\bibitem [{\citenamefont {Ma}\ \emph {et~al.}(2017{\natexlab{a}})\citenamefont
  {Ma}, \citenamefont {Lake}, \citenamefont {Chen},\ and\ \citenamefont
  {Hermele}}]{Ma_2017}%
  \BibitemOpen
  \bibfield  {author} {\bibinfo {author} {\bibfnamefont {H.}~\bibnamefont
  {Ma}}, \bibinfo {author} {\bibfnamefont {E.}~\bibnamefont {Lake}}, \bibinfo
  {author} {\bibfnamefont {X.}~\bibnamefont {Chen}}, \ and\ \bibinfo {author}
  {\bibfnamefont {M.}~\bibnamefont {Hermele}},\ }\href@noop {} {\bibfield
  {journal} {\bibinfo  {journal} {Phys. Rev. B}\ }\textbf {\bibinfo {volume}
  {95}} (\bibinfo {year} {2017}{\natexlab{a}})}\BibitemShut {NoStop}%
\bibitem [{\citenamefont {Pretko}(2017{\natexlab{a}})}]{Pretko_2017a}%
  \BibitemOpen
  \bibfield  {author} {\bibinfo {author} {\bibfnamefont {M.}~\bibnamefont
  {Pretko}},\ }\href {\doibase 10.1103/PhysRevB.95.115139} {\bibfield
  {journal} {\bibinfo  {journal} {Phys. Rev. B}\ }\textbf {\bibinfo {volume}
  {95}},\ \bibinfo {pages} {115139} (\bibinfo {year}
  {2017}{\natexlab{a}})}\BibitemShut {NoStop}%
\bibitem [{\citenamefont {Pretko}(2017{\natexlab{b}})}]{Pretko_2017b}%
  \BibitemOpen
  \bibfield  {author} {\bibinfo {author} {\bibfnamefont {M.}~\bibnamefont
  {Pretko}},\ }\href {\doibase 10.1103/PhysRevB.96.035119} {\bibfield
  {journal} {\bibinfo  {journal} {Phys. Rev. B}\ }\textbf {\bibinfo {volume}
  {96}},\ \bibinfo {pages} {035119} (\bibinfo {year}
  {2017}{\natexlab{b}})}\BibitemShut {NoStop}%
\bibitem [{\citenamefont {Petrova}\ and\ \citenamefont
  {Regnault}(2017)}]{Petrova_2017}%
  \BibitemOpen
  \bibfield  {author} {\bibinfo {author} {\bibfnamefont {O.}~\bibnamefont
  {Petrova}}\ and\ \bibinfo {author} {\bibfnamefont {N.}~\bibnamefont
  {Regnault}},\ }\href {\doibase 10.1103/PhysRevB.96.224429} {\bibfield
  {journal} {\bibinfo  {journal} {Phys. Rev. B}\ }\textbf {\bibinfo {volume}
  {96}},\ \bibinfo {pages} {224429} (\bibinfo {year} {2017})}\BibitemShut
  {NoStop}%
\bibitem [{\citenamefont {Hal\'asz}\ \emph {et~al.}(2017)\citenamefont
  {Hal\'asz}, \citenamefont {Hsieh},\ and\ \citenamefont
  {Balents}}]{Halasz_2017}%
  \BibitemOpen
  \bibfield  {author} {\bibinfo {author} {\bibfnamefont {G.~B.}\ \bibnamefont
  {Hal\'asz}}, \bibinfo {author} {\bibfnamefont {T.~H.}\ \bibnamefont {Hsieh}},
  \ and\ \bibinfo {author} {\bibfnamefont {L.}~\bibnamefont {Balents}},\ }\href
  {\doibase 10.1103/PhysRevLett.119.257202} {\bibfield  {journal} {\bibinfo
  {journal} {Phys. Rev. Lett.}\ }\textbf {\bibinfo {volume} {119}},\ \bibinfo
  {pages} {257202} (\bibinfo {year} {2017})}\BibitemShut {NoStop}%
\bibitem [{\citenamefont {Prem}\ \emph {et~al.}(2019)\citenamefont {Prem},
  \citenamefont {Huang}, \citenamefont {Song},\ and\ \citenamefont
  {Hermele}}]{Prem_2019}%
  \BibitemOpen
  \bibfield  {author} {\bibinfo {author} {\bibfnamefont {A.}~\bibnamefont
  {Prem}}, \bibinfo {author} {\bibfnamefont {S.-J.}\ \bibnamefont {Huang}},
  \bibinfo {author} {\bibfnamefont {H.}~\bibnamefont {Song}}, \ and\ \bibinfo
  {author} {\bibfnamefont {M.}~\bibnamefont {Hermele}},\ }\href {\doibase
  10.1103/PhysRevX.9.021010} {\bibfield  {journal} {\bibinfo  {journal} {Phys.
  Rev. X}\ }\textbf {\bibinfo {volume} {9}},\ \bibinfo {pages} {021010}
  (\bibinfo {year} {2019})}\BibitemShut {NoStop}%
\bibitem [{\citenamefont {Williamson}\ \emph {et~al.}(2019)\citenamefont
  {Williamson}, \citenamefont {Bi},\ and\ \citenamefont
  {Cheng}}]{Williamson_2019}%
  \BibitemOpen
  \bibfield  {author} {\bibinfo {author} {\bibfnamefont {D.~J.}\ \bibnamefont
  {Williamson}}, \bibinfo {author} {\bibfnamefont {Z.}~\bibnamefont {Bi}}, \
  and\ \bibinfo {author} {\bibfnamefont {M.}~\bibnamefont {Cheng}},\ }\href
  {\doibase 10.1103/PhysRevB.100.125150} {\bibfield  {journal} {\bibinfo
  {journal} {Phys. Rev. B}\ }\textbf {\bibinfo {volume} {100}},\ \bibinfo
  {pages} {125150} (\bibinfo {year} {2019})}\BibitemShut {NoStop}%
\bibitem [{\citenamefont {Song}\ \emph {et~al.}(2019)\citenamefont {Song},
  \citenamefont {Prem}, \citenamefont {Huang},\ and\ \citenamefont
  {Martin-Delgado}}]{Song_2019}%
  \BibitemOpen
  \bibfield  {author} {\bibinfo {author} {\bibfnamefont {H.}~\bibnamefont
  {Song}}, \bibinfo {author} {\bibfnamefont {A.}~\bibnamefont {Prem}}, \bibinfo
  {author} {\bibfnamefont {S.-J.}\ \bibnamefont {Huang}}, \ and\ \bibinfo
  {author} {\bibfnamefont {M.~A.}\ \bibnamefont {Martin-Delgado}},\ }\href
  {\doibase 10.1103/PhysRevB.99.155118} {\bibfield  {journal} {\bibinfo
  {journal} {Phys. Rev. B}\ }\textbf {\bibinfo {volume} {99}},\ \bibinfo
  {pages} {155118} (\bibinfo {year} {2019})}\BibitemShut {NoStop}%
\bibitem [{\citenamefont {Schmitz}(2019)}]{Schmitz_2018}%
  \BibitemOpen
  \bibfield  {author} {\bibinfo {author} {\bibfnamefont {A.~T.}\ \bibnamefont
  {Schmitz}},\ }\href {\doibase https://doi.org/10.1016/j.aop.2019.167927}
  {\bibfield  {journal} {\bibinfo  {journal} {Annals of Physics}\ }\textbf
  {\bibinfo {volume} {410}},\ \bibinfo {pages} {167927} (\bibinfo {year}
  {2019})}\BibitemShut {NoStop}%
\bibitem [{\citenamefont {Pai}\ and\ \citenamefont {Hermele}(2019)}]{Pai_2019}%
  \BibitemOpen
  \bibfield  {author} {\bibinfo {author} {\bibfnamefont {S.}~\bibnamefont
  {Pai}}\ and\ \bibinfo {author} {\bibfnamefont {M.}~\bibnamefont {Hermele}},\
  }\href@noop {} {\enquote {\bibinfo {title} {Fracton fusion and statistics},}\
  } (\bibinfo {year} {2019}),\ \Eprint {http://arxiv.org/abs/1903.11625}
  {arXiv:1903.11625 [cond-mat.str-el]} \BibitemShut {NoStop}%
\bibitem [{\citenamefont {Yan}\ \emph {et~al.}(2019)\citenamefont {Yan},
  \citenamefont {Benton}, \citenamefont {Jaubert},\ and\ \citenamefont
  {Shannon}}]{Yan_2019}%
  \BibitemOpen
  \bibfield  {author} {\bibinfo {author} {\bibfnamefont {H.}~\bibnamefont
  {Yan}}, \bibinfo {author} {\bibfnamefont {O.}~\bibnamefont {Benton}},
  \bibinfo {author} {\bibfnamefont {L.~D.~C.}\ \bibnamefont {Jaubert}}, \ and\
  \bibinfo {author} {\bibfnamefont {N.}~\bibnamefont {Shannon}},\ }\href@noop
  {} {\enquote {\bibinfo {title} {Rank-2 $u(1)$ spin liquid on the breathing
  pyrochlore lattice},}\ } (\bibinfo {year} {2019}),\ \Eprint
  {http://arxiv.org/abs/1902.10934} {arXiv:1902.10934 [cond-mat.str-el]}
  \BibitemShut {NoStop}%
\bibitem [{\citenamefont {{Shirley}}\ \emph {et~al.}(2019)\citenamefont
  {{Shirley}}, \citenamefont {{Slagle}},\ and\ \citenamefont
  {{Chen}}}]{Shirley_2019}%
  \BibitemOpen
  \bibfield  {author} {\bibinfo {author} {\bibfnamefont {W.}~\bibnamefont
  {{Shirley}}}, \bibinfo {author} {\bibfnamefont {K.}~\bibnamefont {{Slagle}}},
  \ and\ \bibinfo {author} {\bibfnamefont {X.}~\bibnamefont {{Chen}}},\
  }\href@noop {} {\bibfield  {journal} {\bibinfo  {journal} {arXiv e-prints}\ }
  (\bibinfo {year} {2019})},\ \Eprint {http://arxiv.org/abs/1907.09048}
  {arXiv:1907.09048 [cond-mat.str-el]} \BibitemShut {NoStop}%
\bibitem [{\citenamefont {{Fuji}}(2019)}]{Fuji_2019}%
  \BibitemOpen
  \bibfield  {author} {\bibinfo {author} {\bibfnamefont {Y.}~\bibnamefont
  {{Fuji}}},\ }\href@noop {} {\bibfield  {journal} {\bibinfo  {journal} {arXiv
  e-prints}\ } (\bibinfo {year} {2019})},\ \Eprint
  {http://arxiv.org/abs/1908.02257} {arXiv:1908.02257 [cond-mat.str-el]}
  \BibitemShut {NoStop}%
\bibitem [{\citenamefont {Chamon}\ \emph {et~al.}(2019)\citenamefont {Chamon},
  \citenamefont {Green},\ and\ \citenamefont {Yang}}]{Chamon_2019}%
  \BibitemOpen
  \bibfield  {author} {\bibinfo {author} {\bibfnamefont {C.}~\bibnamefont
  {Chamon}}, \bibinfo {author} {\bibfnamefont {D.}~\bibnamefont {Green}}, \
  and\ \bibinfo {author} {\bibfnamefont {Z.-C.}\ \bibnamefont {Yang}},\
  }\href@noop {} {\enquote {\bibinfo {title} {How to emulate quantum spin
  liquids and build topological qubits with available quantum hardware},}\ }
  (\bibinfo {year} {2019}),\ \Eprint {http://arxiv.org/abs/1908.04791}
  {arXiv:1908.04791 [cond-mat.str-el]} \BibitemShut {NoStop}%
\bibitem [{\citenamefont {Pretko}\ and\ \citenamefont
  {Radzihovsky}(2018)}]{Pretko_2018}%
  \BibitemOpen
  \bibfield  {author} {\bibinfo {author} {\bibfnamefont {M.}~\bibnamefont
  {Pretko}}\ and\ \bibinfo {author} {\bibfnamefont {L.}~\bibnamefont
  {Radzihovsky}},\ }\href {\doibase 10.1103/PhysRevLett.120.195301} {\bibfield
  {journal} {\bibinfo  {journal} {Phys. Rev. Lett.}\ }\textbf {\bibinfo
  {volume} {120}},\ \bibinfo {pages} {195301} (\bibinfo {year}
  {2018})}\BibitemShut {NoStop}%
\bibitem [{\citenamefont {Gromov}(2019)}]{Gromov_2019}%
  \BibitemOpen
  \bibfield  {author} {\bibinfo {author} {\bibfnamefont {A.}~\bibnamefont
  {Gromov}},\ }\href {\doibase 10.1103/PhysRevLett.122.076403} {\bibfield
  {journal} {\bibinfo  {journal} {Phys. Rev. Lett.}\ }\textbf {\bibinfo
  {volume} {122}},\ \bibinfo {pages} {076403} (\bibinfo {year}
  {2019})}\BibitemShut {NoStop}%
\bibitem [{\citenamefont {Pai}\ and\ \citenamefont {Pretko}(2018)}]{Pai_2018}%
  \BibitemOpen
  \bibfield  {author} {\bibinfo {author} {\bibfnamefont {S.}~\bibnamefont
  {Pai}}\ and\ \bibinfo {author} {\bibfnamefont {M.}~\bibnamefont {Pretko}},\
  }\href {\doibase 10.1103/PhysRevB.97.235102} {\bibfield  {journal} {\bibinfo
  {journal} {Phys. Rev. B}\ }\textbf {\bibinfo {volume} {97}},\ \bibinfo
  {pages} {235102} (\bibinfo {year} {2018})}\BibitemShut {NoStop}%
\bibitem [{\citenamefont {Kumar}\ and\ \citenamefont
  {Potter}(2019)}]{Kumar_2019}%
  \BibitemOpen
  \bibfield  {author} {\bibinfo {author} {\bibfnamefont {A.}~\bibnamefont
  {Kumar}}\ and\ \bibinfo {author} {\bibfnamefont {A.~C.}\ \bibnamefont
  {Potter}},\ }\href {\doibase 10.1103/PhysRevB.100.045119} {\bibfield
  {journal} {\bibinfo  {journal} {Phys. Rev. B}\ }\textbf {\bibinfo {volume}
  {100}},\ \bibinfo {pages} {045119} (\bibinfo {year} {2019})}\BibitemShut
  {NoStop}%
\bibitem [{\citenamefont {Pretko}\ \emph {et~al.}(2019)\citenamefont {Pretko},
  \citenamefont {Zhai},\ and\ \citenamefont {Radzihovsky}}]{Pretko_2019}%
  \BibitemOpen
  \bibfield  {author} {\bibinfo {author} {\bibfnamefont {M.}~\bibnamefont
  {Pretko}}, \bibinfo {author} {\bibfnamefont {Z.}~\bibnamefont {Zhai}}, \ and\
  \bibinfo {author} {\bibfnamefont {L.}~\bibnamefont {Radzihovsky}},\ }\href
  {\doibase 10.1103/PhysRevB.100.134113} {\bibfield  {journal} {\bibinfo
  {journal} {Phys. Rev. B}\ }\textbf {\bibinfo {volume} {100}},\ \bibinfo
  {pages} {134113} (\bibinfo {year} {2019})}\BibitemShut {NoStop}%
\bibitem [{\citenamefont {Radzihovsky}\ and\ \citenamefont
  {Hermele}(2019)}]{Radzihovsky_2019}%
  \BibitemOpen
  \bibfield  {author} {\bibinfo {author} {\bibfnamefont {L.}~\bibnamefont
  {Radzihovsky}}\ and\ \bibinfo {author} {\bibfnamefont {M.}~\bibnamefont
  {Hermele}},\ }\href@noop {} {\enquote {\bibinfo {title} {Fractons from vector
  gauge theory},}\ } (\bibinfo {year} {2019}),\ \Eprint
  {http://arxiv.org/abs/1905.06951} {arXiv:1905.06951 [cond-mat.str-el]}
  \BibitemShut {NoStop}%
\bibitem [{\citenamefont {Prem}\ \emph {et~al.}(2017)\citenamefont {Prem},
  \citenamefont {Haah},\ and\ \citenamefont {Nandkishore}}]{Prem_2017}%
  \BibitemOpen
  \bibfield  {author} {\bibinfo {author} {\bibfnamefont {A.}~\bibnamefont
  {Prem}}, \bibinfo {author} {\bibfnamefont {J.}~\bibnamefont {Haah}}, \ and\
  \bibinfo {author} {\bibfnamefont {R.}~\bibnamefont {Nandkishore}},\
  }\href@noop {} {\bibfield  {journal} {\bibinfo  {journal} {Phys. Rev. B}\
  }\textbf {\bibinfo {volume} {95}},\ \bibinfo {pages} {155133} (\bibinfo
  {year} {2017})}\BibitemShut {NoStop}%
\bibitem [{\citenamefont {Pai}\ \emph {et~al.}(2019)\citenamefont {Pai},
  \citenamefont {Pretko},\ and\ \citenamefont {Nandkishore}}]{Pai_2019b}%
  \BibitemOpen
  \bibfield  {author} {\bibinfo {author} {\bibfnamefont {S.}~\bibnamefont
  {Pai}}, \bibinfo {author} {\bibfnamefont {M.}~\bibnamefont {Pretko}}, \ and\
  \bibinfo {author} {\bibfnamefont {R.~M.}\ \bibnamefont {Nandkishore}},\
  }\href {\doibase 10.1103/PhysRevX.9.021003} {\bibfield  {journal} {\bibinfo
  {journal} {Phys. Rev. X}\ }\textbf {\bibinfo {volume} {9}},\ \bibinfo {pages}
  {021003} (\bibinfo {year} {2019})}\BibitemShut {NoStop}%
\bibitem [{\citenamefont {Pretko}(2017{\natexlab{c}})}]{Pretko_2017c}%
  \BibitemOpen
  \bibfield  {author} {\bibinfo {author} {\bibfnamefont {M.}~\bibnamefont
  {Pretko}},\ }\href {\doibase 10.1103/PhysRevD.96.024051} {\bibfield
  {journal} {\bibinfo  {journal} {Phys. Rev. D}\ }\textbf {\bibinfo {volume}
  {96}},\ \bibinfo {pages} {024051} (\bibinfo {year}
  {2017}{\natexlab{c}})}\BibitemShut {NoStop}%
\bibitem [{\citenamefont {Yan}(2019{\natexlab{a}})}]{Yan_2019b}%
  \BibitemOpen
  \bibfield  {author} {\bibinfo {author} {\bibfnamefont {H.}~\bibnamefont
  {Yan}},\ }\href {\doibase 10.1103/PhysRevB.99.155126} {\bibfield  {journal}
  {\bibinfo  {journal} {Phys. Rev. B}\ }\textbf {\bibinfo {volume} {99}},\
  \bibinfo {pages} {155126} (\bibinfo {year} {2019}{\natexlab{a}})}\BibitemShut
  {NoStop}%
\bibitem [{\citenamefont {Yan}(2019{\natexlab{b}})}]{Yan_2019c}%
  \BibitemOpen
  \bibfield  {author} {\bibinfo {author} {\bibfnamefont {H.}~\bibnamefont
  {Yan}},\ }\href@noop {} {\enquote {\bibinfo {title} {Hyperbolic fracton
  model, subsystem symmetry, and holography ii: The dual eight-vertex model},}\
  } (\bibinfo {year} {2019}{\natexlab{b}}),\ \Eprint
  {http://arxiv.org/abs/1906.02305} {arXiv:1906.02305 [hep-th]} \BibitemShut
  {NoStop}%
\bibitem [{\citenamefont {You}\ and\ \citenamefont {von
  Oppen}(2019)}]{You_2019}%
  \BibitemOpen
  \bibfield  {author} {\bibinfo {author} {\bibfnamefont {Y.}~\bibnamefont
  {You}}\ and\ \bibinfo {author} {\bibfnamefont {F.}~\bibnamefont {von
  Oppen}},\ }\href {\doibase 10.1103/PhysRevResearch.1.013011} {\bibfield
  {journal} {\bibinfo  {journal} {Phys. Rev. Research}\ }\textbf {\bibinfo
  {volume} {1}},\ \bibinfo {pages} {013011} (\bibinfo {year}
  {2019})}\BibitemShut {NoStop}%
\bibitem [{\citenamefont {Wang}\ \emph {et~al.}(2019)\citenamefont {Wang},
  \citenamefont {Shirley},\ and\ \citenamefont {Chen}}]{Wang_2019}%
  \BibitemOpen
  \bibfield  {author} {\bibinfo {author} {\bibfnamefont {T.}~\bibnamefont
  {Wang}}, \bibinfo {author} {\bibfnamefont {W.}~\bibnamefont {Shirley}}, \
  and\ \bibinfo {author} {\bibfnamefont {X.}~\bibnamefont {Chen}},\ }\href
  {\doibase 10.1103/PhysRevB.100.085127} {\bibfield  {journal} {\bibinfo
  {journal} {Phys. Rev. B}\ }\textbf {\bibinfo {volume} {100}},\ \bibinfo
  {pages} {085127} (\bibinfo {year} {2019})}\BibitemShut {NoStop}%
\bibitem [{\citenamefont {Ma}\ and\ \citenamefont {Pretko}(2018)}]{Ma_2018}%
  \BibitemOpen
  \bibfield  {author} {\bibinfo {author} {\bibfnamefont {H.}~\bibnamefont
  {Ma}}\ and\ \bibinfo {author} {\bibfnamefont {M.}~\bibnamefont {Pretko}},\
  }\href {\doibase 10.1103/PhysRevB.98.125105} {\bibfield  {journal} {\bibinfo
  {journal} {Phys. Rev. B}\ }\textbf {\bibinfo {volume} {98}},\ \bibinfo
  {pages} {125105} (\bibinfo {year} {2018})}\BibitemShut {NoStop}%
\bibitem [{\citenamefont {Castelnovo}\ \emph {et~al.}(2010)\citenamefont
  {Castelnovo}, \citenamefont {Chamon},\ and\ \citenamefont
  {Sherrington}}]{Castelnovo_2010}%
  \BibitemOpen
  \bibfield  {author} {\bibinfo {author} {\bibfnamefont {C.}~\bibnamefont
  {Castelnovo}}, \bibinfo {author} {\bibfnamefont {C.}~\bibnamefont {Chamon}},
  \ and\ \bibinfo {author} {\bibfnamefont {D.}~\bibnamefont {Sherrington}},\
  }\href {\doibase 10.1103/PhysRevB.81.184303} {\bibfield  {journal} {\bibinfo
  {journal} {Phys. Rev. B}\ }\textbf {\bibinfo {volume} {81}},\ \bibinfo
  {pages} {184303} (\bibinfo {year} {2010})}\BibitemShut {NoStop}%
\bibitem [{\citenamefont {Trebst}\ \emph {et~al.}(2007)\citenamefont {Trebst},
  \citenamefont {Werner}, \citenamefont {Troyer}, \citenamefont {Shtengel},\
  and\ \citenamefont {Nayak}}]{Trebst_2007}%
  \BibitemOpen
  \bibfield  {author} {\bibinfo {author} {\bibfnamefont {S.}~\bibnamefont
  {Trebst}}, \bibinfo {author} {\bibfnamefont {P.}~\bibnamefont {Werner}},
  \bibinfo {author} {\bibfnamefont {M.}~\bibnamefont {Troyer}}, \bibinfo
  {author} {\bibfnamefont {K.}~\bibnamefont {Shtengel}}, \ and\ \bibinfo
  {author} {\bibfnamefont {C.}~\bibnamefont {Nayak}},\ }\href {\doibase
  10.1103/PhysRevLett.98.070602} {\bibfield  {journal} {\bibinfo  {journal}
  {Phys. Rev. Lett.}\ }\textbf {\bibinfo {volume} {98}},\ \bibinfo {pages}
  {070602} (\bibinfo {year} {2007})}\BibitemShut {NoStop}%
\bibitem [{\citenamefont {Vidal}\ \emph {et~al.}(2009)\citenamefont {Vidal},
  \citenamefont {Dusuel},\ and\ \citenamefont {Schmidt}}]{Vidal_2009}%
  \BibitemOpen
  \bibfield  {author} {\bibinfo {author} {\bibfnamefont {J.}~\bibnamefont
  {Vidal}}, \bibinfo {author} {\bibfnamefont {S.}~\bibnamefont {Dusuel}}, \
  and\ \bibinfo {author} {\bibfnamefont {K.~P.}\ \bibnamefont {Schmidt}},\
  }\href {\doibase 10.1103/PhysRevB.79.033109} {\bibfield  {journal} {\bibinfo
  {journal} {Phys. Rev. B}\ }\textbf {\bibinfo {volume} {79}},\ \bibinfo
  {pages} {033109} (\bibinfo {year} {2009})}\BibitemShut {NoStop}%
\bibitem [{\citenamefont {Tupitsyn}\ \emph {et~al.}(2010)\citenamefont
  {Tupitsyn}, \citenamefont {Kitaev}, \citenamefont {Prokof'ev},\ and\
  \citenamefont {Stamp}}]{Tupitsyn_2010}%
  \BibitemOpen
  \bibfield  {author} {\bibinfo {author} {\bibfnamefont {I.~S.}\ \bibnamefont
  {Tupitsyn}}, \bibinfo {author} {\bibfnamefont {A.}~\bibnamefont {Kitaev}},
  \bibinfo {author} {\bibfnamefont {N.~V.}\ \bibnamefont {Prokof'ev}}, \ and\
  \bibinfo {author} {\bibfnamefont {P.~C.~E.}\ \bibnamefont {Stamp}},\ }\href
  {\doibase 10.1103/PhysRevB.82.085114} {\bibfield  {journal} {\bibinfo
  {journal} {Phys. Rev. B}\ }\textbf {\bibinfo {volume} {82}},\ \bibinfo
  {pages} {085114} (\bibinfo {year} {2010})}\BibitemShut {NoStop}%
\bibitem [{\citenamefont {Dusuel}\ \emph {et~al.}(2011)\citenamefont {Dusuel},
  \citenamefont {Kamfor}, \citenamefont {Or\'{u}s}, \citenamefont {Schmidt},\
  and\ \citenamefont {Vidal}}]{Dusuel_2011}%
  \BibitemOpen
  \bibfield  {author} {\bibinfo {author} {\bibfnamefont {S.}~\bibnamefont
  {Dusuel}}, \bibinfo {author} {\bibfnamefont {M.}~\bibnamefont {Kamfor}},
  \bibinfo {author} {\bibfnamefont {R.}~\bibnamefont {Or\'{u}s}}, \bibinfo
  {author} {\bibfnamefont {K.~P.}\ \bibnamefont {Schmidt}}, \ and\ \bibinfo
  {author} {\bibfnamefont {J.}~\bibnamefont {Vidal}},\ }\href {\doibase
  10.1103/PhysRevLett.106.107203} {\bibfield  {journal} {\bibinfo  {journal}
  {Phys. Rev. Lett.}\ }\textbf {\bibinfo {volume} {106}},\ \bibinfo {pages}
  {107203} (\bibinfo {year} {2011})}\BibitemShut {NoStop}%
\bibitem [{\citenamefont {Schulz}\ \emph {et~al.}(2012)\citenamefont {Schulz},
  \citenamefont {Dusuel}, \citenamefont {Or{\'{u}}s}, \citenamefont {Vidal},\
  and\ \citenamefont {Schmidt}}]{Schulz_2012}%
  \BibitemOpen
  \bibfield  {author} {\bibinfo {author} {\bibfnamefont {M.~D.}\ \bibnamefont
  {Schulz}}, \bibinfo {author} {\bibfnamefont {S.}~\bibnamefont {Dusuel}},
  \bibinfo {author} {\bibfnamefont {R.}~\bibnamefont {Or{\'{u}}s}}, \bibinfo
  {author} {\bibfnamefont {J.}~\bibnamefont {Vidal}}, \ and\ \bibinfo {author}
  {\bibfnamefont {K.~P.}\ \bibnamefont {Schmidt}},\ }\href {\doibase
  10.1088/1367-2630/14/2/025005} {\bibfield  {journal} {\bibinfo  {journal}
  {New Journal of Physics}\ }\textbf {\bibinfo {volume} {14}},\ \bibinfo
  {pages} {025005} (\bibinfo {year} {2012})}\BibitemShut {NoStop}%
\bibitem [{\citenamefont {Jahromi}\ \emph
  {et~al.}(2013{\natexlab{a}})\citenamefont {Jahromi}, \citenamefont
  {Kargarian}, \citenamefont {Masoudi},\ and\ \citenamefont
  {Schmidt}}]{Jahromi_2013a}%
  \BibitemOpen
  \bibfield  {author} {\bibinfo {author} {\bibfnamefont {S.~S.}\ \bibnamefont
  {Jahromi}}, \bibinfo {author} {\bibfnamefont {M.}~\bibnamefont {Kargarian}},
  \bibinfo {author} {\bibfnamefont {S.~F.}\ \bibnamefont {Masoudi}}, \ and\
  \bibinfo {author} {\bibfnamefont {K.~P.}\ \bibnamefont {Schmidt}},\ }\href
  {\doibase 10.1103/PhysRevB.87.094413} {\bibfield  {journal} {\bibinfo
  {journal} {Phys. Rev. B}\ }\textbf {\bibinfo {volume} {87}},\ \bibinfo
  {pages} {094413} (\bibinfo {year} {2013}{\natexlab{a}})}\BibitemShut
  {NoStop}%
\bibitem [{\citenamefont {Schulz}\ \emph {et~al.}(2013)\citenamefont {Schulz},
  \citenamefont {Dusuel}, \citenamefont {Schmidt},\ and\ \citenamefont
  {Vidal}}]{Schulz_2013}%
  \BibitemOpen
  \bibfield  {author} {\bibinfo {author} {\bibfnamefont {M.~D.}\ \bibnamefont
  {Schulz}}, \bibinfo {author} {\bibfnamefont {S.}~\bibnamefont {Dusuel}},
  \bibinfo {author} {\bibfnamefont {K.~P.}\ \bibnamefont {Schmidt}}, \ and\
  \bibinfo {author} {\bibfnamefont {J.}~\bibnamefont {Vidal}},\ }\href
  {\doibase 10.1103/PhysRevLett.110.147203} {\bibfield  {journal} {\bibinfo
  {journal} {Phys. Rev. Lett.}\ }\textbf {\bibinfo {volume} {110}},\ \bibinfo
  {pages} {147203} (\bibinfo {year} {2013})}\BibitemShut {NoStop}%
\bibitem [{\citenamefont {Schmidt}(2013)}]{Schmidt_2013}%
  \BibitemOpen
  \bibfield  {author} {\bibinfo {author} {\bibfnamefont {K.~P.}\ \bibnamefont
  {Schmidt}},\ }\href {\doibase 10.1103/PhysRevB.88.035118} {\bibfield
  {journal} {\bibinfo  {journal} {Phys. Rev. B}\ }\textbf {\bibinfo {volume}
  {88}},\ \bibinfo {pages} {035118} (\bibinfo {year} {2013})}\BibitemShut
  {NoStop}%
\bibitem [{\citenamefont {Jahromi}\ \emph
  {et~al.}(2013{\natexlab{b}})\citenamefont {Jahromi}, \citenamefont {Masoudi},
  \citenamefont {Kargarian},\ and\ \citenamefont {Schmidt}}]{Jahromi_2013b}%
  \BibitemOpen
  \bibfield  {author} {\bibinfo {author} {\bibfnamefont {S.~S.}\ \bibnamefont
  {Jahromi}}, \bibinfo {author} {\bibfnamefont {S.~F.}\ \bibnamefont
  {Masoudi}}, \bibinfo {author} {\bibfnamefont {M.}~\bibnamefont {Kargarian}},
  \ and\ \bibinfo {author} {\bibfnamefont {K.~P.}\ \bibnamefont {Schmidt}},\
  }\href {\doibase 10.1103/PhysRevB.88.214411} {\bibfield  {journal} {\bibinfo
  {journal} {Phys. Rev. B}\ }\textbf {\bibinfo {volume} {88}},\ \bibinfo
  {pages} {214411} (\bibinfo {year} {2013}{\natexlab{b}})}\BibitemShut
  {NoStop}%
\bibitem [{\citenamefont {Schulz}\ \emph {et~al.}(2014)\citenamefont {Schulz},
  \citenamefont {Dusuel}, \citenamefont {Misguich}, \citenamefont {Schmidt},\
  and\ \citenamefont {Vidal}}]{Schulz_2014}%
  \BibitemOpen
  \bibfield  {author} {\bibinfo {author} {\bibfnamefont {M.~D.}\ \bibnamefont
  {Schulz}}, \bibinfo {author} {\bibfnamefont {S.}~\bibnamefont {Dusuel}},
  \bibinfo {author} {\bibfnamefont {G.}~\bibnamefont {Misguich}}, \bibinfo
  {author} {\bibfnamefont {K.~P.}\ \bibnamefont {Schmidt}}, \ and\ \bibinfo
  {author} {\bibfnamefont {J.}~\bibnamefont {Vidal}},\ }\href {\doibase
  10.1103/PhysRevB.89.201103} {\bibfield  {journal} {\bibinfo  {journal} {Phys.
  Rev. B}\ }\textbf {\bibinfo {volume} {89}},\ \bibinfo {pages} {201103}
  (\bibinfo {year} {2014})}\BibitemShut {NoStop}%
\bibitem [{\citenamefont {Dusuel}\ and\ \citenamefont
  {Vidal}(2015)}]{Dusuel_2015}%
  \BibitemOpen
  \bibfield  {author} {\bibinfo {author} {\bibfnamefont {S.}~\bibnamefont
  {Dusuel}}\ and\ \bibinfo {author} {\bibfnamefont {J.}~\bibnamefont {Vidal}},\
  }\href {\doibase 10.1103/PhysRevB.92.125150} {\bibfield  {journal} {\bibinfo
  {journal} {Phys. Rev. B}\ }\textbf {\bibinfo {volume} {92}},\ \bibinfo
  {pages} {125150} (\bibinfo {year} {2015})}\BibitemShut {NoStop}%
\bibitem [{\citenamefont {Schuler}\ \emph {et~al.}(2016)\citenamefont
  {Schuler}, \citenamefont {Whitsitt}, \citenamefont {Henry}, \citenamefont
  {Sachdev},\ and\ \citenamefont {L\"auchli}}]{Schuler_2016}%
  \BibitemOpen
  \bibfield  {author} {\bibinfo {author} {\bibfnamefont {M.}~\bibnamefont
  {Schuler}}, \bibinfo {author} {\bibfnamefont {S.}~\bibnamefont {Whitsitt}},
  \bibinfo {author} {\bibfnamefont {L.-P.}\ \bibnamefont {Henry}}, \bibinfo
  {author} {\bibfnamefont {S.}~\bibnamefont {Sachdev}}, \ and\ \bibinfo
  {author} {\bibfnamefont {A.~M.}\ \bibnamefont {L\"auchli}},\ }\href {\doibase
  10.1103/PhysRevLett.117.210401} {\bibfield  {journal} {\bibinfo  {journal}
  {Phys. Rev. Lett.}\ }\textbf {\bibinfo {volume} {117}},\ \bibinfo {pages}
  {210401} (\bibinfo {year} {2016})}\BibitemShut {NoStop}%
\bibitem [{\citenamefont {Hal\'{a}sz}\ \emph {et~al.}(2019)\citenamefont
  {Hal\'{a}sz}, \citenamefont {Hsieh},\ and\ \citenamefont
  {Balents}}]{Schotte_2019}%
  \BibitemOpen
  \bibfield  {author} {\bibinfo {author} {\bibfnamefont {G.~B.}\ \bibnamefont
  {Hal\'{a}sz}}, \bibinfo {author} {\bibfnamefont {T.~H.}\ \bibnamefont
  {Hsieh}}, \ and\ \bibinfo {author} {\bibfnamefont {L.}~\bibnamefont
  {Balents}},\ }\href@noop {} {\bibfield  {journal} {\bibinfo  {journal}
  {arXiv:1909.06284}\ } (\bibinfo {year} {2019})}\BibitemShut {NoStop}%
\bibitem [{\citenamefont {Devakul}\ \emph {et~al.}(2018)\citenamefont
  {Devakul}, \citenamefont {Parameswaran},\ and\ \citenamefont
  {Sondhi}}]{Devakul_2018}%
  \BibitemOpen
  \bibfield  {author} {\bibinfo {author} {\bibfnamefont {T.}~\bibnamefont
  {Devakul}}, \bibinfo {author} {\bibfnamefont {S.~A.}\ \bibnamefont
  {Parameswaran}}, \ and\ \bibinfo {author} {\bibfnamefont {S.~L.}\
  \bibnamefont {Sondhi}},\ }\href {\doibase 10.1103/PhysRevB.97.041110}
  {\bibfield  {journal} {\bibinfo  {journal} {Phys. Rev. B}\ }\textbf {\bibinfo
  {volume} {97}},\ \bibinfo {pages} {041110} (\bibinfo {year}
  {2018})}\BibitemShut {NoStop}%
\bibitem [{\citenamefont {Nandkishore}\ and\ \citenamefont
  {Hermele}(2019)}]{Nandkishore18}%
  \BibitemOpen
  \bibfield  {author} {\bibinfo {author} {\bibfnamefont {R.~M.}\ \bibnamefont
  {Nandkishore}}\ and\ \bibinfo {author} {\bibfnamefont {M.}~\bibnamefont
  {Hermele}},\ }\href {\doibase 10.1146/annurev-conmatphys-031218-013604}
  {\bibfield  {journal} {\bibinfo  {journal} {Annual Review of Condensed Matter
  Physics}\ }\textbf {\bibinfo {volume} {10}},\ \bibinfo {pages} {295}
  (\bibinfo {year} {2019})}\BibitemShut {NoStop}%
\bibitem [{\citenamefont {Vijay}\ \emph
  {et~al.}(2016{\natexlab{b}})\citenamefont {Vijay}, \citenamefont {Haah},\
  and\ \citenamefont {Fu}}]{Vijay16}%
  \BibitemOpen
  \bibfield  {author} {\bibinfo {author} {\bibfnamefont {S.}~\bibnamefont
  {Vijay}}, \bibinfo {author} {\bibfnamefont {J.}~\bibnamefont {Haah}}, \ and\
  \bibinfo {author} {\bibfnamefont {L.}~\bibnamefont {Fu}},\ }\href@noop {}
  {\bibfield  {journal} {\bibinfo  {journal} {Physical Review B}\ }\textbf
  {\bibinfo {volume} {94}},\ \bibinfo {pages} {235157} (\bibinfo {year}
  {2016}{\natexlab{b}})}\BibitemShut {NoStop}%
\bibitem [{\citenamefont {Slagle}\ and\ \citenamefont
  {Kim}(2018)}]{Slagle_2018}%
  \BibitemOpen
  \bibfield  {author} {\bibinfo {author} {\bibfnamefont {K.}~\bibnamefont
  {Slagle}}\ and\ \bibinfo {author} {\bibfnamefont {Y.~B.}\ \bibnamefont
  {Kim}},\ }\href {\doibase 10.1103/PhysRevB.97.165106} {\bibfield  {journal}
  {\bibinfo  {journal} {Phys. Rev. B}\ }\textbf {\bibinfo {volume} {97}},\
  \bibinfo {pages} {165106} (\bibinfo {year} {2018})}\BibitemShut {NoStop}%
\bibitem [{\citenamefont {Shirley}\ \emph {et~al.}(2018)\citenamefont
  {Shirley}, \citenamefont {Slagle}, \citenamefont {Wang},\ and\ \citenamefont
  {Chen}}]{Shirley18}%
  \BibitemOpen
  \bibfield  {author} {\bibinfo {author} {\bibfnamefont {W.}~\bibnamefont
  {Shirley}}, \bibinfo {author} {\bibfnamefont {K.}~\bibnamefont {Slagle}},
  \bibinfo {author} {\bibfnamefont {Z.}~\bibnamefont {Wang}}, \ and\ \bibinfo
  {author} {\bibfnamefont {X.}~\bibnamefont {Chen}},\ }\href {\doibase
  10.1103/PhysRevX.8.031051} {\bibfield  {journal} {\bibinfo  {journal} {Phys.
  Rev. X}\ }\textbf {\bibinfo {volume} {8}},\ \bibinfo {pages} {031051}
  (\bibinfo {year} {2018})}\BibitemShut {NoStop}%
\bibitem [{\citenamefont {Shirley}\ \emph {et~al.}(2019)\citenamefont
  {Shirley}, \citenamefont {Slagle},\ and\ \citenamefont
  {Chen}}]{Shirley_2019b}%
  \BibitemOpen
  \bibfield  {author} {\bibinfo {author} {\bibfnamefont {W.}~\bibnamefont
  {Shirley}}, \bibinfo {author} {\bibfnamefont {K.}~\bibnamefont {Slagle}}, \
  and\ \bibinfo {author} {\bibfnamefont {X.}~\bibnamefont {Chen}},\ }\href
  {\doibase https://doi.org/10.1016/j.aop.2019.167922} {\bibfield  {journal}
  {\bibinfo  {journal} {Annals of Physics}\ }\textbf {\bibinfo {volume}
  {410}},\ \bibinfo {pages} {167922} (\bibinfo {year} {2019})}\BibitemShut
  {NoStop}%
\bibitem [{\citenamefont {Ma}\ \emph {et~al.}(2018)\citenamefont {Ma},
  \citenamefont {Schmitz}, \citenamefont {Parameswaran}, \citenamefont
  {Hermele},\ and\ \citenamefont {Nandkishore}}]{Ma17Entropy}%
  \BibitemOpen
  \bibfield  {author} {\bibinfo {author} {\bibfnamefont {H.}~\bibnamefont
  {Ma}}, \bibinfo {author} {\bibfnamefont {A.}~\bibnamefont {Schmitz}},
  \bibinfo {author} {\bibfnamefont {S.}~\bibnamefont {Parameswaran}}, \bibinfo
  {author} {\bibfnamefont {M.}~\bibnamefont {Hermele}}, \ and\ \bibinfo
  {author} {\bibfnamefont {R.~M.}\ \bibnamefont {Nandkishore}},\ }\href@noop {}
  {\bibfield  {journal} {\bibinfo  {journal} {Physical Review B}\ }\textbf
  {\bibinfo {volume} {97}},\ \bibinfo {pages} {125101} (\bibinfo {year}
  {2018})}\BibitemShut {NoStop}%
\bibitem [{\citenamefont {Lake}(2016)}]{Lake}%
  \BibitemOpen
  \bibfield  {author} {\bibinfo {author} {\bibfnamefont {E.}~\bibnamefont
  {Lake}},\ }\href@noop {} {}\bibinfo {howpublished}
  {\url{http://www.physics.utah.edu/~lake/fractons/summer_summary.pdf}}
  (\bibinfo {year} {2016}),\ \bibinfo {note} {accessed: 2018-08-27}\BibitemShut
  {NoStop}%
\bibitem [{\citenamefont {Ma}\ \emph {et~al.}(2017{\natexlab{b}})\citenamefont
  {Ma}, \citenamefont {Lake}, \citenamefont {Chen},\ and\ \citenamefont
  {Hermele}}]{Ma17}%
  \BibitemOpen
  \bibfield  {author} {\bibinfo {author} {\bibfnamefont {H.}~\bibnamefont
  {Ma}}, \bibinfo {author} {\bibfnamefont {E.}~\bibnamefont {Lake}}, \bibinfo
  {author} {\bibfnamefont {X.}~\bibnamefont {Chen}}, \ and\ \bibinfo {author}
  {\bibfnamefont {M.}~\bibnamefont {Hermele}},\ }\href {\doibase
  10.1103/PhysRevB.95.245126} {\bibfield  {journal} {\bibinfo  {journal} {Phys.
  Rev. B}\ }\textbf {\bibinfo {volume} {95}},\ \bibinfo {pages} {245126}
  (\bibinfo {year} {2017}{\natexlab{b}})}\BibitemShut {NoStop}%
\bibitem [{\citenamefont {Williamson}(2016)}]{Williamson_2016}%
  \BibitemOpen
  \bibfield  {author} {\bibinfo {author} {\bibfnamefont {D.~J.}\ \bibnamefont
  {Williamson}},\ }\href {\doibase 10.1103/PhysRevB.94.155128} {\bibfield
  {journal} {\bibinfo  {journal} {Phys. Rev. B}\ }\textbf {\bibinfo {volume}
  {94}},\ \bibinfo {pages} {155128} (\bibinfo {year} {2016})}\BibitemShut
  {NoStop}%
\bibitem [{foo()}]{footnote1}%
  \BibitemOpen
  \href@noop {} {\bibinfo  {journal} {The plaquette Ising model has gained some
  importance in the statistical mechanics of smooth surfaces and as a lattice
  regularization of string theory
  \cite{Savvidy_1994a,Savvidy_1994b,Savvidy_1996,Pietig_1998,Lipowski_2000,Dimopoulos_2002}}\
  }\BibitemShut {NoStop}%
\bibitem [{\citenamefont {Knetter}\ and\ \citenamefont
  {Uhrig}(2000)}]{Knetter_2000}%
  \BibitemOpen
\bibfield  {journal} {  }\bibfield  {author} {\bibinfo {author} {\bibfnamefont
  {C.}~\bibnamefont {Knetter}}\ and\ \bibinfo {author} {\bibfnamefont {G.~S.}\
  \bibnamefont {Uhrig}},\ }\href {\doibase 10.1007/s100510050026} {\bibfield
  {journal} {\bibinfo  {journal} {Eur. Phys. J. B}\ }\textbf {\bibinfo {volume}
  {13}},\ \bibinfo {pages} {209} (\bibinfo {year} {2000})}\BibitemShut
  {NoStop}%
\bibitem [{\citenamefont {Knetter}\ \emph {et~al.}(2003)\citenamefont
  {Knetter}, \citenamefont {Schmidt},\ and\ \citenamefont
  {Uhrig}}]{Knetter_2003}%
  \BibitemOpen
  \bibfield  {author} {\bibinfo {author} {\bibfnamefont {C.}~\bibnamefont
  {Knetter}}, \bibinfo {author} {\bibfnamefont {K.~P.}\ \bibnamefont
  {Schmidt}}, \ and\ \bibinfo {author} {\bibfnamefont {G.~S.}\ \bibnamefont
  {Uhrig}},\ }\href {\doibase 10.1088/0305-4470/36/29/302} {\bibfield
  {journal} {\bibinfo  {journal} {J. Phys. A: Math. Gen.}\ }\textbf {\bibinfo
  {volume} {36}},\ \bibinfo {pages} {7889} (\bibinfo {year}
  {2003})}\BibitemShut {NoStop}%
\bibitem [{\citenamefont {Coester}\ and\ \citenamefont
  {Schmidt}(2015)}]{Coester_2015}%
  \BibitemOpen
  \bibfield  {author} {\bibinfo {author} {\bibfnamefont {K.}~\bibnamefont
  {Coester}}\ and\ \bibinfo {author} {\bibfnamefont {K.~P.}\ \bibnamefont
  {Schmidt}},\ }\href {\doibase 10.1103/physreve.92.022118} {\bibfield
  {journal} {\bibinfo  {journal} {Phys. Rev. E}\ }\textbf {\bibinfo {volume}
  {92}},\ \bibinfo {pages} {022118} (\bibinfo {year} {2015})}\BibitemShut
  {NoStop}%
\bibitem [{\citenamefont {Guttmann}(1989)}]{Guttmann1989}%
  \BibitemOpen
  \bibfield  {author} {\bibinfo {author} {\bibfnamefont {A.~C.}\ \bibnamefont
  {Guttmann}},\ }\href@noop {} {\emph {\bibinfo {title} {Phase Transitions and
  Critical Phenomena}}},\ edited by\ \bibinfo {editor} {\bibfnamefont
  {C.}~\bibnamefont {Domb}}\ and\ \bibinfo {editor} {\bibfnamefont
  {J.}~\bibnamefont {Lebowitz}},\ Vol.~\bibinfo {volume} {13}\ (\bibinfo
  {publisher} {Academic Press},\ \bibinfo {address} {New York},\ \bibinfo
  {year} {1989})\BibitemShut {NoStop}%
\bibitem [{\citenamefont {R\"ochner}\ \emph {et~al.}(2016)\citenamefont
  {R\"ochner}, \citenamefont {Balents},\ and\ \citenamefont
  {Schmidt}}]{Roechner2016}%
  \BibitemOpen
  \bibfield  {author} {\bibinfo {author} {\bibfnamefont {J.}~\bibnamefont
  {R\"ochner}}, \bibinfo {author} {\bibfnamefont {L.}~\bibnamefont {Balents}},
  \ and\ \bibinfo {author} {\bibfnamefont {K.~P.}\ \bibnamefont {Schmidt}},\
  }\href {\doibase 10.1103/PhysRevB.94.201111} {\bibfield  {journal} {\bibinfo
  {journal} {Phys. Rev. B}\ }\textbf {\bibinfo {volume} {94}},\ \bibinfo
  {pages} {201111} (\bibinfo {year} {2016})}\BibitemShut {NoStop}%
\end{thebibliography}%

\end{document}